\documentclass[a4paper,11pt]{article}
\pdfoutput=1 

\usepackage{jcappub} 

\usepackage[T1]{fontenc} 

\usepackage{hyperref}
\usepackage[table]{xcolor}
\usepackage{amsmath, bm}
\usepackage{amssymb}
\usepackage{mathtools}
\usepackage{mathrsfs}
\usepackage{enumerate}
\usepackage{feynmp-auto}
\usepackage{cleveref}
\usepackage{float}
\usepackage{subcaption}
\usepackage{empheq}
\usepackage[tikz]{bclogo}  
\usepackage[most]{tcolorbox}  
\usepackage[normalem]{ulem}

\definecolor{myblue}{RGB}{0, 102, 204}  
\definecolor{lightgray}{gray}{0.9}  

\newcommand{\mx}{M_{\rm X}}
\newcommand{\mn}{M_{\rm N}}
\newcommand{\mdm}{M_{\rm \chi}}
\newcommand{\tdom}{T_{\rm dom}}
\newcommand{\hdom}{H_{\rm dom}}
\newcommand{\tdec}{T_{\rm dec}}
\newcommand{\tf}{T_{\rm f}}

\newcommand{\nt}{n_{\rm T}}
\newcommand{\trh}{T_{\rm RH}}
\newcommand{\keq}{k_\text{eq}}
\newcommand{\krh}{k_\text{RH}}
\newcommand{\kdec}{k_\text{dec}}
\newcommand{\kdecs}{k_\text{dec,S}}
\newcommand{\krhs}{k_\text{RH,S}}
\newcommand{\sigmav}{\langle\sigma v\rangle_{\rm ann}}
\newcommand{\br}{{\rm Br}_{{\rm N}\to {\rm \chi}}}

\title{Primordial gravitational waves as complementary probe of dark matter indirect detection}

\author[a]{Anish Ghoshal,}
\author[b]{Debarun Paul,}
\author[b]{Supratik Pal}

\affiliation[a]{\,Institute of Theoretical Physics, Faculty of Physics, \\ University of Warsaw, ul. Pasteura 5, 02-093 Warsaw, Poland}
\affiliation[b]{\,Physics and Applied Mathematics Unit, Indian Statistical Institute, \\ 203 B.T. Road, Kolkata 700108, India }

\emailAdd{anish.ghoshal@fuw.edu.pl}
\emailAdd{debarun31paul@gmail.com}
\emailAdd{supratik@isical.ac.in}

\abstract{
We propose a novel cosmological probe of dark matter (DM) through inflationary primordial gravitational wave (GW) measurements in a wide range of GW missions, highlighting its complementarity with traditional indirect detection. In scenarios like early matter domination (EMD), the thermal DM relic is diluted and then replenished via non-thermal production. We derive the critical branching ratio required for this mechanism, and find that even very a tiny value of this may yield the correct relic abundance of DM, that leaves a characteristic imprints on the primordial GW spectrum, inducing frequency-dependent suppressions in the GW amplitudes. 
By analysing signal-to-noise ratio (SNR) and employing Fisher forecast, we show that upcoming GW experiments have good potential to probe the DM parameter space involving its mass and annihilation cross-section, which is allowed by the current indirect searches.
We show, for instance, LISA will be sensitive to DM mass range $[2\times 10^2-10^5]$ GeV. Furthermore, we identify a significant overlap of the GW missions' sensitivity reaches with the projected reach of future indirect searches like CTA with gamma rays, ANTARES, KM3NeT with neutrinos.
In those overlapping regions of interests, we forecast on the GW experiments to estimate the precision of measurements.
We show, for instance, that DM mass of $10^5$ GeV with an annihilation cross-section of $10^{-24}~{\rm cm}^3{\rm /s}$, and a mass of $10^4$ GeV with an annihilation cross-section of $2\times10^{-25}~{\rm cm}^3{\rm /s}$, lie within the projections of CTA. We find that whilst the former can be probed by ET with $\sim 1\%$ uncertainties, the latter can be probed by $\mu$-ARES with $\sim 7 \%$ uncertainties. Similarly, DM mass of $10^5$ GeV, with cross-section $10^{-23}~{\rm cm}^3{\rm /s}$ lies within the projection of ANTARES and KM3NeT, which can be probed by ET with $\sim 1\%$ uncertainties. Our work opens up a new avenue to explore where GW observations and indirect detection together offer complementary probes of DM.
}

\begin{document}
\maketitle
\flushbottom

\section{Introduction}
\label{sec:intro}

The Cosmic inflation provides seeds of initial density fluctuations for large-scale structures and quite naturally fixes the flatness and horizon issues~\cite{Brout:1977ix,Sato:1980yn,Guth:1980zm,Linde:1981mu,Starobinsky:1982ee} in the form of minute temperature fluctuations in the cosmic microwave background (CMB) measurements \cite{Planck:2018vyg}.
From several cosmological observations the content of our present Universe in the form of dark matter (DM), dark energy (DE) and radiation is now well established leading to what is known as the ``Standard Model of Cosmology''. Furthermore, we are able to identify the class of inflation models using the latest constraints on the existence of tensor modes in the CMB and the improved observations of the scalar perturbation modes. This being said, nevertheless still the thermal and non-thermal history of the Universe from the end of cosmic inflation to the period of hot big bang period of the cosmic history eludes any sort of experimental constraints. 
All we know is the big bang nucleosynthesis (BBN) data that constrains the Standard Model (SM) particles without much information about the evolution history of the Universe prior to this period.
Consequently, the evolution of the metric perturbation modes from their formation during inflation to the present is challenging to examine and is mostly unknown. 

In the {\it standard model} of Cosmology, unless otherwise stated explicitly it is always assumed that the cosmic inflation era culminates into a radiation-dominated (RD) era, which is also sometime known as the hot big bang epoch of the cosmic history. 
The metric perturbations during inflation gets stretched and goes out of the horizon which remain frozen until perturbation modes began to rise linearly with the expansion. If the initial inflationary spectrum produces a nearly scale-invariant spectrum of first-order tensor perturbations, these tensors are expected to remain subdominant compared to the sensitivity of present GW detectors~\cite{Chen:2024roo}, therefore it is quite likely that, in the scenario when the Universe enters a phase of the radiation-dominated era immediately  
after the end of cosmic inflation will not lead to any large detectable GW signal in the near future. However, this is not always true as we will describe below, in several scenarios involving non-standard cosmic evolution
 as has been extensively discussed in the literature  \cite{Bernal:2020ywq, Ghoshal:2023sfa, Ghoshal:2022ruy, Chen:2024roo, Ghoshal:2024gai}. 
Let us understand some of the challenges to the minimal scenario:
the inflaton must decay instantly into standard radiation species to release all of its energy density, as soon as it leaves {\em slow roll} phase of inflation. In order facilitate this one needs a very fast decay of the inflaton field which in turn requires the inflaton to significantly carry large interaction terms between the inflaton field and SM  fields. However, as we know, significant interactions between the inflaton and the SM are not motivated for a number of reasons. For example, they destroy the potential's flatness or the inflationary dynamics \cite{Buchmuller:2014pla,Buchmuller:2015oma,Argurio:2017joe,Heurtier:2019eou}.
Moreover in order for the decay to occur properly after inflation ends, the inflaton field typically also needs to spend quite a bit of time oscillating around the minimum of its potential. This is because only when it does this its coherent oscillations can quickly get transferred to SM particles via particle production. Overall this scenario assumes the fact that the inflation potential minimum lives closeby which is generically not true in all inflationary scenarios. In fact often one encounters inflationary paradigms where inflaton cannot lead to oscillation around a minimum and one needs a separate species like a second reheaton (or curvaton) field or other mechanisms like gravitational or primordial black hole reheating. Frequently, in several well studied scenarios the inflaton concludes with an early matter dominance phase, which corresponds to an equation-of-state (EoS) parameter that is less stiff than the radiation-dominated one, which we denote with the barotropic parameter $w = 0$. For convenience, we will call this time period the \textit{`matter-like'} era in the following.

Besides the not-so-well-known pre-BBN history of the Universe, another puzzle of modern particle physics and cosmology is the the origin and composition of DM~\cite{Jungman:1995df,Bertone:2004pz,Feng:2010gw}. Since the last 30 years there have been significant progress and efforts no the experimental frontiers in DM searches  ~\cite{Lee:1977ua,Scherrer:1985zt,Srednicki:1988ce,Gondolo:1990dk}. Searches of DM include direct-detection experiments which can detect signals when a DM comes and scatters off nuclei ~\cite{PandaX-II:2017hlx,PandaX:2018wtu,XENON:2020kmp,XENON:2018voc,LUX-ZEPLIN:2018poe,DARWIN:2016hyl}, indirect detection via DM annihilation into various final measurable final states like gamma rays and neutrinos~\cite{HESS:2016mib,MAGIC:2016xys} and through direct production in high energy particle colliders (\textit{e.g.}~at the Large Hadron Collider (LHC)~\cite{ATLAS:2017bfj,CMS:2017zts}). Inspite of these huge efforts in several experiments, nonetheless some of the most simple and elegant models attributing to a dark matter particle, for instance, one that freezes out from the SM plasma in early universe also known as the ``WIMP miracle'' (Weakly Interacting Massive Particle) have eluded detection. It is in this situation that has led to several other ideas for the DM formation mechanisms such as the having DM freeze-out in modified cosmic expansion history like for instance in an early matter dominated era. Alongside this, several other open problems in the SM of particle physics, e.g., the microscopic origin of the matter-antimatter asymmetry, the flavour puzzle in the SM matter sector, or the ultraviolet (IV) dynamics in the Higgs scalar sector in early universe, (see \textit{e.g.} Ref.~\cite{Gouttenoire:2022gwi} for a review) poses significant challenge to the standard thermal history of the universe, that is, the universe being radiation-dominated in pre-BBN era. Some well-known examples are those of long-lived heavy meta-stable particle generating a period of early matter domination~\cite{McDonald:1989jd,Moroi:1999zb, Visinelli:2009kt,Erickcek:2015jza,  Nelson:2018via,Cheek:2023fht,Cirelli:2018iax,Gouttenoire:2019rtn,Allahverdi:2020bys, Allahverdi:2021grt,Allahverdi:2022zqr}, a rapidly scalar field generating a kination era \cite{Spokoiny:1993kt,Joyce:1996cp,Peebles:1998qn,Poulin:2018dzj,Gouttenoire:2021jhk,Gouttenoire:2021wzu,Co:2021lkc,Ghoshal:2022ruy,Heurtier:2022rhf,Chen:2025awt}, or a period of supercooled phase transition \cite{Guth:1980zk,Witten:1980ez,Creminelli:2001th,Randall:2006py,Konstandin:2011dr,Baratella:2018pxi,Baldes:2020kam,Baldes:2021aph}.

The LIGO-Virgo collaboration \cite{LIGOScientific:2016aoc, LIGOScientific:2016sjg} detected gravitational waves (GWs) from mergers of black hole mergers few years ago. Very recently an observation of a very strong evidence for a stochastic GW background (SGWB) from several pulsar timing array (PTA) collaborations was reported~\cite{Carilli:2004nx, Janssen:2014dka, Weltman:2018zrl, EPTA:2015qep, EPTA:2015gke, NANOGrav:2023gor, NANOGrav:2023hvm}. These discoveries have inspired a wide range of beyond standard model (BSM) scenarios involving possibilities of cosmological GW production in the early Universe, including those of primordial origins of GWs, for instance arising during the cosmic inflationary epoch.  Our analysis here proposes to form a synergy between GW observations and dark matter indirect detection searches. We will particularly study the tensor fluctuations of the metric during cosmic inflation which are stretched out of the Hubble horizon and later on re-enter and propagate as primordial gravitational waves (PGWs) in the post-inflationary era~\cite{Grishchuk:1974ny, Starobinsky:1979ty, Rubakov:1982df, Guzzetti:2016mkm}. These PGWs while they propagate their amplitude and frequency spectrum is very sensitive to the thermal cosmic history, due to red-shifting and entropy injection, therefore PGWs serves as a record of the cosmic expansion history of our Universe in early times~\cite{Seto:2003kc, Boyle:2005se, Boyle:2007zx, Kuroyanagi:2008ye, Nakayama:2009ce, Kuroyanagi:2013ns, Jinno:2013xqa, Saikawa:2018rcs, Chen:2024roo}. Moreover they are scale-invariant. Therefore, any tiny deviation from the traditional radiation-dominated pre-BBN era can be judiciously investigated by studying the features in the PGW spectral shapes, in the present day~\cite{Nakayama:2008ip, Nakayama:2008wy, Kuroyanagi:2011fy, Buchmuller:2013lra, Buchmuller:2013dja, Jinno:2014qka, Kuroyanagi:2014qza, Bernal:2020ywq, Ghoshal:2023sfa, Ghoshal:2022ruy, Chen:2024roo, Ghoshal:2024gai, Maity:2024cpq, Haque_2021, Chakraborty:2023ocr}. In this regard, we consider an epoch where a heavy meta-stable particle attributes to the significant and dominant energy component of the Universe, and then decays. The DM formation occurs during this era. Since this affects the post-inflationary evolution, it consequently leads to characteristic GW spectral shapes which can be seen in various current and upcoming GW detectors. Such primordial features in the GW spectrum usually tells us about two important quantities: \textit{(a)} the time of onset of the early matter domination; this is basically determined by the highest frequency of the PGW which goes away from a flat spectrum, and \textit{(b)} the time duration of early matter-domination era, this is inferred from the width of the characteristic feature in frequency space~\cite{Berbig:2023yyy, Borboruah:2024eha, Cheek:2025gvx, Borboruah:2024eal,  Bernal:2020ywq, Datta:2022tab, Datta:2023vbs, Chianese:2024nyw, Ghoshal:2022ruy, Datta:2025yow, Borboruah:2025hai, Cheek:2025gvx}. We will show that these two quantities in our case will reveal about DM mass and its annihilation cross-section rate.  

Dark matter detection in several experiments only constrain the mass and annihilation cross-section in some regions of the parameter space due to the possible experimental sensitivities and the limited scope of reaching very high energies. It is thus very important and necessary to find complementary sources of knowledge about such dark matter searches. We propose, in this paper, that primordial features in the SGWB from the inflationary paradigm will be useful in obtaining independent evidence for such information on DM physics. This suggests a potential synergy between GW observatories and DM search experiments in testing such a beyond-the-standard-model (BSM) scenario.

\textit{The paper is organized as follows:} In Sec.~\ref{sec:EMD} we discuss how EMD is generated via long-lived particles. We illustrates its impact on the DM relic abundance. In Sec.~\ref{sec:pgw}, we review the imprints of the EMD era on the primordial GW spectrum and discuss how these signals can be detected by future GW missions. Sec.~\ref{sec:fisher} presents the Fisher matrix forecast analysis for estimating the uncertainties on the measurements for the various GW detectors. After that in Sec.~\ref{sec:complementary}, we highlight complementarity between GW experiments and traditional DM indirect search in probing DM parameter space. Finally, we summarize our findings and conclude in Sec.~\ref{sec:conclusion}.

\section{Early matter domination with long-lived particles}
\label{sec:EMD}
An epoch of early matter domination (EMD) can naturally emerge due to the presence of any long-lived standard model (SM) singlet particle, which we denote here by $N$~\cite{Allahverdi:2021grt,Allahverdi:2022zqr}. We assume that $N$ initially thermalizes through interactions with the SM thermal bath \footnote{See Ref. \cite{Borboruah:2025hai} for detailed justification in the parameter space for this assumption needed.}. As the Universe cools down, $N$ becomes non-relativistic and begins to behave as pressure-less matter. If it remains sufficiently long-lived, its energy density can eventually exceed that of radiation, giving rise to a phase of EMD. This matter-dominated phase concludes when $N$ decays, restoring radiation domination and setting the stage for successful Big Bang Nucleosynthesis (BBN).

\subsection{Conditions for early matter domination}
\label{sec:EMD_condition}
To characterize this thermal history quantitatively, we consider the coupled dynamics of $N$, radiation, and dark matter $\chi$ , with $\mdm$ and $\sigmav$ being the mass and annihilation cross-section of the DM, respectively. In addition to its interaction with the thermal bath, $N$ also couples to the dark sector, and its decay contributes to the DM abundance. Taking into account both annihilation and decay processes, the relevant Boltzmann equations for the energy and number densities of $N$, DM, and radiation are given by
\begin{eqnarray}
\frac{d \rho_N}{dt} &=& -3H \rho_N-\Gamma_N \rho_N
\label{eq:boltz_N} \;,  \\
\frac{d n_{\rm \chi}}{dt} &=& -3H n_{\rm \chi} + \br \Gamma_N n_{\rm \chi} -\sigmav  \left[ n_{\rm \chi}^2 - \left({n_{\rm \chi}^{eq}}\right)^2\right]\; , \label{eq:boltz_DM}\\
\frac{d \rho_R}{dt} &=& -4H \rho_R +  (1-\br)\Gamma_N n_N  + 2 \mdm\sigmav \left[ n_{\rm \chi}^2 - \left({n_{\rm \chi}^{eq}}\right)^2\right]
\label{eq:boltz_R} \;,
\end{eqnarray}
where the subscript ``eq'' denotes equilibrium number density, $H$ is the Hubble parameter, and $\br$ denotes the branching ratio of $N$ decays into DM which is defined as the ratio between decay width of $N$ into DM ($\Gamma_{N\to\chi}$) and the total decay width of $N$, $\Gamma_N$, \textit{i.e.} $\br\equiv\frac{\Gamma_{N\to\chi}}{\Gamma_N}$. Here, $\rho_i$ and $n_i$ represent the energy and number densities of species $i$, respectively, and $n_{\chi}^{eq}$ is the equilibrium number density of the DM. 

\begin{enumerate}[a)]
    \item \textbf{Onset of early matter domination} 
    
    The onset of EMD is marked by the transition of $N$ into a non-relativistic regime, typically around $T \sim \mn$ \cite{Berbig:2023yyy}. At this stage, the energy density of $N$ redshifts more slowly than that of radiation, scaling as $a^{-3}$ compared to $a^{-4}$. As a result, $N$ can eventually come to dominate the total energy density of the Universe. The temperature at which this transition occurs is denoted by $\tdom$, and the corresponding Hubble expansion rate is $\hdom$. For EMD to be realized, $N$ must be long-lived enough that it does not decay before reaching this point, which requires $\Gamma_N < \hdom$. The condition $\rho_N \simeq \rho_R$, on-setting the EMD, can be used to calculate $\hdom$ as~\cite{Allahverdi:2021grt}
    \begin{equation}
    \hdom \simeq \frac{4 g_\ast(\tdom)}{g_\ast(T = \mn)} H(T = \mn).
    \end{equation}
    During the period in which $H$ decreases from $H(T = \mn)$ to $\hdom$, the comoving entropy is approximately conserved, allowing us to relate $\tdom$ to $\hdom$ through~\cite{Allahverdi:2021grt}
    \begin{equation}
    \label{eq:tdom}
    \tdom^3 \simeq \frac{g_{\ast s}(T = \mn)}{g_{\ast s}(\tdom)} \mn^3 \left(\frac{\hdom}{H(T = \mn)}\right)^{3/2}.
    \end{equation}

    \item \textbf{Conclusion of early matter domination}
    
    Following this matter-dominated phase, the Universe must go back to radiation domination in order for BBN to proceed unimpeded, as mentioned earlier. This transition is triggered by the decay of $N$, which occurs when $H \simeq \Gamma_N$. The temperature at which this decay takes place is denoted by $\tdec$, and can be approximated as
    \begin{equation}
    \label{eq:tdec}
    \tdec \simeq \left(\frac{90}{\pi^2 g_\ast(\tdec)}\right)^{1/4} \sqrt{\Gamma_N M_P}.
    \end{equation}
    Here, $M_P\simeq 2.435\times10^{18}~{\rm GeV}$ is the reduced Planck mass. To ensure consistency with BBN, $N$ must decay before nucleosynthesis begins. This imposes a lower bound on the decay temperature, typically $\tdec \gtrsim 4$ MeV, corresponding to $\Gamma_N \gtrsim H_{\rm BBN} \sim 10~\text{s}^{-1}$.

\end{enumerate}

\noindent Thus, the two critical conditions for the successful realization of an EMD epoch are: (i) the decay rate of $N$ must be small enough to allow it to dominate the energy density before decaying ($\Gamma_N < \hdom$), and (ii) it must decay early enough to not disrupt BBN ($\Gamma_N > H_{\rm BBN}$).

During the EMD epoch, i.e., for $\tdom > T > \tdec$, the Hubble parameter scales as $H \propto a^{-3/2}$, characteristic of matter domination. However, the thermal evolution differs from standard radiation domination: the SM temperature evolves as $T \propto a^{-3/8}$ due to continuous entropy injection from $N$ decays, in contrast to $T \propto a^{-1}$ for pure radiation, in the absence of such injection.

After $N$ has decayed completely, the Universe is reheated, and the entropy associated with the decay products dilutes the pre-existing relics. This dilution is quantified by the entropy dilution factor, defined as the ratio of comoving entropy before and after the decay
\begin{equation}
\label{eq:dilution_factor}
\Delta_s \equiv \frac{S(\tdec)}{S(\tdom)} \simeq \frac{g_{\ast s}(\tdec)}{g_{\ast s}(\tdom)} \frac{g_\ast(\tdom)}{g_\ast(\tdec)} \frac{\tdom}{\tdec},
\end{equation}
where $S(T) \equiv s(T)a(T)^3$ denotes the comoving entropy density.

For clarity, we define the lifetime of $N$ as $\tau_N \equiv \Gamma_N^{-1}$. The conditions discussed above not only determine the dynamics of the EMD epoch but also constrain the viable parameter space for any scenario involving late-decaying particles coupled to both visible and dark sectors.

\medskip
\subsection{Impact of early matter domination on dark matter annihilation}
\label{subsec:DM_ann}

If the dark matter (DM) particle, with mass $\mdm$, freezes out at a temperature $\tf$ before the onset of early matter domination (EMD), \textit{i.e.}, $\tf > \tdom$, its relic abundance is subject to dilution by the entropy released during the EMD epoch. In contrast, if freeze-out occurs after EMD ends (\textit{i.e.}, $\tf < \tdec$), the standard radiation-dominated (RD) thermal history resumes, and the relic abundance is unaffected by the earlier EMD phase.

We consider, in this work, the scenario where $\tf > \tdom$, which dilutes the DM abundance. This dilution is compensated by a non-thermal contribution from the decay of long-lived particles $N$, which produce DM during EMD. To ensure that $N$ decays into DM and not the reverse process, we consider $\mn > \mdm$. The observed DM abundance emerges from the interplay between this non-thermal injection and subsequent DM annihilation~\cite{Allahverdi:2022zqr}.

Following the decay of $N$ at temperature $\tdec$, the injected DM may annihilate depending on its number density and the annihilation cross-section $\sigmav$, ~\cite{Dutta:2009uf,Acharya:2009zt}. Unlike earlier works, our scenario includes both contributions to the DM number density: \textit{(i)} thermally frozen-out DM diluted by EMD, and \textit{(ii)} DM produced from $N$ decay. The total DM number density at $T = \tdec$ is denoted by $n_{\rm DM}$, capturing the combined effect.

At the time of $N$ decay, we assume that the annihilation rate of DM is comparable to the Hubble rate, \textit{i.e.},
\begin{equation}
\label{eq:non_therm_DM_clean}
n_{\rm DM}(\tdec)\, \sigmav \simeq H(\tdec),
\end{equation}
which ensures that annihilation efficiently reduces any DM overproduction. Since the temperature $\tdec$ marks the onset of radiation domination, the Hubble rate scales as $H(\tdec) \sim \tdec^2$, while the entropy density behaves as $s(\tdec) \sim \tdec^3$. Consequently, the DM-to-entropy ratio at the end of the EMD epoch, and hence the DM relic abundance, scales as 
\begin{equation}
\label{eq:non_therm_DM_abundance_clean}
\Omega_{\rm DM}(\tdec) \sim \frac{n_{\rm DM}(\tdec)}{s(\tdec)} \sim \frac{H(\tdec)}{\sigmav \tdec^3} \sim \frac{1}{\sigmav \tdec}.
\end{equation}
Beyond $\tdec$, no further entropy injection or DM production occurs, and the comoving DM abundance remains fixed.

By contrast, in the standard cosmological history characterized by immediate radiation domination after reheating, DM undergoes thermal freeze-out with an  annihilation cross-section of around $\sigmav \simeq 3 \times 10^{-26}\, \text{cm}^3/\text{s}$, which yields the observed relic density of DM~\cite{Planck:2018vyg}. In this case, the relic abundance scales as
\begin{equation}
\label{eq:standard_DM_abundance_clean}
\Omega_{\rm DM}^{\rm std}(\tf) \sim \frac{1}{(3 \times 10^{-26}\, \text{cm}^3/\text{s}) \, \tf}.
\end{equation}
Since freeze-out marks the point at which the DM comoving density becomes conserved, $\Omega_{\rm DM}^{\rm std}(\tf)$ equals the present-day observed value, $\Omega_{\rm DM}^{\rm obs}$.

A comparison of Eqs.~\eqref{eq:non_therm_DM_abundance_clean} and \eqref{eq:standard_DM_abundance_clean} yields the relation
\begin{equation}
\label{eq:dm_underabundance_clean}
\Omega_{\rm DM} = \left( \frac{\tf}{\tdec} \right) \left( \frac{3 \times 10^{-26}\, \text{cm}^3/\text{s}}{\sigmav} \right) \Omega_{\rm DM}^{\rm obs},
\end{equation}
from which we infer the condition for achieving the observed relic abundance of DM in this scenario, to be
\begin{equation}
\label{eq:sigmav_tdec_relation}
\tdec = \left( \frac{3 \times 10^{-26}\, \text{cm}^3/\text{s}}{\sigmav} \right) \tf.
\end{equation}
This relation demonstrates that a smaller $\sigmav$ must be compensated by a later $N$ decay (smaller $\tdec$) in order to reproduce the correct relic density.

The consistency of Eq.\eqref{eq:non_therm_DM_clean} requires two simultaneous conditions: (i) the branching ratio $\br$ of $N$ decays into DM must be sufficiently large to generate the required non-thermal DM population, and (ii) the annihilation cross-section must exceed the standard thermal value, i.e., $\sigmav > 3 \times 10^{-26}\, \text{cm}^3/\text{s}$ \cite{Allahverdi:2022zqr}. The first requirement introduces a critical branching ratio, $\br^c$, above which DM production is sufficient to trigger efficient annihilation:
\begin{equation}
\label{eq:br_critical_clean}
\br^c = \frac{4}{3} \left( \frac{\mn}{\tdec} \right) \left( \frac{\alpha_\ast}{\alpha_{\ast s}} \right) \frac{1}{M_P \sigmav} \left[ \frac{1}{\tdec} - \frac{\tdec}{\tdom} \frac{1}{\tilde{\tf}} \right],
\end{equation}
with $\alpha_{\ast}\equiv\sqrt{\frac{\pi^2g_{\ast}(\tdec)}{90}}$ and $\alpha_{\ast s}\equiv\frac{2\pi^2}{45}g_{\ast s}(\tdec)$, where $g_{\ast}$ and $g_{\ast s}$ are the number of relativistic degrees of freedom contributing to energy density and entropy density, respectively.
Here $\tilde{\tf}$ denotes the freeze-out temperature corresponding to an arbitrary $\sigmav$, whereas $\tf$ refers specifically to the value yielding the observed relic density in the absence of EMD. A more detailed derivation of $\br^c$ is provided in Appendix~\ref{app:critical_br}. As an illustrative example, for $(\tau_N, \mn) = (0.1\, \text{s}, 5\, \text{TeV})$ and $(\mdm, \sigmav) = (100\, \text{GeV}, 10^{-22}\, \text{cm}^3/\text{s})$, we find that $\br^c \sim 10^{-5}$ which is a relatively small value, suggesting that non-thermal DM production can be significant even with modest branching fractions.

\begin{figure}[H]
    \centering
        \includegraphics[height=5.5cm,width=7cm]{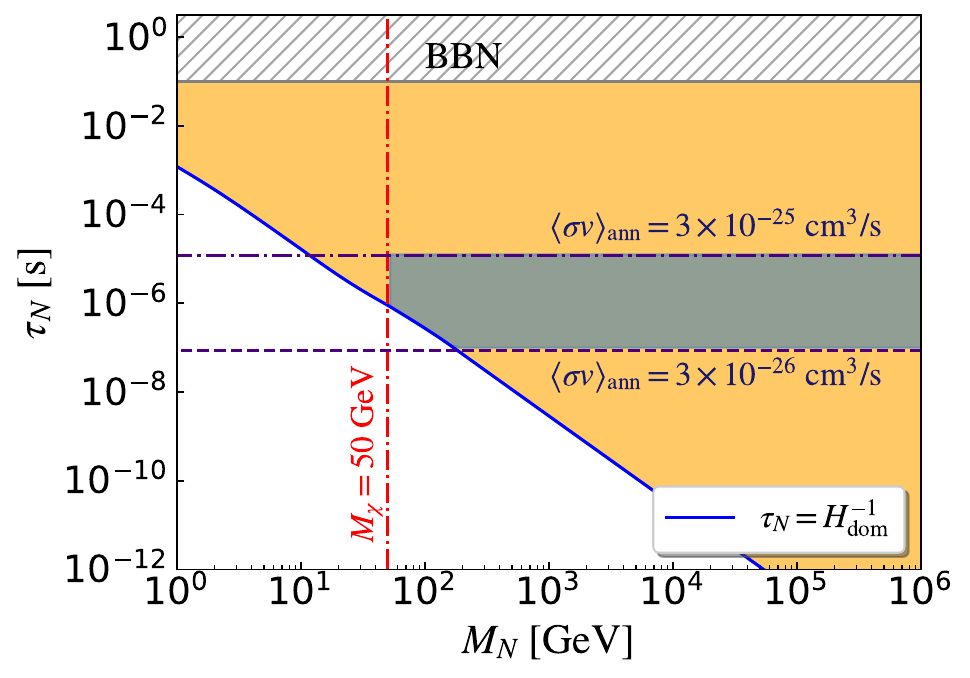}
        \includegraphics[height=5.5cm,width=8cm]{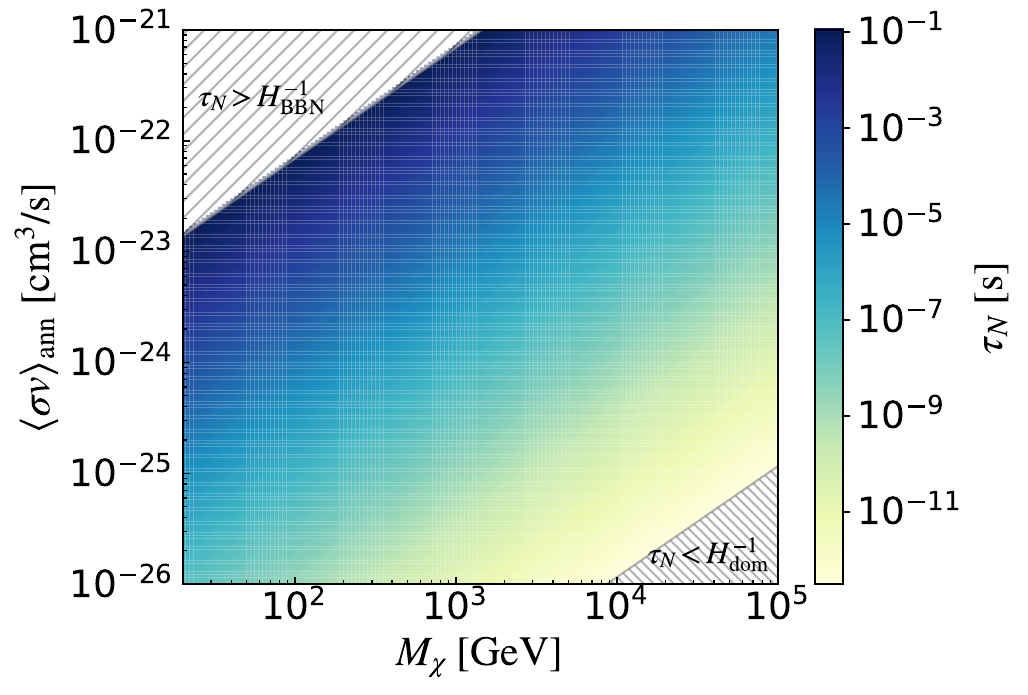}
    \caption{\it Illustration of the parameter space consistent with the observed relic abundance of dark matter, measured by Planck-18~\cite{Planck:2018vyg}. \textbf{Left panel:} Allowed region in the $\mn$–$\tau_N$ plane for a fixed dark matter mass of $50$ GeV, where the thermally averaged annihilation cross-section lies within $\sigmav \in [3\times10^{-26}, 3\times10^{-25}]~{\rm cm}^3/{\rm s}$, in the presence of an EMD epoch. \textbf{Right panel:} Viable relic abundance region in the $\mdm-\sigmav$ plane for fixed $\mn = 10^5$ GeV, showing the impact of varying $\tau_N$ during EMD.}
    \label{fig:sigmav_mdm_contour}
\end{figure}

Recalling that the freeze-out temperature scales with the DM mass as $\tf \approx \mdm/20$, Eq.\eqref{eq:sigmav_tdec_relation} establishes a direct link between the DM parameters $(\mdm, \sigmav)$ and the EMD timescale set by $\tau_N = \Gamma_N^{-1}$, provided $\mn > \mdm$. This connection is illustrated in Fig.\ref{fig:sigmav_mdm_contour}. The left panel of the figure shows the region consistent with the observed DM relic abundance in the $\mn-\tau_N$ plane, which is in accordance with Planck 2018~\cite{Planck:2018vyg}. The light-orange colour denotes the region where EMD occurs and the steel-blue region highlights the combination of $\mn$ and $\tau_N$ for which a DM particle with mass $50$ GeV yields the correct relic abundance for $\sigmav\in[3\times10^{-26},3\times10^{-25}]~{\rm cm}^3{\rm /s}$. The right panel of the figure presents the corresponding viable region in the $\mdm-\sigmav$ plane, which displays contours of constant $\tau_N$ in the $\mdm$–$\sigmav$ plane for $\mn = 10^5$ GeV. Each point on the plane corresponds to a viable combination of DM mass and cross-section that reproduces the observed relic abundance for a specific $N$ lifetime. Since $\tdec \sim \tau_N^{-1/2}$ and $\tf \sim \mdm$, the correct relic density is obtained when $\mdm$ and $\sigmav$ vary proportionally for fixed $\tau_N$, as reflected in the figure.

As we plan to demonstrate in the present article, the primordial gravitational waves from inflation provides a novel probe of both DM physics and the early Universe. In the following section, we explore how the EMD-induced features, particularly those tied to $N$ decay can be imprinted on the GW spectrum, offering a new route to investigating the imprints of DM,  complementary to conventional indirect DM searches.
\medskip


\section{Imprints of early matter domination on primordial gravitational waves}
\label{sec:pgw}
Primordial gravitational waves (PGWs), corresponding to tensor perturbations, are often considered as a unique feature of the early Universe, originating from quantum fluctuation during the inflation \footnote{However, it is important to note that PGWs can also arise from secondary sources, such as scalar-induced tensor modes or other post-inflationary dynamics.~\cite{Baumann:2007zm,Kohri:2018awv,Bhaumik:2025kuj}.}. In principle, PGWs encode information about the entire history of the Universe from the inflation to the present time, including any non-standard intermediate cosmic epochs. Consequently, the aforementioned EMD phase have an impact on PGW, which we will explore in this section. The tensor fluctuations in Fourier space are typically characterized by the tensor power spectrum
~\cite{Planck:2018jri}:
\begin{align}
    \label{eq:pri_tensor}
    P_T^\text{prim}(k) = A_T \left(\frac{k}{k_{\ast}}\right)^{\nt}.
\end{align}
Here $A_T$ is the amplitude of the tensor power spectra and $\nt$ is the tensor spectral index. The amplitude is characterised by observed quantity tensor-to-scalar ratio ($r$), which is measured by cosmic microwave background (CMB) observations (Planck-18) at a particular pivot scale ($k_{\ast}=0.05$ Mpc$^{-1}$). Current constraints from Planck-18, along with BICEP2/$Keck$ 2018 data, impose an upper bound on $r$ which reads as $r_{0.05}<0.036$~\cite{BICEP:2021xfz,BICEP2:2018kqh}. The tilt of the spectrum is determined by $\nt$. In standard single-field, slow-roll, canonical inflation, $\nt$ is typically negative (red-tilted), satisfying consistency relation $\nt\simeq-r/8$~\cite{Liddle:1993fq}. However, there are several scenarios beyond vanilla slow-roll inflation which allow the possibility of a blue-tilted spectrum (\textit{i.e.} $\nt>0$). These include string motivated models~\cite{Brandenberger:2006xi,Calcagni:2013lya}, modified gravity~\cite{Fujita:2018ehq}, particle production during inflation~\cite{Cook:2011hg,Mukohyama:2014gba}, super-inflation modes~\cite{Baldi:2005gk}, G-inflation~\cite{Kobayashi:2010cm}, natural inflation on steep potentials~\cite{Anber:2009ua}, Higgsed chromo-natural inflation~\cite{Adshead:2016omu}, Higgsed gauge-flation~\cite{Adshead:2017hnc}, Axion gauge field inflation~\cite{Caldwell:2017chz,Dimastrogiovanni:2018xnn}, Einstein-Gauss-Bonnet inflation~\cite{Oikonomou:2021kql}. Notably, recent results from NANAOGrav have been well described by a blue-tilted spectrum~\cite{Kuroyanagi:2020sfw,NANOGrav:2023hvm}.  Throughout our analysis we are not restricting ourselves in vanilla slow-roll, canonical inflation which allows us to treat $\nt$ as a free parameter.

However, the blue-tilted spectrum can overproduce the GW spectrum which contributes to the effective relativistic degrees of freedom ($N_{\rm eff}$) of the Universe, with its deviation from the Standard Model, $\Delta N_{\rm eff}$, serving as an important parameter. Thus, the GW spectrum is subject to an upper bound arising from $\Delta N_{\rm eff}$,  given by~\cite{Maggiore:1999vm}
\begin{eqnarray}\label{eq:omega_delta_nu}
    \int^{\infty}_{f_{\rm min}} \frac{df}{f} \, \Omega_{\rm GW}(f) h^2 \;\leq\; 5.6\times 10^{-6}\, \Delta N_{\rm eff},
\end{eqnarray}
where $f_{\rm min}$ is the frequency cutoff, typically $\simeq 10^{-10}\,\mathrm{Hz}$ for BBN and $\simeq 10^{-18}\,\mathrm{Hz}$ for CMB analyses. $h$ represents the reduced Hubble parameter, defined as $H_0/100$, with $H_0$ being the Hubble parameter at the present epoch.
Current cosmological observations impose upper limits on $\Delta N_{\rm eff}$, which in turn constrain the GW spectrum via Eq.~\eqref{eq:omega_delta_nu}. These constraints read
\begin{eqnarray}
\label{eq:neff_current}
    \Delta N_{\rm eff} <
    \begin{cases}
        0.28 & \ \text{Planck 2018 + BAO~\cite{Planck:2018vyg}}, \\
        0.4  & \ \text{BBN~\cite{Cyburt:2015mya}}.
    \end{cases}
\end{eqnarray}
Besides these established bounds, forthcoming  CMB experiments are expected to significantly improve sensitivity reach of $\Delta N_{\rm eff}$, thereby impacting the GW searches. Experiments such as COrE~\cite{CORE:2017oje}, SPT-3G~\cite{SPT-3G:2014dbx}, and the Simons Observatory~\cite{SimonsObservatory:2018koc} are projected to achieve a sensitivity of $\Delta N_{\rm eff} < 0.12$, while improved senstivity reaches are expected from CMB-S4~\cite{Abazajian:2019eic,TopicalConvenersKNAbazajianJECarlstromATLee:2013bxd} and PICO~\cite{NASAPICO:2019thw}, both targetting $\Delta N_{\rm eff} < 0.06$. CMB Bharat is expected to improve this limit to $\Delta N_{\rm eff} < 0.05$~\cite{CMB-bharat}. The most sensitive future constraint is projected for CMB-HD, which aims to reach $\Delta N_{\rm eff}<0.027$~\cite{CMB-HD:2022bsz}. In our study, we have incorporated these current bounds and projected future sensitivity reaches to analyze their implications as depicted in Figs. \ref{fig:GW_spectrum} and \ref{fig:GW_EMD_DM} below.  

The primordial tensor perturbations lead to the observed GW spectrum, at the present time, which reads as~\footnote{Here, mode ($k$) is related with frequency ($f$) as $k=2\pi f$.}
\begin{align}
    \label{eq:omega}
    \Omega_{\rm GW}(k) = \frac{1}{12}\left(\frac{k}{a_0 H_0}\right)^2 T_T^2(k) \;P_T^{\rm prim}(k),
\end{align}
with $H_0$ and $a_0$ being the Hubble parameter and scale factor, respectively, at the present time~\footnote{In our analysis, we considered $a_0=1$ and $H_0=67~{\rm km/s/Mpc}$, based on the Planck 2018~\cite{Planck:2018vyg} for the vanilla $\Lambda$CDM background.}. $T_T(k)$ is the transfer function which takes care of the evolution of the GW spectrum through different scales. 
In general, $T^2_T(k)$ can be expressed as~\cite{Turner:1993vb,Chongchitnan:2006pe,Nakayama:2008wy,Nakayama:2009ce,Kuroyanagi:2011fy,Kuroyanagi:2014nba}
\begin{align}
\label{eq:transfer}
    T_T^2(k) \equiv \Omega_{m,0}^2 \left(\frac{g_*(T_\text{in})}{g_*^0}\right)\left(\frac{g_{\ast s}^0}{g_{\ast s}(T_{\rm in})}\right)^\frac{4}{3} \left(\frac{3j_1(k\tau_0)}{k\tau_0}\right)^2 F(k),
\end{align}
with $g_{\ast s}$ and $g_{*}$ being the total number of degrees of freedom contributing to entropy and energy density, respectively. $j_1(k\tau_0)$ is the first order spherical Bessel function, with $\tau_0=2 / H_0$~\cite{Datta:2022tab,Planck:2018vyg} being the conformal time today. $\Omega_{m,0}$ represents the total matter density at the present epoch~\footnote{According to the Planck 2018~\cite{Planck:2018vyg}, the constraint on $\Omega_{m,0}$ reads as $\Omega_{m,0}=0.3111\pm 0.0056$ for $\Lambda$CDM scenario, where the mean value is considered throughout the analysis.}. In the pre-BBN period, $F(k)$ contains the signatures of any non-standard epoch. The shape of $F(k)$ is as follows in the standard scenario (\textit{i.e.} just radiation domination at the post inflationary scenario):
\begin{align}\label{eq:stand}
    F(k)\bigg |_\text{standard} \equiv T_1^2\left(\frac{k}{\keq}\right)T_2^2\left(\frac{k}{\krh}\right),
\end{align}
where $k_{\rm eq}$ and $k_{\rm RH}$ are the modes that re-enter the horizon at matter-radiation equality and the end of reheating, respectively,
\begin{eqnarray}
    \keq &=& 7.1\times 10^{-2} \Omega_m h^2\,{\rm Mpc}^{-1}, \label{eq:keq}\\
    \krh &=& 1.7\times 10^{14}\left(\frac{g_{\ast s}(\trh)}{g_{\ast s}^0}\right)^\frac{1}{6} \left(\frac{\trh}{10^7\;\text{GeV}}\right)\,{\rm Mpc}^{-1} \label{eq:krh}.
\end{eqnarray}
and $T_1$ and $T_2$ are the transfer functions which we show below. In presence of EMD, $F(k)$ takes the form
\begin{eqnarray}
\label{eq:IMD}
    F(k)\bigg |_{\rm EMD} \equiv  
      T_1^2\left(\frac{k}{\keq}\right)T_2^2\left(\frac{k}{\kdec}\right)T_3^2\left(\frac{k}{\kdecs}\right)T_2^2\left(\frac{k}{\krhs}\right).
\end{eqnarray}
where one more transfer function $T_3$ comes into play due to this non-trivial phase. The fitting functions
$T_1, T_2$ and $T_3$ 
can be expressed as~\cite{Kuroyanagi:2014nba}
\begin{eqnarray}
    T_1^2(x) &\equiv& 1+1.57 x +3.42 x^2,\label{eq:T1}\\
    T_2^2(x) &\equiv& (1-0.22x^{3/2}+0.65x^2)^{-1}\label{eq:T2}\\
    T_3^2(x) &\equiv& 1+0.59 x +0.65 x^2\label{eq:T3}.
\end{eqnarray} 
The scales corresponding to the end of EMD ($\kdec$) can be expressed as 
\begin{eqnarray}
\label{eq:kdec}
    \kdec &= 1.7\times 10^{14}\left(\frac{g_{\ast s}(\tdec)}{g_{\ast s}^0}\right)^\frac{1}{6} \left(\frac{\tdec}{10^7\;\text{GeV}}\right) {\rm Mpc}^{-1} .
\end{eqnarray}
 The characteristic scales $\kdecs$ and $\krhs$ carry the information of EMD period, \textit{i.e.} the onset and end of EMD period and the entropy dilution due to the period, can be expressed as (the subscript $S$  representing the comoving entropy density)
\begin{eqnarray}
    \kdecs &\equiv& \kdec \Delta_s^{2/3},\label{eq:kdecs}\\
    \krhs &\equiv& \krh \Delta_s^{-1/3}\label{eq:krhs}.
\end{eqnarray}
where $\Delta_s$ is defined as in Eq.~\eqref{eq:dilution_factor}.
Thus, utilizing Eqs.~\eqref{eq:pri_tensor}-\eqref{eq:krhs}, the GW spectrum can be estimated, incorporating the impact of EMD epoch.

\subsection{Signatures in stochastic gravitational wave backgrounds}
\label{subsec:detection_prospects}
Having set the theoretical stage, let us now explore the prospects of detecting the GW signal, outlined in the previous subsection, focusing on their observability in current and upcoming GW missions. To this end, we plan to evaluate the capability of various GW missions in searching for the imprints of the specific early Universe scenario as described above.
The relevant experimental landscape can be broadly classified into the following categories:

\begin{enumerate}[a)]
    \item \textbf{Ground based interferometers:} \textit{Laser Interferomenter Gravitational-wave Observatory} (LIGO)~\cite{LIGOScientific:2016aoc,LIGOScientific:2016sjg,LIGOScientific:2017bnn,LIGOScientific:2017vox,LIGOScientific:2017ycc,LIGOScientific:2017vwq}, \textit{Advanced} LIGO (a-LIGO)~\cite{LIGOScientific:2014pky,LIGOScientific:2019lzm},  \textit{Einstein Telescope} (ET)~\cite{Punturo_2010,Hild:2010id}, \textit{Cosmic Explorer} (CE)~\cite{Reitze:2019iox}.
    \item \textbf{Space based interferometers: }$\mu$-ARES~\cite{Sesana:2019vho}, \textit{Laser Interferometer Space Antenna} (LISA)~\cite{amaroseoane2017laser,Baker:2019nia}, \textit{Big-Bang Observer} (BBO)~\cite{Corbin:2005ny,Harry_2006}, \textit{Deci-Hertz Interferometer Gravitaitonal-wave Observatory} (DECIGO)~\cite{Yagi:2011yu}, \textit{Upgraded} DECIGO (U-DECIGO)~\cite{Seto:2001qf,Kawamura_2006,Yagi:2011wg}.
    \item \textbf{Pulsar Timing Arrays (PTA):} \textit{European Pulsar Timing Array} (EPTA)~\cite{Kramer:2013kea,Lentati:2015qwp,Babak:2015lua}, \textit{Square Kilometre Array} (SKA)~\cite{Janssen:2014dka,Weltman:2018zrl,Carilli:2004nx}, \textit{North American Nanohertz Observatory for Gravitational Waves} (NANOGrav)~\cite{McLaughlin:2013ira,NANOGRAV:2018hou,Aggarwal:2018mgp,Brazier:2019mmu,NANOGrav:2020bcs}.
\end{enumerate} 
\begin{figure}[!ht]
    \centering
    \includegraphics[scale=0.45]{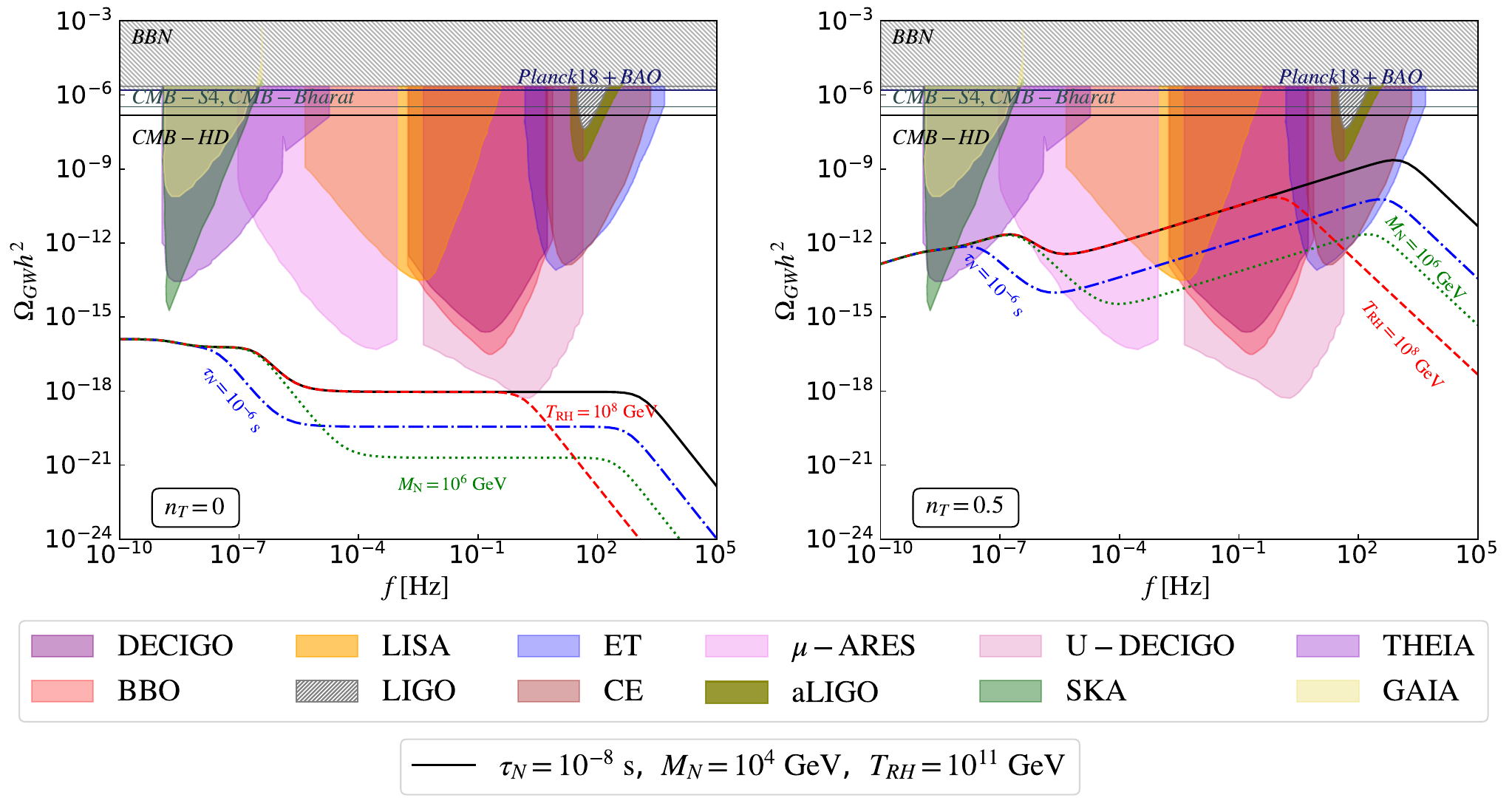}
    \caption{\it Illustration of the GW spectrum for different values of $\tau_N$, $M_N$, and $\trh$. In each spectrum (dashed, dot-dashed, and dotted lines), only the indicated parameter is varied, while the other two are fixed to the values used in the black solid curve. The \textbf{left plot} corresponds to $\nt = 0$, while the \textbf{right plot} shows results for $\nt = 0.5$. The parameters $M_N$ and $\trh$ are given in GeV, and $\tau_N$ in seconds. Both plots assume $r = 0.036$.}
    \label{fig:GW_spectrum}
\end{figure}
Fig.~\ref{fig:GW_spectrum} presents the detection prospects of the GW spectrum at various GW detectors, considering $r=0.036$ (the maximum possible value allowed by Planck-18+BICEP/$Keck$ 2018~\cite{BICEP:2021xfz,BICEP2:2018kqh}), for different $\nt$. The left panel of the Fig.~\ref{fig:GW_spectrum} illustrates the scenario for $\nt=0$, indicating that the prospect of detecting the GW spectrum by the current and upcoming missions is low. However, a blue-tilted spectrum can enhance the GW spectrum as illustrated in the right panel of Fig.~\ref{fig:GW_spectrum} for $\nt=0.5$. At this point we want to remind the reader that we are not restricting ourselves in vanilla slow-roll, canonical inflation and hence the consistency relation, $\nt \approx -r/8$  is not respected, as discussed in detailed in the previous section.

In both panels of Fig.~\ref{fig:GW_spectrum}, black solid line represents a  GW spectrum for benchmark values $\tau_N=10^{-8}$ s, $\mn=10^4$ GeV and $\trh=10^{11}$ GeV. The other three curves (blue dot-dashed, green dotted and red dashed) show how one of the parameter varies, keeping the other parameters unaltered, as shown by corresponding colour in the figure. For example, blue dot-dashed line indicates that $\mn$ and $\trh$ are unchanged from black solid line but $\tau_N=10^{-6}$ s. Both panels depict two characteristic suppressions in the GW spectrum, where the higher frequency suppression arises from inflationary reheating and the lower one marks the decay of $N$ \textit{i.e.} the signature of EMD, with $f_{\rm RH}>f_{\rm dec}$. The lower frequency suppression arises when the entropy dilution, due to $N$ decay, dominates which is characterised by $\kdecs$ (Eq.~\eqref{eq:kdecs}). The end of the suppression is marked by $\tdec$ and occurs at $\kdec$ (Eq.~\eqref{eq:kdec}). Since matter predominates in the universe during this time, the GW spectrum behaves as $f^{-2}$. For instance, the black solid line in both panels of the figure, exhibits low frequency suppression around $f_{\rm dec}\approx 1.87\times 10^{-7}$ Hz which lasts up to $f_{\rm dec,S}\approx 1.55\times10^{-6}$ Hz. The enhanced suppression at lower frequencies is a direct consequence of a prolonged EMD phase, whose duration is governed by the EMD parameters, $\mn$ and $\tau_N$. Both parameters independently extend the EMD period and thereby intensify the suppression. For fixed $\mn$, a larger $\tau_N$ increases the lifetime of the $N$ particles, delaying their decay and thus decreasing $\tdec$ (see Eq.\eqref{eq:tdec}). This extends the EMD duration and leads to a more substantial suppression through increased entropy dilution (see Eq.\eqref{eq:dilution_factor}). Similarly, increasing $\mn$ raises $\tdom$ (see Eq.\eqref{eq:tdom}), further lengthening its duration. This also enhances the entropy dilution, compounding the suppression effect (see Eq.\eqref{eq:dilution_factor}). These dependencies are clearly manifested in Fig.~\ref{fig:GW_spectrum}.

\begin{figure}[!ht]
    \centering
    \includegraphics[scale=0.65]{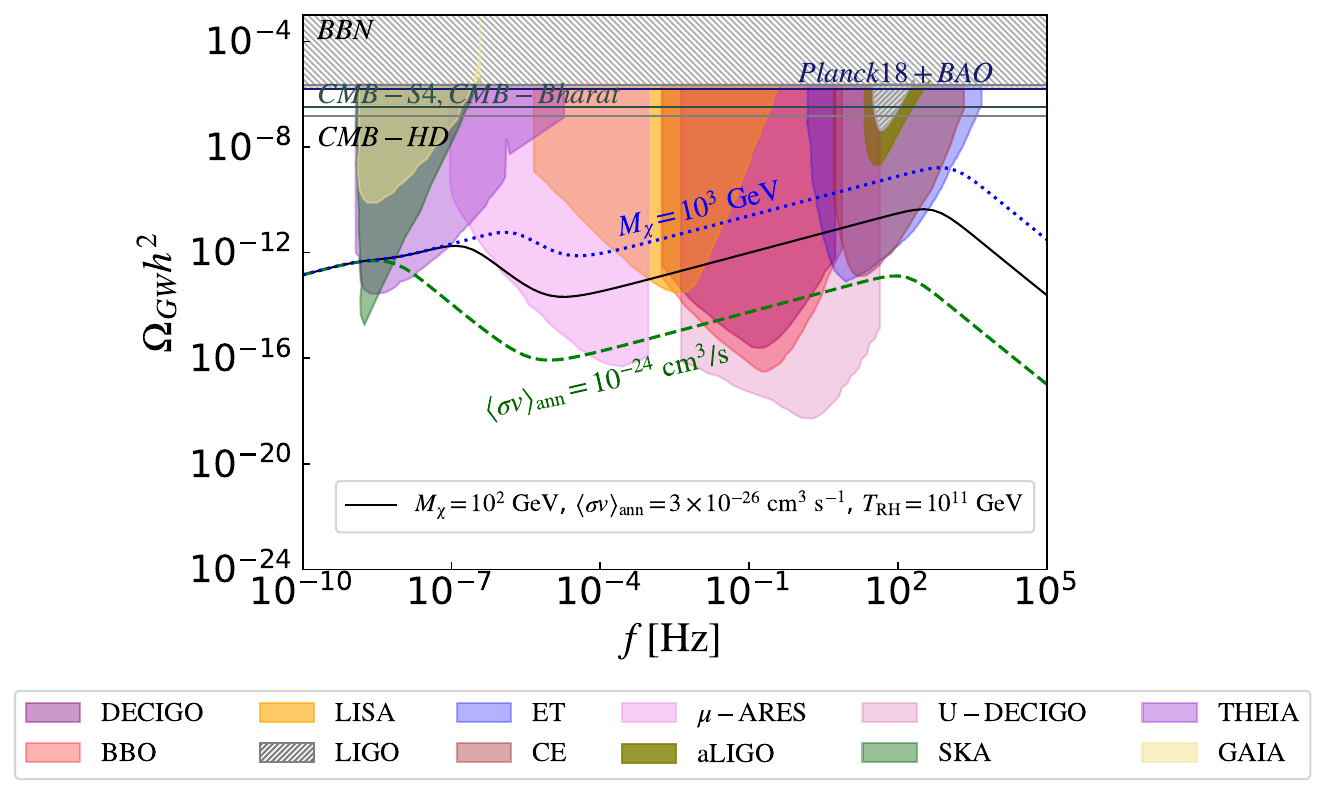}
    \caption{\it Gravitational wave spectra for different values of dark matter parameters ($\mdm$ and $\sigmav$). In each spectrum (dashed, dot-dashed, and dotted lines), only the specified parameter is varied, while the other two remain fixed as in the black solid line. $\mdm$ and $\sigmav$ are expressed in GeV and cm$^3$,s$^{-1}$, respectively. The plot assumes $M_N = 10^5$ GeV, $\trh = 10^{11}$ GeV, and $r = 0.036$.}
    \label{fig:GW_EMD_DM}
\end{figure}
Our primary objective is to demonstrate how GW observations can probe the DM parameter space. Fig.~\ref{fig:sigmav_mdm_contour} illustrates the dependence of the EMD parameter $\tau_N$ on the DM mass, $\mdm$ and annihilation cross-section, $\sigmav$. Building on this, in Fig.~\ref{fig:GW_EMD_DM}, we show the resulting GW spectrum for different DM parameter choices, assuming a reheating temperature $\trh = 10^{11}$ GeV, and a spectral index tilt $\nt = 0.5$. 
The black solid line in the figure represents a benchmark scenario with $\mdm = 10^2$ GeV and $\sigmav = 3 \times 10^{-26} , \mathrm{cm}^3/\mathrm{s}$. Since DM parameters ($\mdm$, $\sigmav$) are directly related with $\tau_N$, through $\tdec$ (Eq.~\eqref{eq:sigmav_tdec_relation}), any variation in the DM parameters modifies the suppression of the GW spectrum during EMD.
Notably, increasing $\mdm$ reduces $\tau_N$ (discussed in detailed in Sec.~\ref{subsec:DM_ann}), thereby shortening the EMD phase and reducing the suppression in the GW spectrum. Conversely, larger values of $\sigmav$ increase $\tau_N$ (see Sec.~\ref{subsec:DM_ann} for detailed discussion), enhancing the duration and impact of the suppression. These trends are reflected in the plot: larger $\sigmav$ (green dashed) leads to a more pronounced suppression, while heavier DM (blue dotted) results in milder suppression.
Overall, the figure highlights how DM parameters, in presence of EMD, have a direct impact on GW spectrum, which can be probed via future GW detectors, by observing the characteristic suppression patterns in the GW spectrum.


\subsection{Signal-to-noise ratio for  GW detectors}
\label{subsec:snr}
In any experimental setup, measurements are inevitably influenced by noise originating from various sources, necessitating a precise characterization of the noise spectrum. To assess the potential for detecting a signal in a given experiment, the signal-to-noise ratio (SNR) serves as a crucial diagnostic tool. For GW experiments, the SNR is defined as~\cite{Thrane:2013oya,Caprini:2015zlo}
\begin{eqnarray}
\label{eq:SNR}
{\rm SNR} \equiv \sqrt{\tau_{\rm obs} \int_{f_{\rm min}}^{f_{\rm max}} \text{d}f \left(\frac{ \Omega_{\rm GW}(f,{\{\theta\}}) h^2}{\Omega_{\rm GW}^{\rm noise}(f) h^2}\right)^2 },
\end{eqnarray}
where $\Omega_{\rm GW}(f,{\{\theta\}})$ denotes the predicted GW energy density spectrum, parametrized by the set $\{\theta\}$, and $\tau_{\rm obs}$ is the detector's observation time. The function $\Omega_{\rm GW}^{\rm noise}(f)$ represents the noise power spectral density of the detector, while $f_{\rm min}$ and $f_{\rm max}$ define its operational frequency range.

\begin{table}[!ht]
    \centering
    \renewcommand{\arraystretch}{1.3}
    \begin{tabular}{|c|c|c|c|}
    \hline
    \hline
       Detectors  & Frequency range & $\tau_{\rm obs}$\\
    \hline
       ET~\cite{Punturo_2010,Hild:2010id}  & $\left[1-10^3\right]$ Hz & $5$ years\\
       BBO~\cite{Corbin:2005ny,Harry_2006} & $\left[10^{-3}-7\right]$ Hz & $4$ years\\
       LISA~\cite{amaroseoane2017laser,Baker:2019nia}  & $\left[10^{-4}-1\right]$ Hz & $4$ years\\
       $\mu$-ARES~\cite{Sesana:2019vho} & $\left[10^{-7}-1\right]$ Hz & $4$ years\\
    \hline
    \hline
    \end{tabular}
    \caption{\it Specifications of the GW detectors used in this study, including their operating frequency ranges and assumed observation durations. }
    \label{tab:detector_spec}
\end{table}

To achieve a wide spectral coverage, we incorporate a suite of GW detectors, $\mu$-ARES, LISA, BBO, and ET, together covering the frequency range from $10^{-7}$ to $10^3$ Hz. A detailed discussion of the noise profiles associated with each detector is provided in Appendix~\ref{app:noise}. For a consistent comparison of sensitivities, we adopt a common detection threshold of ${\rm SNR} = 10$ across all experiments. The specifications of the detectors considered are summarized in Table~\ref{tab:detector_spec}.

\begin{figure}[!ht]
    \centering
    \includegraphics[scale=0.47]{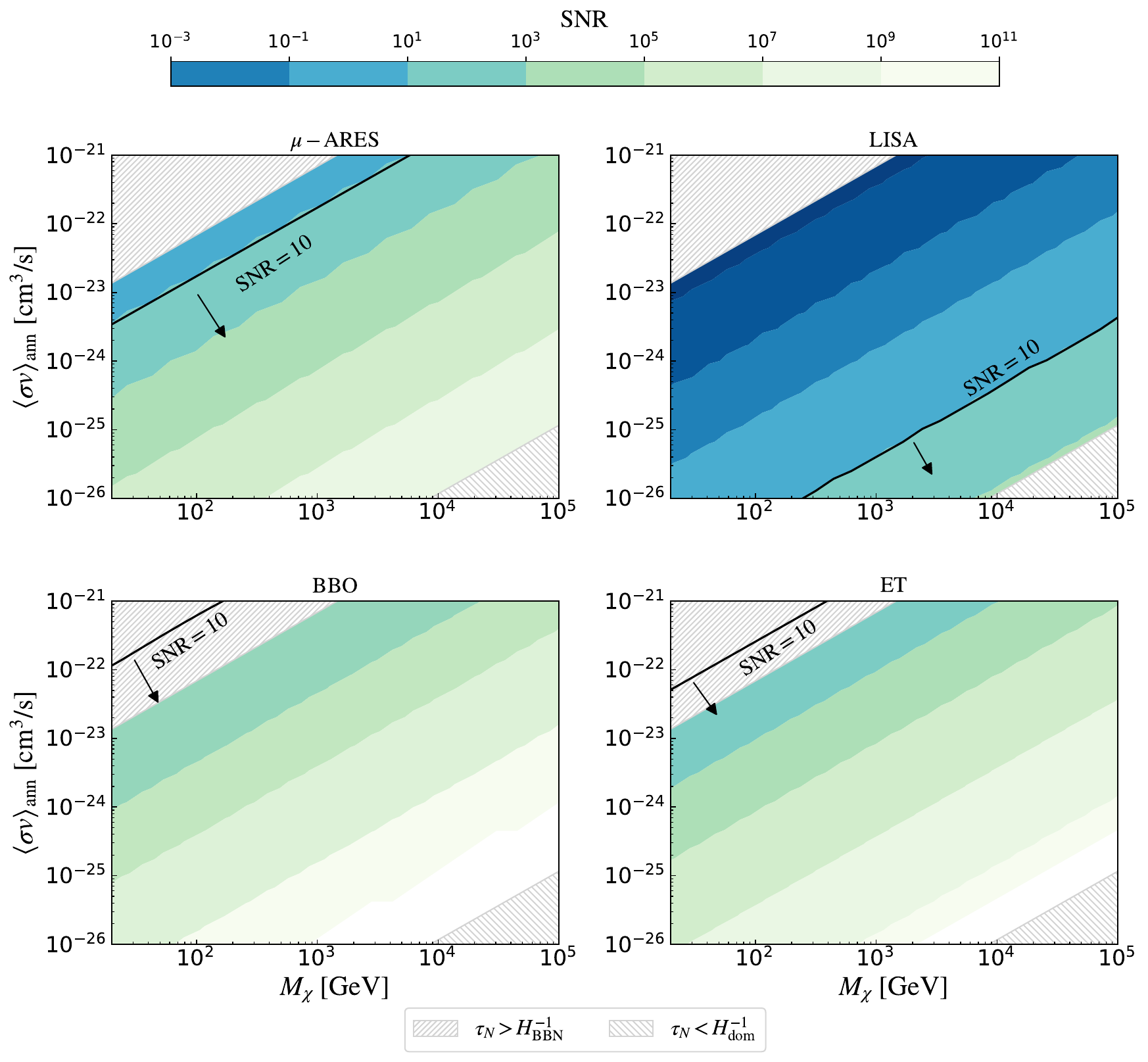}
    \caption{\it Illustration of SNR plots in $\mdm$-$\sigmav$ plane for the GW detectors, $\mu$-ARES, LISA, BBO and ET. The black solid lines indicates ${\rm SNR}=10$, with the region below this line corresponding to ${\rm SNR}>10$. The figure is plotted for $\mn=10^5$ GeV, $\nt=0.5$ and $r=0.036$.}
    \label{fig:snr}
\end{figure}

Fig.~\ref{fig:snr} displays the variation of the SNR in the $\mdm$-$\sigmav$ plane for $\mu$-ARES, LISA, BBO, and ET. The computation assumes benchmark values $\mn = 10^5~\mathrm{GeV}$, $\nt = 0.5$, and $r = 0.036$. The black solid line marks the contour where $\mathrm{SNR} = 10$, marking the threshold for significant detection. The region below this line corresponds to $\mathrm{SNR} > 10$, indicating stronger detection prospects.
Two regions of physical relevance are highlighted, in each panel of the figure. The grey-hatched region is excluded by BBN constraints on late-decay of $N$ particles, whereas the blue-dotted region denotes the regime where $\tau_N < H_{\mathrm{dom}}^{-1}$, indicating that an EMD epoch does not occur.
Our findings reveal that for fixed $\sigmav$, the SNR increases with increasing $\mdm$. Conversely, for a fixed $\mdm$, the SNR decreases as $\sigmav$ increases. This behaviour stems from the fact that smaller $\mdm$ or larger $\sigmav$ values enhance the dominance of the early matter phase, which in turn suppresses the amplitude of the GW spectrum within the sensitivity bands of the detectors. These features are consistent with the trends observed in the sensitivity curves shown in Fig.~\ref{fig:GW_EMD_DM}, and are discussed in detail in Sec.~\ref{subsec:detection_prospects}.
Moreover, among the detectors considered, $\mu$-ARES, BBO, and ET are capable of probing a wide region of the dark matter parameter space for the chosen benchmark values.

\medskip

\section{Forecast on upcoming GW missions}
\label{sec:fisher}

Although the SNR analysis identifies regions of parameter space where detection is feasible, it does not quantify how precisely those parameters can be measured. For this purpose, we perform Fisher forecast analysis, which quantifies expected uncertainties by evaluating the curvature of the log-likelihood near its maximum.

The analysis is performed over $N_b$ logarithmically spaced frequency bins, with each bin contributing an effective number of measurements given by~\cite{Gowling:2021gcy}
\begin{eqnarray}\label{eq:frequencybin}
    n_b \equiv \left[(f_b - f_{b-1})\tau_{\rm obs}\right].
\end{eqnarray}
This framework allows us to errors on the parameters grounded in the detector's noise characteristics.
In the following, we describe the construction of the likelihood and Fisher matrix relevant to our GW signal analysis.

The likelihood function depends on the detector's noise, characterized by the noise energy density $\Omega_{\rm GW}^{\rm noise}(f)$ (see Sec.~\ref{subsec:snr}). Assuming Gaussian statistics, the likelihood of observing a signal $\Omega_{\rm sig}(f_b, \theta)$ across $N_b$ frequency bins is given by~\cite{Dodelson:2003ft} 
\begin{eqnarray}
    \mathscr{L} (\theta) = \prod_{b=1}^{N_b} \sqrt{\frac{n_b}{2\pi\Omega_{\rm GW}^{\rm noise}(f_b)^2}}\,\, {\rm exp}\left( - \frac{n_b\left(\Omega_{\rm GW}(f_b,\{\theta\}) - \Omega_{fid}(f_b)\right)^2}{\Omega_{\rm GW}^{\rm noise}(f_b)^2}\right).
\end{eqnarray}
$\Omega_{fig}(f_b)\equiv \Omega_{\rm GW} (f_b,\{\theta_{fid}\})$ represents the fiducial setup which is based on fiducial parameters ($\{\theta_{fid}\}$). For computational convenience, we maximize the log-likelihood function, so-called chi-squared distribution, as 
\begin{eqnarray}
\label{eq:loglikelihood}
    \mathcal{L} (\theta) \equiv ln\,(\mathscr{L} (\theta)),
\end{eqnarray}
which leads to the Fisher matrix~\cite{Dodelson:2003ft}
\begin{eqnarray}
    F_{ij} \equiv \left\langle-\,\frac{\partial^2 \mathcal{L} (\theta)}{\partial \theta_i \partial \theta_j}\right\rangle,
\end{eqnarray}
with angular brackets indicating the expectation value over the observational data with $\{\theta\}$ being the model parameters. Under the Gaussian approximation (valid for $2n_b \gtrsim \mathcal{O}(10^2)$), the Fisher matrix simplifies to~\cite{Dodelson:2003ft}
\begin{eqnarray}
    F_{ij} \;\approx\; \tau_{\rm obs}\sum_{b=1}^{N_b}\frac{2\Delta f_b}{\Omega_{\rm GW}^{\rm noise}(f_b)^2}\left.\frac{\partial \Omega_{\rm GW}}{\partial \theta_i}\right|_{f_b}\left.\frac{\partial \Omega_{\rm GW}}{\partial \theta_j}\right|_{f_b} ,
\end{eqnarray}
where $\Delta f_b \equiv f_b - f_{b-1}$. In order to ensure this, we set $N_b=500$. 
The inverse of the Fisher matrix, $[C_{ij}] \equiv [F_{ij}]^{-1}$ is the covariance matrix, where the square roots of its diagonal elements, $\sqrt{C_{ii}}$, represent the $1\sigma$ uncertainties associated with the model parameters $\theta_i$. In the following section, we will forecast on the error-estimation of the parameters for the future GW detectors.

\begin{figure}[!ht]
    \centering
    \includegraphics[scale=0.52]{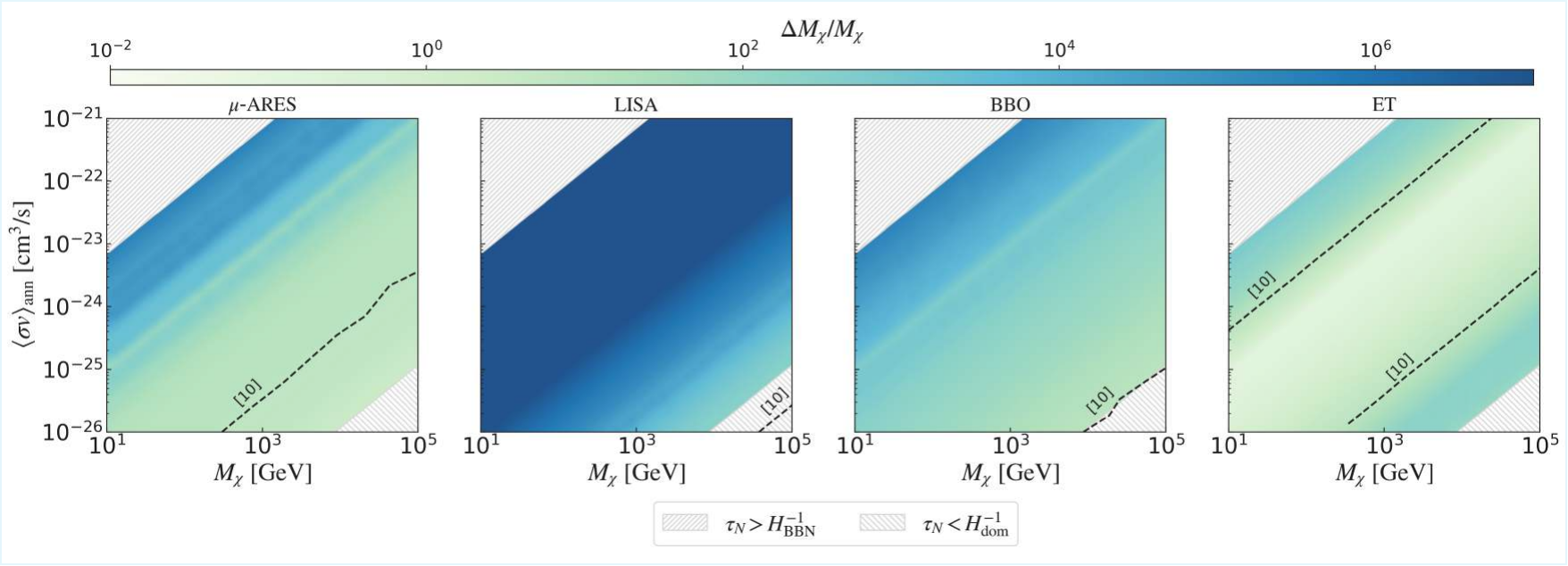}
    \caption{\it Relative uncertainties on $\mdm$, derived from Fisher forecast analysis, for $\mu$-ARES, LISA, BBO, ET, with $\mn=10^5$ GeV and $\trh=10^{11}$ GeV.}
    \label{fig:fisher_mdm_gw}
\end{figure}

In our study, the Fisher analysis for the future GW missions is performed for $\theta\equiv\{\mdm,\sigmav,\nt\}$, considering $r=0.036$, $\mn=10^5$ GeV and $\trh=10^{11}$ GeV. In Fig.~\ref{fig:fisher_mdm_gw}, we present the projected uncertainties on the measurement of $\mdm$, across the $\mdm$-$\sigmav$ plane, for $\nt=0.5$. The results are shown for various GW observations, $\mu$-ARES, LISA, BBO and LISA. In each panel of the figure, the colour shading indicates the level of precision achievable in measuring $\mdm$, with lighter regions corresponding to smaller relative uncertainties, illustrating higher precision in measuring $\mdm$. The black dashed line denotes the boundary where the relative uncertainty in $\mdm$ reaches $10$. We observe that for $\nt=0.5$ and $\mn=10^5$ GeV, the region of high precision fall within the parameter space consistent with EMD epoch, for $\mu$-ARES and ET.

\begin{figure}[!ht]
    \centering
    \includegraphics[scale=0.52]{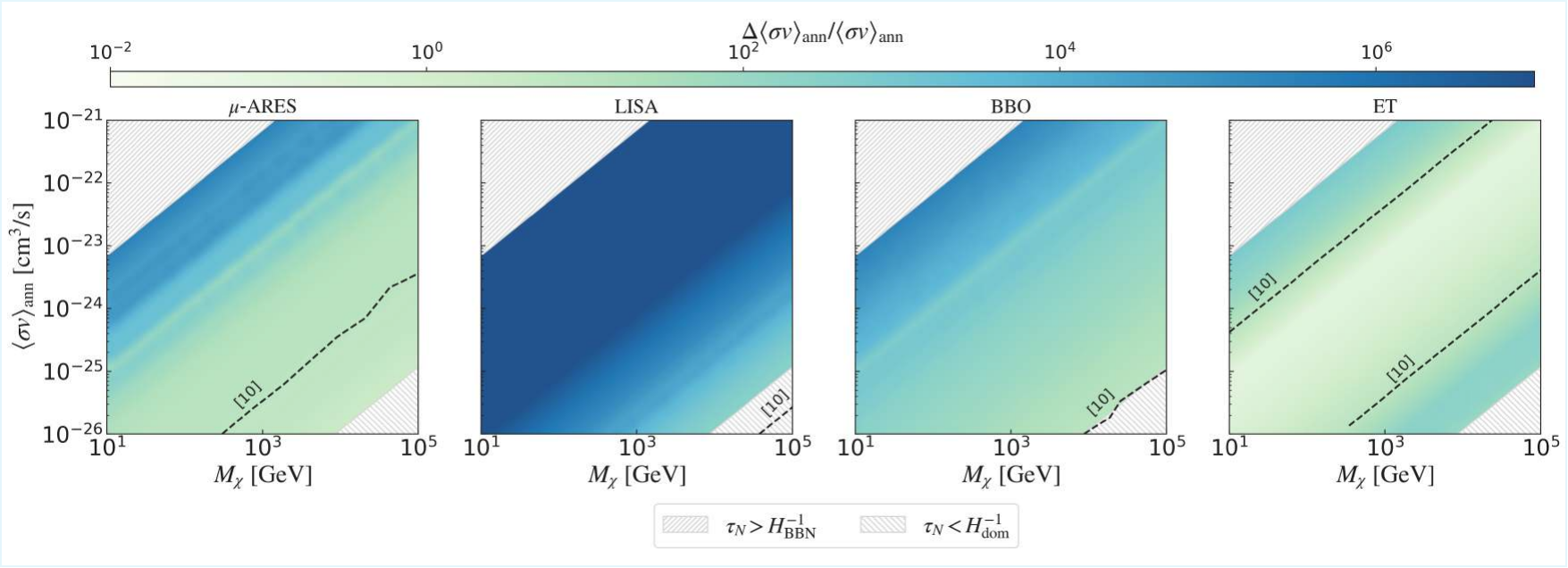}
    \caption{\it Relative uncertainties on $\sigmav$, derived from Fisher forecast analysis, for $\mu$-ARES, LISA, BBO, ET, with $\mn=10^5$ GeV and $\trh=10^{11}$ GeV.}
    \label{fig:fisher_sigmav_gw}
\end{figure}

Fig.~\ref{fig:fisher_sigmav_gw} presents a similar analysis, this time focusing on the projected uncertainties for $\sigmav$. Together, the figures demonstrate the parameter space where future observations can achieve precise measurements of DM properties. In the following section we will emphasize the complementarity between the GW missions and DM indirect searches on probing the DM.

\medskip

\section{Complementary test with indirect detection of DM and GW missions}
\label{sec:complementary}
\begin{figure}[!ht]
    \centering
    \includegraphics[scale=0.5]{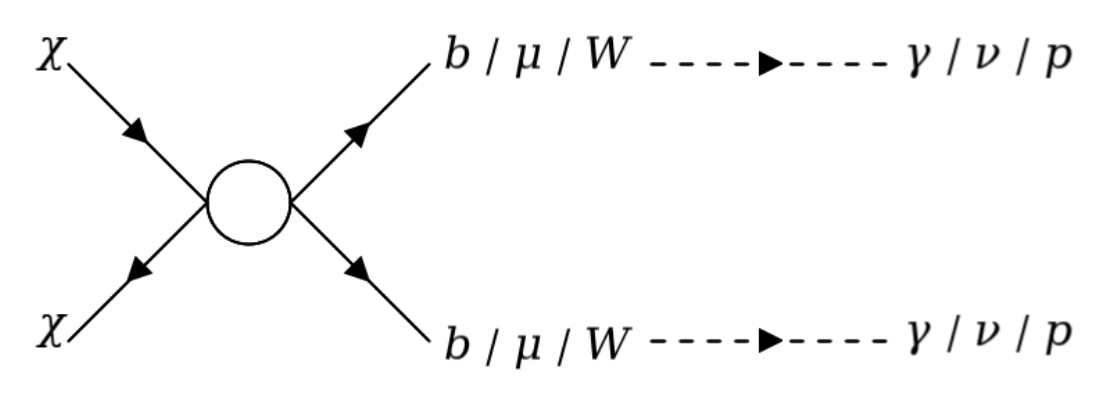}
    \caption{\it Feynman diagram illustrating the annihilation of DM ($\chi$) into $b$, $\mu$, and $W$ final states, which subsequently decay into $\gamma$-rays, neutrinos, and cosmic rays.}
    \label{fig:feynman}
\end{figure}

Annihilating DM particles can produce various primary SM final states, such as quarks, muons, or $W$ bosons, which subsequently decay or hadronize into secondary particles, notably $\gamma$-rays. DM can annihilates into photons directly as well. These $\gamma$-rays serve as primary observables in indirect detection searches. A schematic of this process is shown in Fig.~\ref{fig:feynman}.

The $\gamma$-ray spectra produced by annihilating DM can be probed in astrophysical environments with high DM densities, such as the Galactic centre, galaxy clusters, and dwarf spheroidal galaxies. The spectral features and intensity depend on both dominant annihilation channel (\textit{e.g.}, $q\bar{q}$, $b\bar{b}$, $\mu^+\mu^-$, or $\gamma\gamma$) and the astrophysical environment. 
 
\begin{figure}[!ht]
    \includegraphics[scale=1.1]{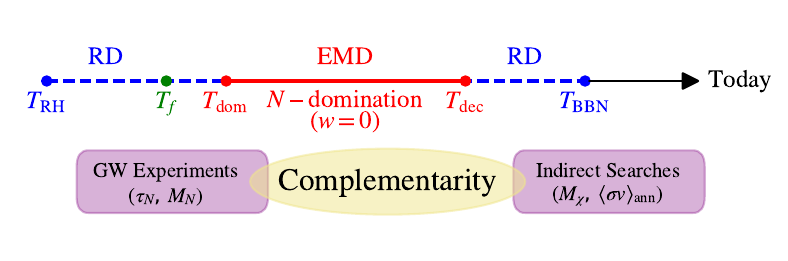}
    \caption{\it Schematic diagram of the timeline of the event. Future GW experiments can directly probe EMD phase ($\tau_N$ and $\mn$). $\tau_N$ can be translated into DM parameters ($\mdm$, $\sigmav$), which can also be probed via indirect searches, indicating the complementarity between two searches.
   }
    \label{fig:schematic_complementarity}
\end{figure}

GW experiments, in contrast, do not rely on such astrophysical modelling\footnote{Detection of a stochastic GW background relies on removing several astrophysical foregrounds which is a topic of active research \cite{Bertone:2019irm,Miller:2025yyx}}. The stochastic GW background is sensitive primarily to the thermal and expansion history of the early Universe, offering a complementary probe of the dark matter physics which we will discuss below.

At this point, let us emphasize that by complementary probes of dark matter, we precisely refer to the ability of two otherwise differently motivated experiments, namely the indirect detection of DM and the GW observations, to probe the same region of parameter space of DM involving its mass ($\mdm$) and annihilation cross-section ($\sigmav$). For instance, as discussed  in Sec.~\ref{subsec:DM_ann}, the parameters characterising EMD epoch, particularly $\tau_N$, is used to map and reconstruct these two DM parameters which is subsequently shown to be probed via the indirect searches of DM as we will specify in detail later on. Moreover, due to this mapping, we are able to identify the overlapping regions in parameter space that can be simultaneously tested by both indirect detection and GW observations. This approach where we study the DM physics in two separate experiments, is what we refer to as ``complementarity between GW experiments and DM searches'' throughout the entire manuscript. Fig.~\ref{fig:schematic_complementarity} schematically illustrates this complementarity between GW observations and DM indirect searches, along with the cosmic timeline. To make the scenario more concrete, we further analyse the SNR for the GW experiments, demonstrating their potential to test the DM parameters. Furthermore, we perform the Fisher forecast analysis to quantify how precisely the upcoming GW missions will be able to measure these parameters. 

\subsection{Current DM searches with future GW missions}
\label{subsec:current_status}

\begin{figure}
    \centering
    \includegraphics[height=6.5cm,width=15cm]{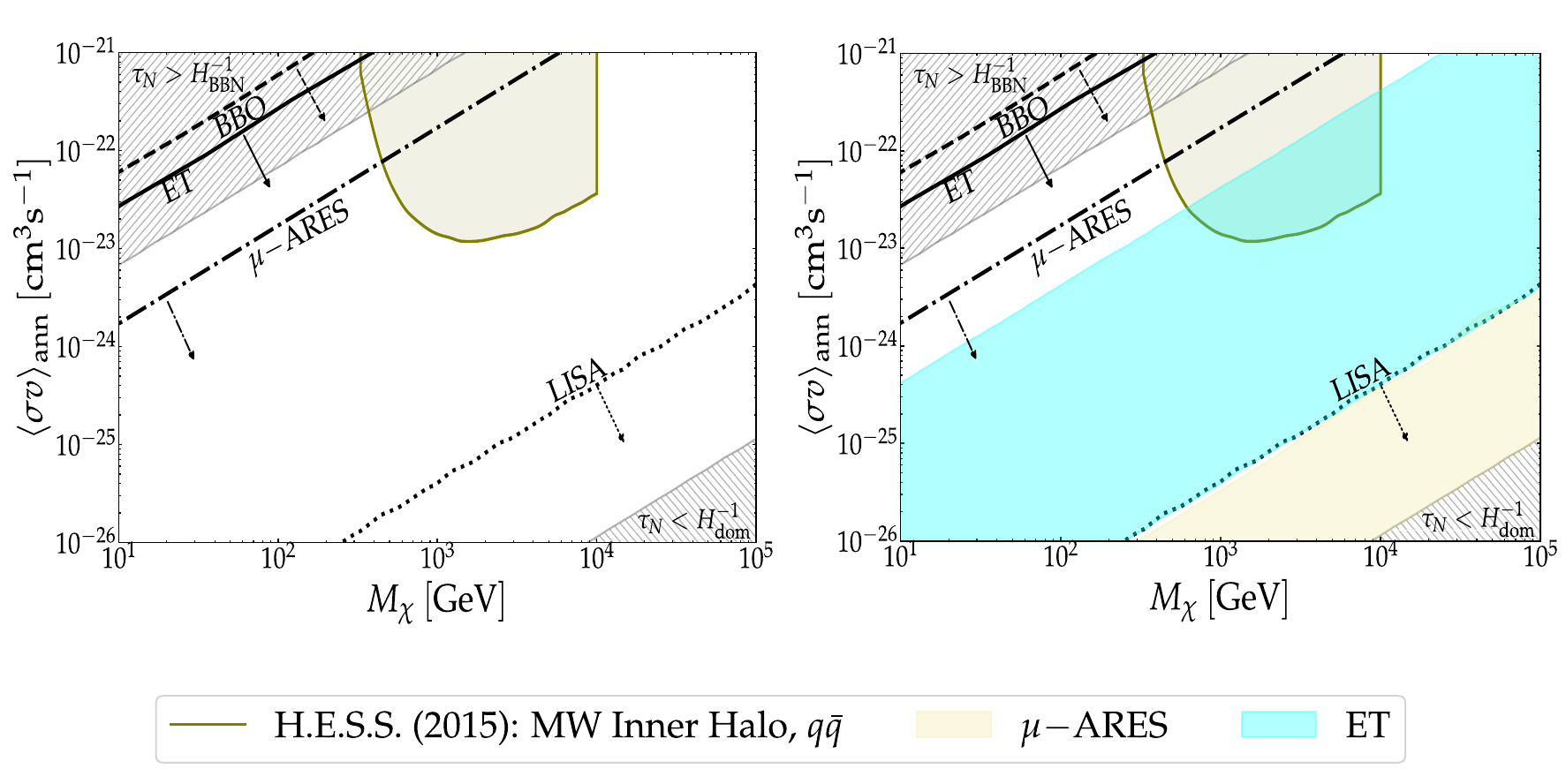}
    \caption{{\it Illustration of current experimental constraints from DM indirect searches, for annihilation into $q\bar{q}$. The SNR contours (\textbf{left} panel) and the projected Fisher results (\textbf{right} panel) from the GW experiments are overlaid to demonstrate the complementarity between indirect detection and GW experiments. Indirect detection bounds are plotted using} \href{https://github.com/moritzhuetten/dmbounds/tree/main?tab=readme-ov-file}{\texttt{github/moritzhuetten}}.}
    \label{fig:current_qq}
\end{figure}

\begin{figure}
    \centering
    \includegraphics[scale=0.5]{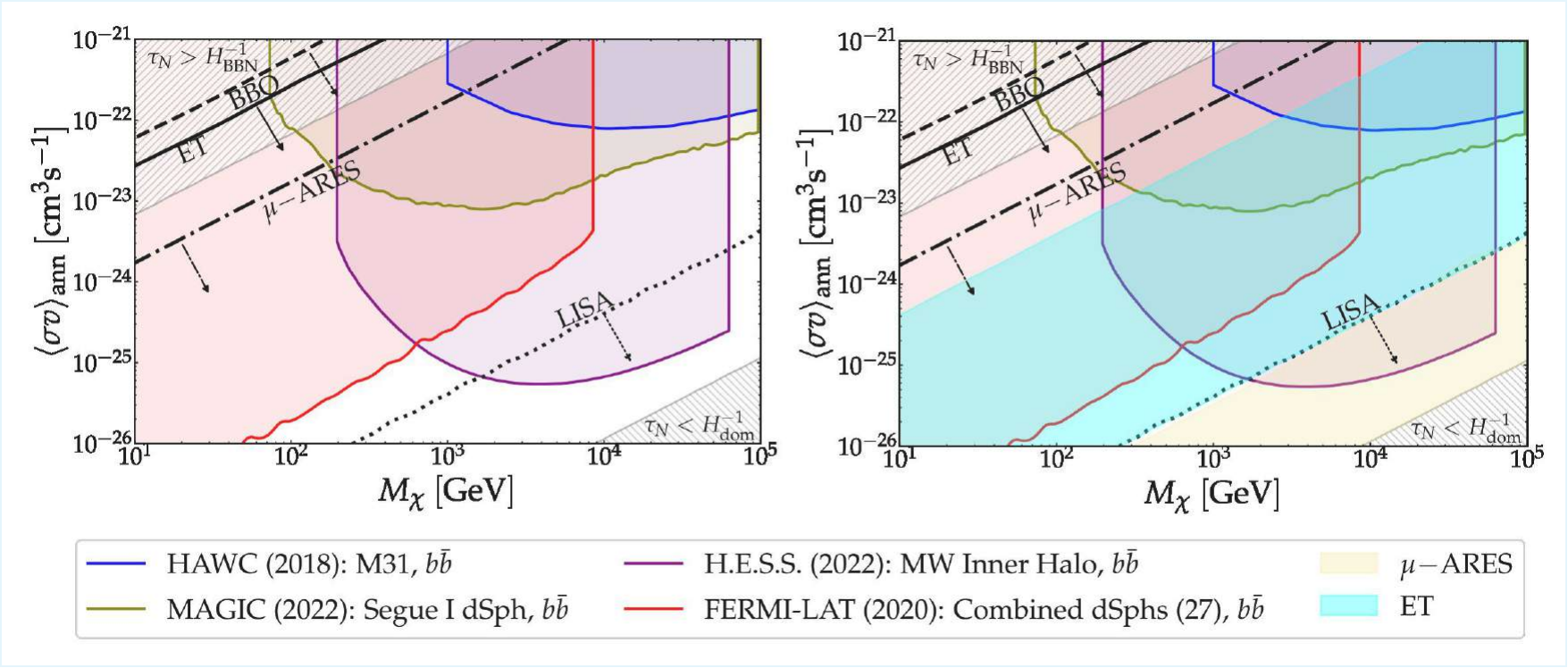}
    \caption{\it The same level of information as in Fig.~\ref{fig:current_qq}, this time for DM annihilation into $b\bar{b}$.}
    \label{fig:current_bb}
\end{figure}

\begin{figure}
    \centering
    \includegraphics[scale=0.5]{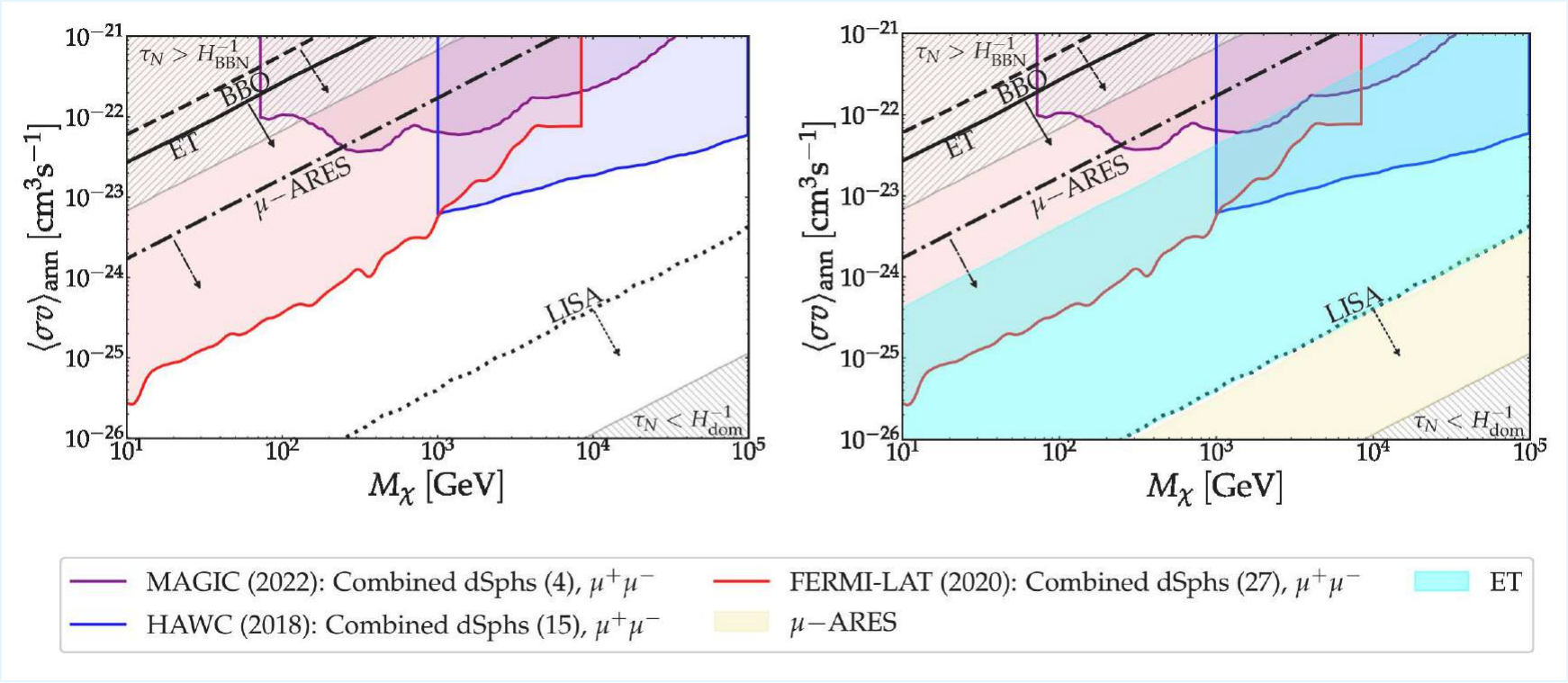}
    \caption{\it The same level of information as in Fig.~\ref{fig:current_qq}, for DM annihilation into $\mu^{\pm}$.}
    \label{fig:current_mumu}
\end{figure}

\begin{figure}
    \centering
    \includegraphics[scale=0.5]{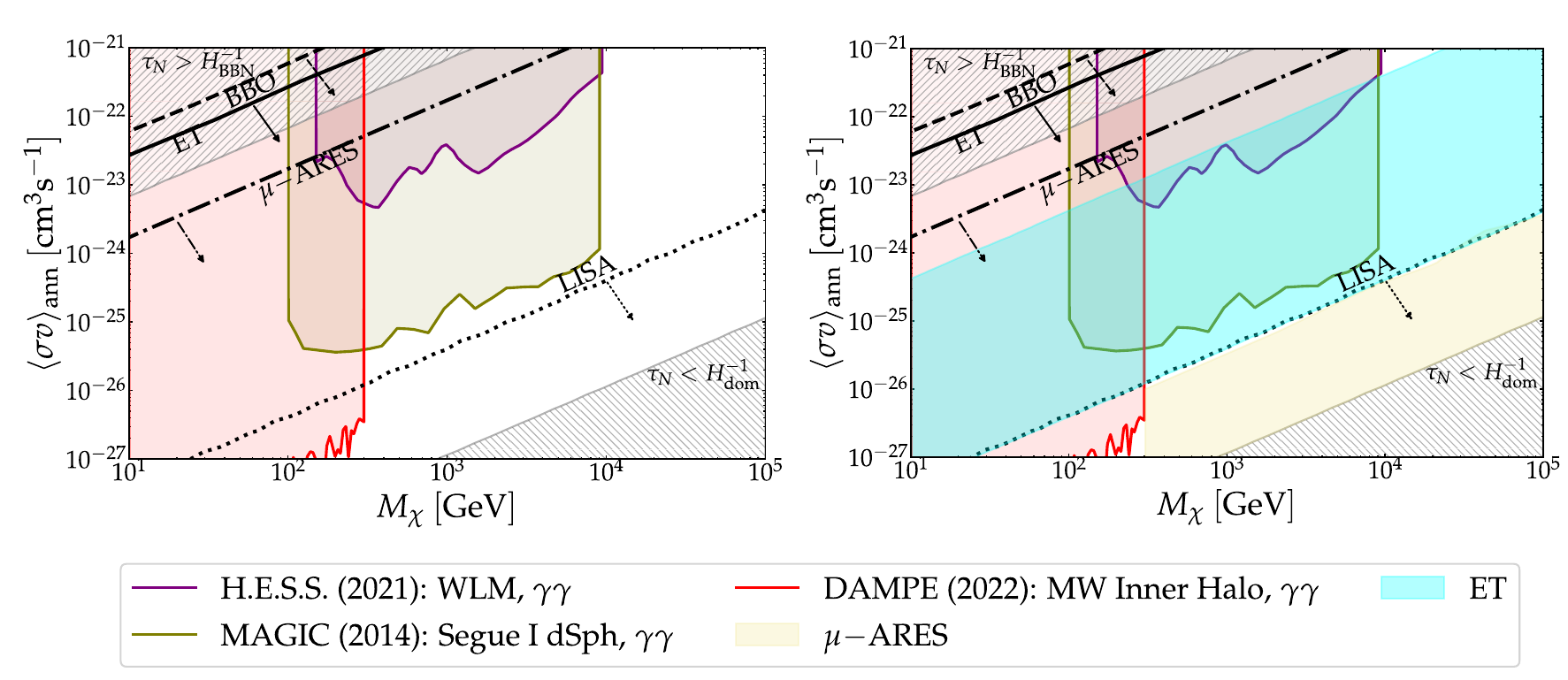}
    \caption{\it The same level of information as in Fig.~\ref{fig:current_qq}, however for DM annihilation into $\gamma\gamma$.}
    \label{fig:current_gammagamma}
\end{figure}

Observations from indirect detection experiments, such as HESS~\cite{HESS:2021zzm,HESS:2020zwn,HESS:2016mib,HESS:2015cda,HESS:2014zqa}, MAGIC~\cite{MAGIC:2021mog,MAGIC:2020ceg,MAGIC:2016xys,Aleksic:2013xea}, HAWC~\cite{HAWC:2018eaa,HAWC:2017mfa}, DAMPE~\cite{DAMPE:2021hsz}, Fermi-LAT~\cite{Thorpe-Morgan:2020czg,Hoof:2018hyn,Fermi-LAT:2016uux,Fermi-LAT:2015xij,Fermi-LAT:2015att}, \textit{etc.} have put fairly strong bounds on the DM annihilation cross-section over a wide range of DM masses, assuming the Navarro-Frenk-White (NFW) density profile. High-mass regions are predominantly constrained by HESS, MAGIC, and HAWC using Galactic centre observations, while Fermi-LAT provides tighter bounds at lower masses through its analysis of dwarf spheroidal galaxies. 

To illustrate the current experimental landscape of the DM indirect searches, Figs.~\ref{fig:current_qq}-\ref{fig:current_gammagamma} present exclusion limits from DM indirect searches for four representative annihilation channels: $q\bar{q}$, $b\bar{b}$, $\mu^+\mu^-$, and $\gamma\gamma$, respectively. The grey-hatched regions are excluded based on BBN constraints on late-decay of $N$ particles ($\tau_N>H_{\rm BBN}^{-1}$) and the requirement for the occurrence of an EMD epoch ($\tau_N<H_{\rm dom}^{-1}$). Beyond these limits, the white regions are allowed by the current DM indirect searches and are consistent with the EMD phase. The presence of such EMD epoch leads to a characteristic feature in the GW background, as discussed in Sec.~\ref{sec:pgw}. We utilize this GW signal to further probe the regions allowed by the indirect searches. 
Accordingly, each panel of the Figs.~\ref{fig:current_qq}-\ref{fig:current_gammagamma} also overlays the projected sensitivities of future GW observatories, $\mu$-ARES, LISA, BBO, and ET, assuming benchmark parameters $\mn = 10^5$ GeV, $\nt = 0.5$, and $r = 0.036$. The black curves represent contours of SNR equals to $10$ for each GW detectors: solid for ET, dashed for BBO, dotted for LISA, and dash-dotted for $\mu$-ARES, with the arrows indicating SNR $>10$ which delineates the potentially detectable regions by the upcoming GW missions. For example, a DM candidate with mass $10^5$ GeV and annihilation cross-section $10^{-24}~{\rm cm}^3{/s}$ lies within the region allowed by the all current indirect detections and can be detected via the future GW missions. 

Furthermore, we also perform the Fisher forecast analysis to demonstrate the precision with which these parameters can be measured. The cyan and light-orange shaded regions in the figures correspond to the $1\sigma$ forecasted confidence regions for ET and $\mu$-ARES, respectively. We show, for instance, that the projected uncertainties on DM mass and annihilation cross-section are nearly $7\%$, while probing $\mdm=10^5$ GeV and $\sigmav=10^{-24}~{\rm cm}^3{/s}$ with $\mu$-ARES. These results  demonstrate that the future GW observations will be able to probe the DM parameters which remain viable under the current DM indirect detection constraints.
In the following section, we will show that the sensitivity reach of the upcoming GW experiments overlaps with the projected reach of the future DM indirect searches, thereby highlighting the complementarity between these two searches in exploring the DM parameter space.

\subsection{Future DM searches with future GW missions}
\label{subsec:future_prospect}

In what follows we'll take one future indirect detection mission, namely, 
the Cherenkov Telescope Array (CTA)~\cite{CherenkovTelescopeArray:2023aqu} as a representative example to demonstrate the prospects of complementarity with upcoming GW missions. The procedure can be repeated with other future DM searches as well.
CTA is poised to significantly advance our ability to detect $\gamma$-ray signals from DM annihilation, particularly in the multi-TeV mass regime. CTA will be able to detect mass of DM ($\sim10^2-10^5~{\rm GeV}$) with cross-sections close to the thermal relic value ($\simeq 3\times 10^{-26}~{\rm cm}^3{\rm /s}$) for several final states of DM annihilation.

\begin{figure}
    \centering
    \includegraphics[scale=0.5]{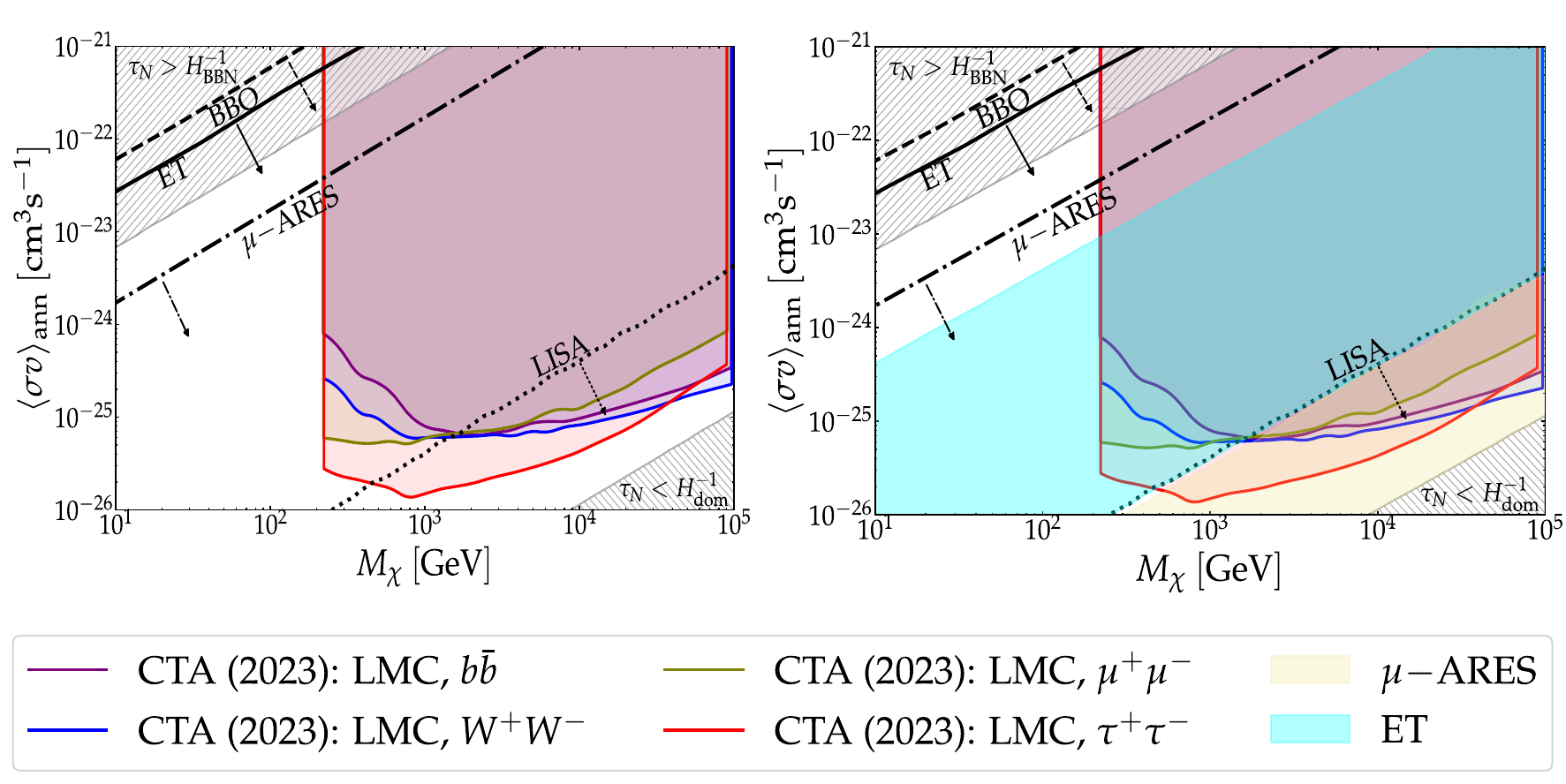}
     \caption{\it Illustration of future experimental constraints from DM indirect searches, for various annihilation channels. The SNR contours (\textbf{left} panel) and the projected Fisher results (\textbf{right} panel) from the GW experiments are overlaid to demonstrate the complementarity between indirect detection and GW experiments.}
    \label{fig:annDM_future}
\end{figure}

The left panel of Fig.~\ref{fig:annDM_future} illustrates the projected sensitivities of CTA, assuming the NFW profile and considering various primary annihilation channels. The SNR for the GW detectors, $\mu$-ARES, LISA, BBO, and ET, respectively, are overlaid on the figures for benchmark values of $\mn=10^5$ GeV, $\nt=0.5$ and $r=0.036$.
In each panel, the solid black contour denotes the SNR threshold of $10$ for the corresponding GW detector. The accompanying solid arrow indicates the region where a detection is expected (SNR $> 10$). The figure shows that BBO, ET, and $\mu$-ARES can probe a wide DM mass range from $\sim 10$ to $10^5$ GeV, while LISA is sensitive to a narrower mass window, approximately $\mdm \in (3 \times 10^2, 10^5)$ GeV. These detection regions exhibit significant overlap with the CTA sensitivity bands, particularly for $\mu$-ARES and ET, highlighting a promising complementarity between indirect $\gamma$-ray searches and GW observations in probing the DM parameter space.

Having understood the parameter regions where GW experiments play a significant role in the complementarity, we employ the Fisher matrix analysis for the GW experiments to quantify the precise measurements of the DM parameters, in the right panel of Fig.~\ref{fig:annDM_future}. The cyan and light-orange shaded regions represent where the relative uncertainties are less than $10$, corresponding to ET and $\mu$-ARES, respectively.  
Our results reveal that both $\mu$-ARES and ET exhibit the capability to precisely probe DM parameters, substantially overlapping with CTA projections, for $\mn=10^5$ GeV, $\nt=0.5$ and $r=0.036$. For instance, when probing $\mdm=10^3$ GeV and $\sigmav=10^{-24}~{\rm cm}^3{/s}$ with ET, the projected uncertainties are approximately $1\%$ on both the DM mass and annihilation cross-section. For $\mu$-ARES while probing $\mdm=10^4$ GeV and $\sigmav=10^{-25}~{\rm cm}^3{/s}$, these are $7\%$. We perform similar analysis for the scenarios where DM annihilates into final states as neutrinos, as discussed in detailed in Sec.~\ref{subsec:neutrino_detectors}. We illustrate that, just like the $\gamma$-ray searches of DM indirect detection, the projected sensitivities of ANTARES~\cite{ANTARES:2022aoa,Albert:2016emp} and Cubic Kilometre Neutrino Telescope (KM3NeT)~\cite{KM3NeT:2024xca} have a significant overlap with the future GW experiments in probing the DM parameters. These alignments underscore the complementarity of future GW and DM indirect observations, offering a powerful multi-messenger approach to probing the microscopic nature of DM.

\subsection{DM indirect detection via neutrino searches}
\label{subsec:neutrino_detectors}
\begin{figure}[!ht]
    \centering
    \includegraphics[height=6.5cm,width=15cm]{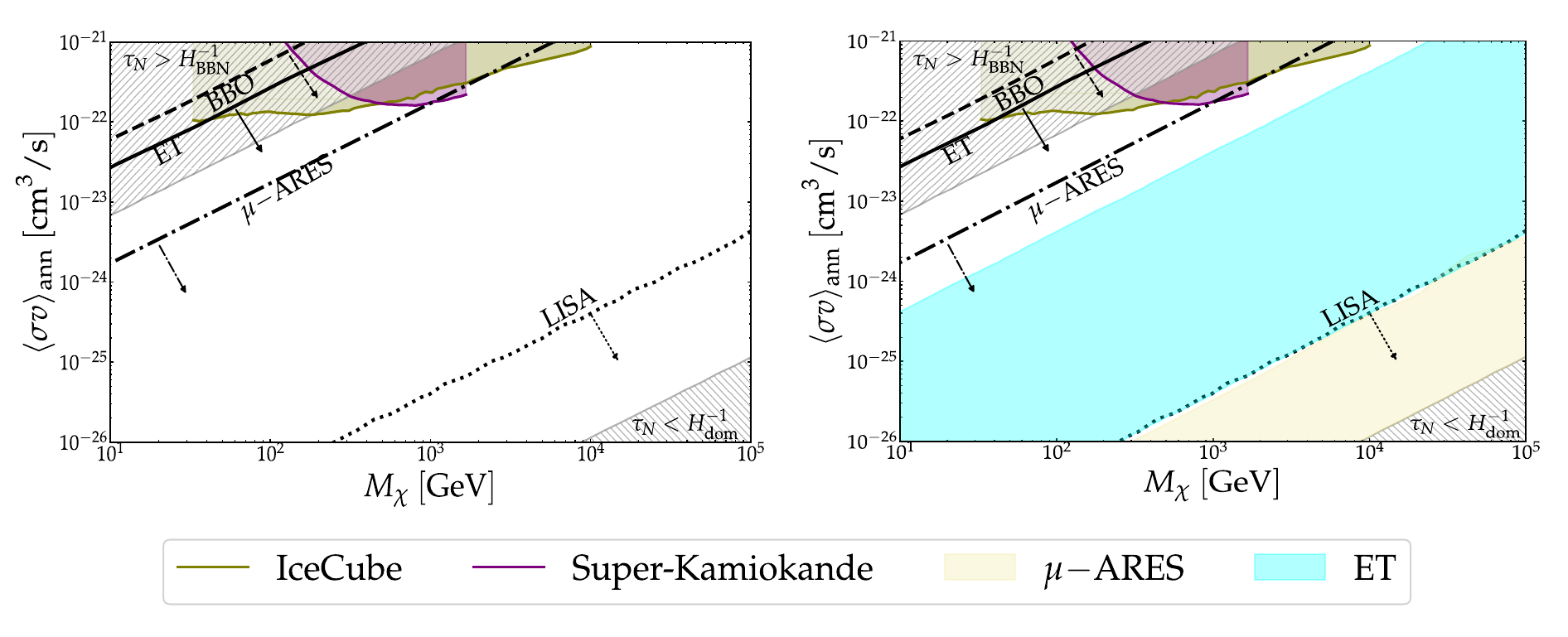}
    \caption{\it Illustration of current experimental constraints from neutrino searches of DM indirect detection, for annihilation into $b\bar{b}$. The SNR contours (\textbf{left} panel) and the projected Fisher results (\textbf{right} panel) from the GW experiments are overlaid to demonstrate the complementarity between indirect detection and GW experiments.}
    \label{fig:current_bb_neutrino}
\end{figure}

\begin{figure}[!ht]
    \centering
    \includegraphics[height=6.5cm,width=15cm]{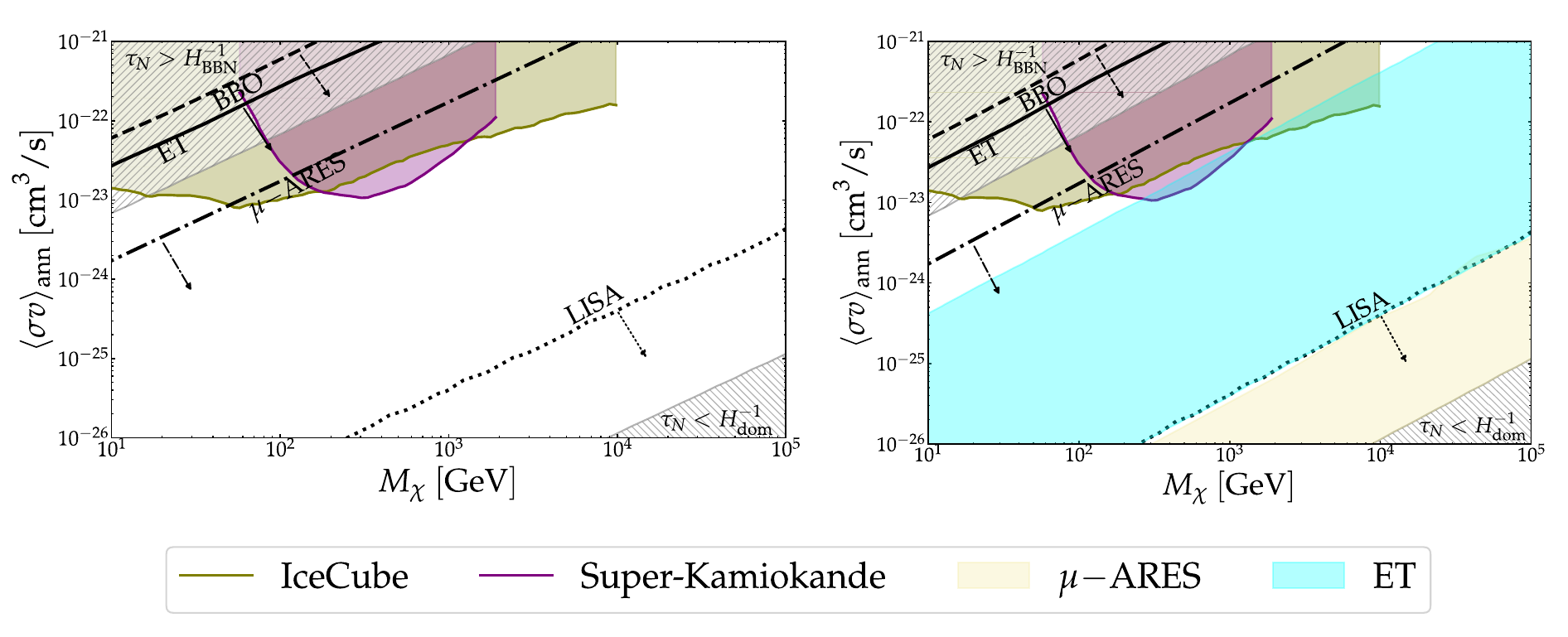}
    \caption{\it The same level of information as in Fig.~\ref{fig:current_bb_neutrino}, but for annihilation into $\mu^{\pm}$.}
    \label{fig:current_mumu_neutrino}
\end{figure}

Observations from neutrino-based DM indirect searches, such as IceCube~\cite{2023PhRvD.108j2004A, 2017EPJC...77..627A}, Super-Kamiokande~\cite{Super-Kamiokande:2020sgt,Frankiewicz:2015zma}, \textit{etc.} also place significant constraints on the DM annihilation cross-section over a wide range of DM masses, assuming the NFW profile. The current experimental landscape of the DM indirect searches via neutrino detectors is presented in Figs.~\ref{fig:current_bb_neutrino} and \ref{fig:current_mumu_neutrino}, focusing two representative annihilation channels: $b\bar{b}$ and $\mu^{+}\mu^{-}$, respectively. The grey-hatched regions, in these plots, are excluded by BBN constraints on late-decay of $N$ particles and the absence of EMD epoch, as also illustrated in Figs.~\ref{fig:current_qq}-\ref{fig:current_gammagamma}. The white regions, in contrast, remains allowed according to the current neutrino search limits of DM indirect detections. Thus, performing the same exercise as shown in Sec.~\ref{subsec:current_status}, we utilize the GW experiments in order to probe these regions, allowed by the indirect searches. The projected sensitivities of future GW experiments, $\mu$-ARES, LISA, BBO and ET, considering $\mn=10^5$ GeV, $\nt=0.5$ and $r=0.036$ are overlaid on DM indirect searches. The black lines indicate the SNR $=10$ contours where the arrows represents SNR $>10$ regions, indicating the potentially detectable regions by the future GW experiments. The analysis shows, for instance, that $\mx=10^3$ GeV and $\sigmav=10^{-25}~{\rm cm}^3{\rm /s}$ which is allowed by neutrino-based DM indirect searches, can be detected via $\mu$-ARES, ET and BBO.

Furthermore, the Fisher analysis on the GW experiments is also performed to estimate the precision with which the aforementioned regions (allowed by indirect searches) can be detected by GW experiments. The cyan and light-orange regions in the right panel of the Figs.~\ref{fig:current_bb_neutrino} and \ref{fig:current_mumu_neutrino} are the $1\sigma$ confidence region for ET and $\mu$-ARES, respectively. We show, for instance, that the ET will be able to probe $\mx=10^3$ GeV and $\sigmav=10^{-25}~{\rm cm}^3{/s}$ with approximately $1\%$ uncertainties. In the following, we show the overlapping between the sensitivity reach of the future GW missions and the future DM indirect searches to illustrate the complementarity between these two searches in exploring the DM parameter space.

\begin{figure}[!ht]
    \centering
    \includegraphics[height=6.5cm,width=15cm]{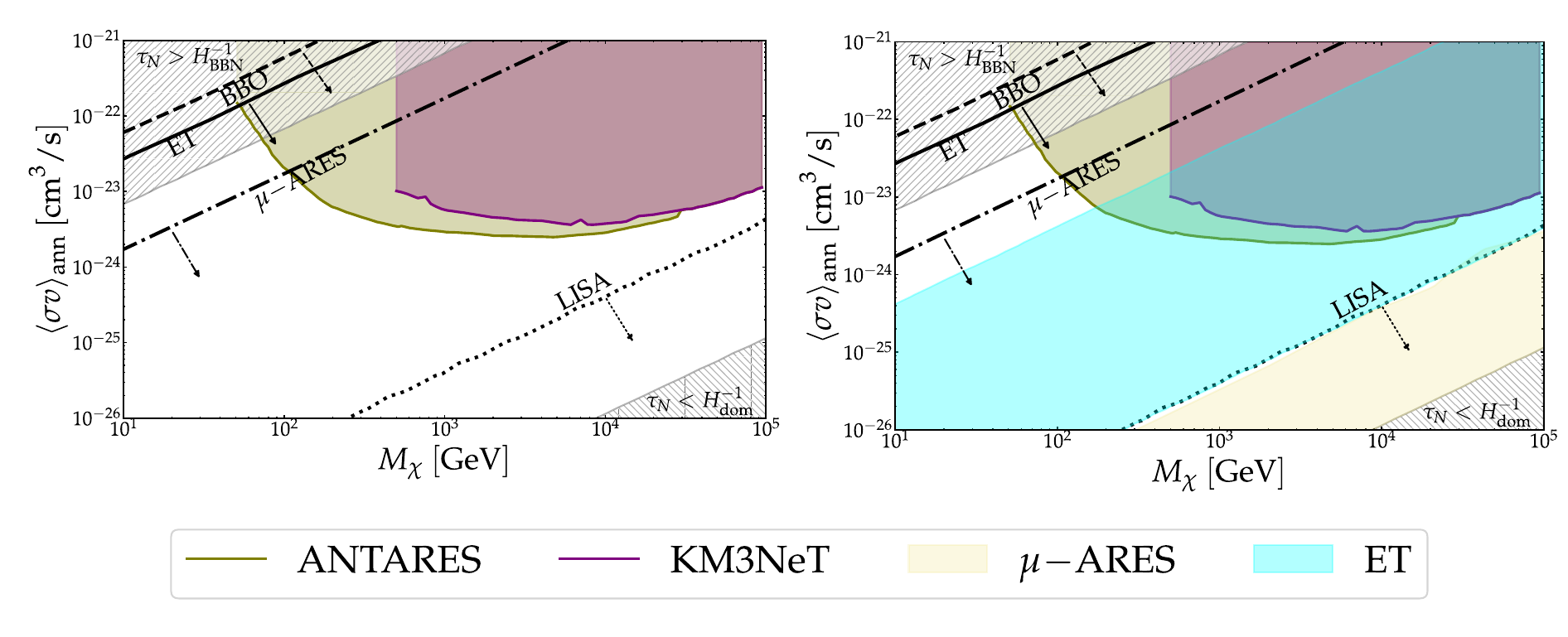}
    \caption{\it Illustration of future experimental constraints from neutrino experiment of DM indirect searches, for $\mu^{+}\mu^{-}$ channel. The SNR contours (\textbf{left} panel) and the projected Fisher results (\textbf{right} panel) from the GW experiments are overlaid to demonstrate the complementarity between indirect detection and GW experiments.}
    \label{fig:future_neutrino}
\end{figure}

To demonstrate the complementarity we consider the future neutrino-based DM indirect searches, namely, ANTARES and Cubic Kilometre Neutrino Telescope (KM3NeT) as representative examples. We execute the same analysis outlined in Sec.~\ref{subsec:future_prospect} to demonstrate the complementarity. The left panel of Fig.~\ref{fig:future_neutrino} illustrates the projected sensitivities of DM indirect searches with the SNR contours of the upcoming GW experiments overlaid for $\mn=10^5$ GeV, $\nt=0.5$ and $r=0.036$. The black contours represent the SNR $=10$ for the GW missions, with the arrows indicating the SRN $>10$ region, corresponding to the parameter space detectable by the future GW missions. The figures (Figs.~\ref{fig:current_bb_neutrino} and \ref{fig:current_mumu_neutrino}) illustrate the significant overlap between the sensitivity reach of ANTARES and KM3NeT, with GW observations, thereby highlighting the complementarity in probing the DM parameter space. Inline with the earlier analysis, as presented in Sec.~\ref{subsec:future_prospect}, we perform the Fisher forecast analysis where GW experiments play a crucial role in complementarity. In the right panel of Fig.~\ref{fig:future_neutrino}, we overlay the relative uncertainties on the DM parameters while measured by ET (cyan-coloured region) and $\mu$-ARES (light-orange region). We show, for instance, a DM candidate having mass $10^5$ GeV and annihilation cross-section $10^{-23}~{\rm cm}^3{\rm /s}$ falls within the projected reach of ANTARES and KM3NeT, and the projected uncertainties on probing these parameters are approximately $\sim 1\%$ while probed by ET. Thus, in conjunction with the $\gamma$-ray searches, the future GW missions are poised to play a crucial role in complementarity with upcoming neutrino searches in probing the DM parameter space.

\medskip

\section{Discussions and conclusions}
\label{sec:conclusion}
Primordial gravitational waves from cosmic inflation offer us a unique probe of the heavy and high energy particle physics that is otherwise inaccessible to traditional particle and astro-particle experiments~\cite{Dasgupta:2022isg, Bhaumik:2022pil, Barman:2022yos, Ghoshal:2022jdt, Dunsky:2021tih, Bernal:2020ywq, Ghoshal:2020vud}. In this analysis, we show that GWs can probe dark matter (DM) formation in presence of a brief period of early matter domination before the onset of the BBN. This probe is executed by investigating the GW spectral shape features which carry the imprints in the inflationary GW spectrum in the form of a characteristic dip in the amplitude of the GW spectrum (see Fig.~\ref{fig:GW_spectrum}). Therefore, the overall setup can be expressed by a minimal set of independent parameters: mass ($\mn$) and life-time ($\tau_N$) of the N-particles, tensor-to-scalar ratio ($r$), spectral tilt ($\nt$) and reheating temperature ($\trh$). By fixing $r$ to be the maximal allowed by the Planck-18 data, we have found a novel complementary involving DM parameters,  $\mdm$ and $\sigmav$, in GW searches like LISA, ET, $\mu-$ARES and BBO, see Fig.~\ref{fig:GW_EMD_DM}. In particular, we have highlighted that ensuring the observed DM density requires a minimum branching fraction ($\br$) of the long-lived particle into DM, captured by Eq.~\eqref{eq:br_critical_clean}, which guarantees sufficient residual annihilation near the end of EMD. To illustrate, for a representative benchmark with $(\tau_N,\mn)=(0.1,5000)~({\rm in ~ GeV})$ and $(\mdm,\sigmav)=(100~{\rm GeV}, 10^{-22}~{\rm cm}^3/s)$, we have found that $\br^c\sim10^{-5}$, a relatively small value, suggesting that non-thermal DM production is significant even with modest branching fractions. This condition plays a pivotal role in establishing a predictive link between cosmological evolution of the thermal history and DM phenomenology, consequently enabling GW observables to directly constrain the structure of the dark sector.

Importantly, the same EMD epoch imprints distinct, detectable features on the GW spectrum see Fig.~\ref{fig:GW_spectrum}. These manifest as characteristic suppressions in the GW amplitude, the shape and scale of which depend sensitively on both the EMD parameters and the DM properties, particularly the mass and annihilation cross-section. To assess the detectability of these features, we have considered a blue-tilted spectrum with $\nt=0.5$ and $r=0.036$, and analysed their detection prospects in $\mu$-ARES, LISA, BBO, and ET. For instance, with $\mn=10^5$ GeV, LISA has been shown to be sensitive to higher DM masses in the range $\mdm\sim 2\times 10^2-10^5$ GeV, and smaller annihilation cross-sections $\sigmav\sim10^{-26}-4\times10^{-24}$ cm$^3$/s (see Fig.~\ref{fig:snr}).

In Figs.~\ref{fig:current_qq}-\ref{fig:current_gammagamma}, we have demonstrated that GW interferometers offer complementary reach compared to current constraints from $\gamma$-ray observatories such as HESS, Fermi-LAT, MAGIC, DAMPE and HAWC. We have shown  that the GW missions are promising to probe the DM parameters which is allowed  by the current DM indirect searches. We have defined the complementarity as the ability of indirect detection and GW experiments to probe the same DM parameter space. The future sensitivity of both the experiments allows us to identify overlapping regions testable by both experiments. In Fig.\ref{fig:annDM_future}, we have highlighted the significant overlap between the projected sensitivities of the CTA and GW missions. This clearly illustrates a complementary approach to DM searches, where interferometric GW observations provide a clean, cosmology-based alternative to traditional indirect detection strategies. 
Unlike $\gamma$-ray or neutrino-based methods, which are subject to astrophysical foregrounds and modelling uncertainties, GW signals directly reflect the thermal history and expansion dynamics of the early Universe, linking them more robustly to DM-related processes. This synergy significantly enhances discovery prospects and broadens the observational window into the dark sector.

Having identified the parameter regions where the GW experiments play a crucial role in complementarity, we have employed  Fisher analysis for the GW experiments. Through Fisher forecasting for the GW missions, we have demonstrated that GW observations can provide precise measurements of DM parameters such as the mass and annihilation cross-section, see for an instance Fig.~\ref{fig:fisher_mdm_gw} and \ref{fig:fisher_sigmav_gw}. We have observed that for $\nt=0.5$ and $\mn=10^5$ GeV, the regions of high precision for both $\mu$-ARES and ET fall within the parameter space compatible with an EMD epoch. Our findings reveal the overlap between the projections future indirect mission (CTA) and future GW missions, depicting the complementarity between these two searches (see Fig.~\ref{fig:annDM_future}). We have shown,  for instance, both $(\mdm,\sigmav)=(10^3~{\rm GeV},10^{-24}~{\rm cm}^3{\rm /s})$ and $(\mdm,\sigmav)=(10^4~{\rm GeV},2\times10^{-25}~{\rm cm}^3{\rm /s})$ lie within the projection of CTA, however, the formal one can be  probed by ET with $\sim1\%$ uncertainty, whereas the later can be probed via $\mu$-ARES with $\sim7\%$. Furthermore, We demonstrate that the future GW missions have a substantial overlap with the neutrino-based DM indirect detection experiments, namely, ANTARES and KM3NeT. For example, our analysis show that $(\mdm,\sigmav)=(10^5~{\rm GeV},10^{-23}~{\rm cm}^3{\rm /s})$ fall within the projected sensitivity of ANTARES and KM3NeT, and can be probed by ET with a relative uncertainties of approximately $1\%$.

As the final conclusion, we reiterate that hunting for apparently disjointed signals in laboratory experiments and cosmological probes will help us have  a better idea about the yet-unexplored and non-trivial evolution history of the Universe beyond the minimal setup of cosmology, such as early matter domination, involving multiple energy scales and heavy particles like dark matter, thus providing a possible pathway to the community a novel and promising way to search for dark matter physics in complementary manner in the upcoming years.

\section*{Acknowledgement}
We sincerely thank Rouzbeh Allahverdi for valuable insights and initial discussions. Thanks also to Amit Dutta Banik, Nicolao Fornengo, Suvam Pal and Md Riajul Haque for fruitful discussions. We acknowledge the computational facilities of Technology Innovation Hub, Indian Statistical Institute (ISI), Kolkata and of the Pegasus cluster of the High Performance Computing (HPC) facility at IUCAA, Pune. DP thanks ISI, Kolkata for financial support through Senior Research Fellowship. SP thanks the ANRF, Govt. of India for partial support through Project No. CRG/2023/003984.

\appendix
\section{Noise modelling for the GW detectors}
\label{app:noise}
The expected sensitivity of a gravitational wave (GW) detector can be characterized by the dimensionless GW energy density spectrum, expressed as~\cite{Gowling:2021gcy}
\begin{eqnarray}
\label{eq:omegagw}
    \Omega_{\rm GW}^{\rm noise} (f) \;=\; \left(\frac{4\pi^2}{3H_0^2}\right)f^3 S(f),
\end{eqnarray}
where $S(f)$ denotes the noise power spectral density of the detector. In the following, we summarize the noise power spectral densities for the GW observatories considered in this work. These will be used to compute the SNR and to forecast the associated uncertainties in probing the model parameters.
\begin{itemize}
\item
\textbf{BBO}

BBO is primarily made up of four triangular sets of detectors, each of which is $5\times 10^4$ km in arm length. The BBO's non-sky-averaged instrumental noise power spectral density is~\cite{Yagi:2011yu} 
\begin{align}
S_{\rm BBO}(f) &= \left[1.8 \times 10^{-49} (f / 1{\rm Hz})^2+2.9 \times 10^{-49} + 9.2 \times 10^{-52} (f / 1{\rm Hz})^{-4}\right]~{\rm Hz^{-1}}.
\label{eq:SeffBBO}
\end{align}
This expression for $S_{\rm BBO}(f)$ can be directly used in Eq.~\eqref{eq:omegagw} to compute the GW energy density spectrum.
\item
\textbf{$\mu$-ARES}

For the $\mu$-ARES mission, we adopt the instrumental noise curve as provided in Ref.\cite{Sesana:2019vho}.

\item
\textbf{LISA}

The three drag-free spacecraft that make up LISA, which has a quadrupolar antenna design, are primarily equipped with free-falling mirrors. There are three primary reasons why instrumental noise in LISA is likely to occur~\cite{Gowling:2021gcy}: \textit{(a)} optical metrology noise (omn), \textit{(b)} shot noise (sn), and \textit{(c)} test mass acceleration noise (accn) as a result of test mass disturbance.

The noise spectral density for the aforementioned contributions can be expressed as~\cite{Gowling:2021gcy}: 
\begin{align}
S_{\rm acc}(f) &= 9 \times 10^{-30} \frac{1}{(2\pi f / 1{\rm Hz})^4} \left( 1 + \frac{10^{-4}}{f / 1{\rm Hz}} \right)~{\rm m^2Hz^{-1}},\\
S_{\rm sn}(f) &= 2.96 \times 10^{-23}~{\rm m^2Hz^{-1}}, \\
S_{\rm oms}(f) &= 2.65 \times 10^{-23}~{\rm m^2Hz^{-1}}.
\end{align}
The resultant instrumental noise for LISA can be expressed as 
\begin{align}
S_{\rm LISA}(f) &= \frac{40}{3} \frac{4S_{\rm acc}(f) + S_{\rm sn}(f) + S_{\rm oms}(f)}{L^2} \left[1 + \left( \frac{f}{0.41c/2L} \right)^2\right].
\label{eq:SeffLISA}
\end{align}
The constellation arm-length of the instrument is $L = 2.5 \times 10^8$~m. Therefore, Eq.~\eqref{eq:omegagw} is used to compute the GW energy density power spectrum.

\item
\textbf{ET}

For the Einstein Telescope, we utilize the instrumental noise curve as presented in Ref.\cite{Chowdhury:2022pnv}.
\end{itemize}

\section{Calculation of critical branching ratio}
\label{app:critical_br}
As discussed in Sec.~\ref{subsec:DM_ann}, our scenario features two distinct DM components. One originates from a pre-existing DM relic that undergoes freeze-out before the EMD epoch ($n_{\rm Th}$), while the other is produced non-thermally from $N$-decay ($n_{\rm NT}$). Consequently, the total number density of the DM can be expressed as 
\begin{eqnarray}
\label{eq:resultant_DM}
    n_{\rm DM} = n_{\rm Th} + n_{\rm NT}.
\end{eqnarray}
A sufficient amount of non-thermal production of DM triggers a residual annihilation. At the end of EMD (\textit{i.e.} at $T=\tdec$), the condition for sufficient annihilation can be estimated as 
\begin{eqnarray}
\label{eq:anni_cond}
    n_{\rm DM} \sigmav \Big|_{\tdec} \geq H(\tdec),
\end{eqnarray}
which imposes a condition on $n_{\rm DM}$, for a specific annihilation cross-section, as
\begin{eqnarray}
\label{eq:cond_ndm}
    (n_{\rm Th} + n_{\rm NT}) \Big|_{\tdec} \geq \frac{H(\tdec)}{\sigmav}.
\end{eqnarray}
Since, $n_{\rm Th}$ originates from the freeze-out process, it can be defined as 
\begin{eqnarray}
\label{eq:ndm_th}
    n_{\rm Th} = \frac{H(\tilde\tf)}{\sigmav},
\end{eqnarray}
with $\tilde\tf$ representing the freeze-out temperature of the DM with cross-section being $\sigmav$. Since, this component of DM has already frozen-out, we always have $n_{\rm Th} \Big|_{\tdec} < \frac{H(\tdec)}{\sigmav}$, which implies a critical condition on $n_{\rm NT}$ to satisfy \eqref{eq:anni_cond}. The non-thermal component, $n_{\rm NT}$, arises from the decay of $N$ and can be expressed at $T=\tdec$ as
\begin{eqnarray}
    n_{\rm NT} (\tdec) = \br \, n_N(\tdec),
\end{eqnarray}
where $n_N(\tdec)$ is the number density of $N$ at $T=\tdec$ and $\br$ is the branching ratio for the production of DM particles ($\chi$) from $N$. Therefore, the abundance at $\tdec$ can be calculated as 
\begin{eqnarray}
\label{eq:abundance_nth}
    \frac{n(\tdec)}{s(\tdec)} = \br \frac{n_N(\tdec)}{s(\tdec)} = \br\frac{3}{4} \frac{\tdec}{m_N} .
\end{eqnarray}
Hence, \eqref{eq:cond_ndm} leads to
\begin{eqnarray}
    n_{\rm NT}(\tdec) &\geq& \frac{H(\tdec)}{\sigmav} - n_{\rm Th} (\tdec)\\
    \implies \frac{n_{\rm NT}(\tdec)}{s(\tdec)} &\geq& \frac{H(\tdec)}{\sigmav s(\tdec)} - \frac{n_{\rm Th}(\tdec)}{s(\tdec)}\nonumber\\
    &\geq& \left(\frac{\alpha_{\ast}}{\alpha_{\ast s}}\right)\frac{\tdec^2}{\sigmav\tdec^3 M_P} - \left(\frac{\tdec}{\tdom}\right)\left(\frac{\alpha_{\ast}}{\alpha_{\ast s}}\right)\frac{1}{\sigmav \tilde\tf M_P}\nonumber\\
    &\geq& \left(\frac{\alpha_{\ast}}{\alpha_{\ast s}}\right)\frac{1}{\sigmav M_P}\left[\frac{1}{\tdec}-\frac{\tdec}{\tdom}\tilde\tf\right]\nonumber.
\end{eqnarray}
Hence, we can estimate the condition on $\br$ for which \eqref{eq:anni_cond} holds which reads as
\begin{eqnarray}
\label{eq:br_critical}
    \br &\geq& \br^c\equiv \frac{4}{3} \left( \frac{m_N}{\tdec} \right) \left( \frac{\alpha_{\ast}}{\alpha_{\ast s}} \right) \frac{1}{M_P \sigmav} \left[ \frac{1}{\tdec} - \frac{\tdec}{\tdom} \frac{1}{\tilde\tf} \right].
\end{eqnarray}

Here, $\alpha_{\ast}\equiv\sqrt{\frac{\pi^2g_{\ast}(\tdec)}{90}}$ and $\alpha_{\ast s}\equiv\frac{2\pi^2}{45}g_{\ast s}(\tdec)$, with $g_{\ast}$ and $g_{\ast s}$ are the number of relativistic degrees of freedom contributing to energy density and entropy density, respectively.


\bibliographystyle{JHEP}
\bibliography{ref}

\providecommand{\href}[2]{#2}\begingroup\raggedright\begin{thebibliography}{100}

\bibitem{Brout:1977ix}
R.~Brout, F.~Englert and E.~Gunzig, \emph{{The Creation of the Universe as a Quantum Phenomenon}}, \href{https://doi.org/10.1016/0003-4916(78)90176-8}{\emph{Annals Phys.} {\bfseries 115} (1978) 78}.

\bibitem{Sato:1980yn}
K.~Sato, \emph{{First Order Phase Transition of a Vacuum and Expansion of the Universe}}, {\emph{Mon. Not. Roy. Astron. Soc.} {\bfseries 195} (1981) 467}.

\bibitem{Guth:1980zm}
A.H.~Guth, \emph{{The Inflationary Universe: A Possible Solution to the Horizon and Flatness Problems}}, \href{https://doi.org/10.1103/PhysRevD.23.347}{\emph{Phys. Rev. D} {\bfseries 23} (1981) 347}.

\bibitem{Linde:1981mu}
A.D.~Linde, \emph{{A New Inflationary Universe Scenario: A Possible Solution of the Horizon, Flatness, Homogeneity, Isotropy and Primordial Monopole Problems}}, \href{https://doi.org/10.1016/0370-2693(82)91219-9}{\emph{Phys. Lett. B} {\bfseries 108} (1982) 389}.

\bibitem{Starobinsky:1982ee}
A.A.~Starobinsky, \emph{{Dynamics of Phase Transition in the New Inflationary Universe Scenario and Generation of Perturbations}}, \href{https://doi.org/10.1016/0370-2693(82)90541-X}{\emph{Phys. Lett. B} {\bfseries 117} (1982) 175}.

\bibitem{Planck:2018vyg}
{\scshape Planck} collaboration, \emph{{Planck 2018 results. VI. Cosmological parameters}}, \href{https://doi.org/10.1051/0004-6361/201833910}{\emph{Astron. Astrophys.} {\bfseries 641} (2020) A6} [\href{https://arxiv.org/abs/1807.06209}{{\ttfamily 1807.06209}}].

\bibitem{Chen:2024roo}
C.~Chen, K.~Dimopoulos, C.~Er\"oncel and A.~Ghoshal, \emph{{Enhanced primordial gravitational waves from a stiff postinflationary era due to an oscillating inflaton}}, \href{https://doi.org/10.1103/PhysRevD.110.063554}{\emph{Phys. Rev. D} {\bfseries 110} (2024) 063554} [\href{https://arxiv.org/abs/2405.01679}{{\ttfamily 2405.01679}}].

\bibitem{Bernal:2020ywq}
N.~Bernal, A.~Ghoshal, F.~Hajkarim and G.~Lambiase, \emph{{Primordial Gravitational Wave Signals in Modified Cosmologies}}, \href{https://doi.org/10.1088/1475-7516/2020/11/051}{\emph{JCAP} {\bfseries 11} (2020) 051} [\href{https://arxiv.org/abs/2008.04959}{{\ttfamily 2008.04959}}].

\bibitem{Ghoshal:2023sfa}
A.~Ghoshal, Y.~Gouttenoire, L.~Heurtier and P.~Simakachorn, \emph{{Primordial black hole archaeology with gravitational waves from cosmic strings}}, \href{https://doi.org/10.1007/JHEP08(2023)196}{\emph{JHEP} {\bfseries 08} (2023) 196} [\href{https://arxiv.org/abs/2304.04793}{{\ttfamily 2304.04793}}].

\bibitem{Ghoshal:2022ruy}
A.~Ghoshal, L.~Heurtier and A.~Paul, \emph{{Signatures of non-thermal dark matter with kination and early matter domination. Gravitational waves versus laboratory searches}}, \href{https://doi.org/10.1007/JHEP12(2022)105}{\emph{JHEP} {\bfseries 12} (2022) 105} [\href{https://arxiv.org/abs/2208.01670}{{\ttfamily 2208.01670}}].

\bibitem{Ghoshal:2024gai}
A.~Ghoshal, D.~Paul and S.~Pal, \emph{{Primordial gravitational waves as probe of dark matter in interferometer missions: Fisher forecast and MCMC}}, \href{https://doi.org/10.1007/JHEP12(2024)150}{\emph{JHEP} {\bfseries 12} (2024) 150} [\href{https://arxiv.org/abs/2405.06741}{{\ttfamily 2405.06741}}].

\bibitem{Buchmuller:2014pla}
W.~Buchmuller, E.~Dudas, L.~Heurtier and C.~Wieck, \emph{{Large-Field Inflation and Supersymmetry Breaking}}, \href{https://doi.org/10.1007/JHEP09(2014)053}{\emph{JHEP} {\bfseries 09} (2014) 053} [\href{https://arxiv.org/abs/1407.0253}{{\ttfamily 1407.0253}}].

\bibitem{Buchmuller:2015oma}
W.~Buchmuller, E.~Dudas, L.~Heurtier, A.~Westphal, C.~Wieck and M.W.~Winkler, \emph{{Challenges for Large-Field Inflation and Moduli Stabilization}}, \href{https://doi.org/10.1007/JHEP04(2015)058}{\emph{JHEP} {\bfseries 04} (2015) 058} [\href{https://arxiv.org/abs/1501.05812}{{\ttfamily 1501.05812}}].

\bibitem{Argurio:2017joe}
R.~Argurio, D.~Coone, L.~Heurtier and A.~Mariotti, \emph{{Sgoldstino-less inflation and low energy SUSY breaking}}, \href{https://doi.org/10.1088/1475-7516/2017/07/047}{\emph{JCAP} {\bfseries 07} (2017) 047} [\href{https://arxiv.org/abs/1705.06788}{{\ttfamily 1705.06788}}].

\bibitem{Heurtier:2019eou}
L.~Heurtier and F.~Huang, \emph{{Inflaton portal to a highly decoupled EeV dark matter particle}}, \href{https://doi.org/10.1103/PhysRevD.100.043507}{\emph{Phys. Rev. D} {\bfseries 100} (2019) 043507} [\href{https://arxiv.org/abs/1905.05191}{{\ttfamily 1905.05191}}].

\bibitem{Jungman:1995df}
G.~Jungman, M.~Kamionkowski and K.~Griest, \emph{{Supersymmetric dark matter}}, \href{https://doi.org/10.1016/0370-1573(95)00058-5}{\emph{Phys. Rept.} {\bfseries 267} (1996) 195} [\href{https://arxiv.org/abs/hep-ph/9506380}{{\ttfamily hep-ph/9506380}}].

\bibitem{Bertone:2004pz}
G.~Bertone, D.~Hooper and J.~Silk, \emph{{Particle dark matter: Evidence, candidates and constraints}}, \href{https://doi.org/10.1016/j.physrep.2004.08.031}{\emph{Phys. Rept.} {\bfseries 405} (2005) 279} [\href{https://arxiv.org/abs/hep-ph/0404175}{{\ttfamily hep-ph/0404175}}].

\bibitem{Feng:2010gw}
J.L.~Feng, \emph{{Dark Matter Candidates from Particle Physics and Methods of Detection}}, \href{https://doi.org/10.1146/annurev-astro-082708-101659}{\emph{Ann. Rev. Astron. Astrophys.} {\bfseries 48} (2010) 495} [\href{https://arxiv.org/abs/1003.0904}{{\ttfamily 1003.0904}}].

\bibitem{Lee:1977ua}
B.W.~Lee and S.~Weinberg, \emph{{Cosmological Lower Bound on Heavy Neutrino Masses}}, \href{https://doi.org/10.1103/PhysRevLett.39.165}{\emph{Phys. Rev. Lett.} {\bfseries 39} (1977) 165}.

\bibitem{Scherrer:1985zt}
R.J.~Scherrer and M.S.~Turner, \emph{{On the Relic, Cosmic Abundance of Stable Weakly Interacting Massive Particles}}, \href{https://doi.org/10.1103/PhysRevD.33.1585}{\emph{Phys. Rev. D} {\bfseries 33} (1986) 1585}.

\bibitem{Srednicki:1988ce}
M.~Srednicki, R.~Watkins and K.A.~Olive, \emph{{Calculations of Relic Densities in the Early Universe}}, \href{https://doi.org/10.1016/0550-3213(88)90099-5}{\emph{Nucl. Phys. B} {\bfseries 310} (1988) 693}.

\bibitem{Gondolo:1990dk}
P.~Gondolo and G.~Gelmini, \emph{{Cosmic abundances of stable particles: Improved analysis}}, \href{https://doi.org/10.1016/0550-3213(91)90438-4}{\emph{Nucl. Phys. B} {\bfseries 360} (1991) 145}.

\bibitem{PandaX-II:2017hlx}
{\scshape PandaX-II} collaboration, \emph{{Dark Matter Results From 54-Ton-Day Exposure of PandaX-II Experiment}}, \href{https://doi.org/10.1103/PhysRevLett.119.181302}{\emph{Phys. Rev. Lett.} {\bfseries 119} (2017) 181302} [\href{https://arxiv.org/abs/1708.06917}{{\ttfamily 1708.06917}}].

\bibitem{PandaX:2018wtu}
{\scshape PandaX} collaboration, \emph{{Dark matter direct search sensitivity of the PandaX-4T experiment}}, \href{https://doi.org/10.1007/s11433-018-9259-0}{\emph{Sci. China Phys. Mech. Astron.} {\bfseries 62} (2019) 31011} [\href{https://arxiv.org/abs/1806.02229}{{\ttfamily 1806.02229}}].

\bibitem{XENON:2020kmp}
{\scshape XENON} collaboration, \emph{{Projected WIMP sensitivity of the XENONnT dark matter experiment}}, \href{https://doi.org/10.1088/1475-7516/2020/11/031}{\emph{JCAP} {\bfseries 11} (2020) 031} [\href{https://arxiv.org/abs/2007.08796}{{\ttfamily 2007.08796}}].

\bibitem{XENON:2018voc}
{\scshape XENON} collaboration, \emph{{Dark Matter Search Results from a One Ton-Year Exposure of XENON1T}}, \href{https://doi.org/10.1103/PhysRevLett.121.111302}{\emph{Phys. Rev. Lett.} {\bfseries 121} (2018) 111302} [\href{https://arxiv.org/abs/1805.12562}{{\ttfamily 1805.12562}}].

\bibitem{LUX-ZEPLIN:2018poe}
{\scshape LZ} collaboration, \emph{{Projected WIMP sensitivity of the LUX-ZEPLIN dark matter experiment}}, \href{https://doi.org/10.1103/PhysRevD.101.052002}{\emph{Phys. Rev. D} {\bfseries 101} (2020) 052002} [\href{https://arxiv.org/abs/1802.06039}{{\ttfamily 1802.06039}}].

\bibitem{DARWIN:2016hyl}
{\scshape DARWIN} collaboration, \emph{{DARWIN: towards the ultimate dark matter detector}}, \href{https://doi.org/10.1088/1475-7516/2016/11/017}{\emph{JCAP} {\bfseries 11} (2016) 017} [\href{https://arxiv.org/abs/1606.07001}{{\ttfamily 1606.07001}}].

\bibitem{HESS:2016mib}
{\scshape H.E.S.S.} collaboration, \emph{{Search for dark matter annihilations towards the inner Galactic halo from 10 years of observations with H.E.S.S}}, \href{https://doi.org/10.1103/PhysRevLett.117.111301}{\emph{Phys. Rev. Lett.} {\bfseries 117} (2016) 111301} [\href{https://arxiv.org/abs/1607.08142}{{\ttfamily 1607.08142}}].

\bibitem{MAGIC:2016xys}
{\scshape MAGIC, Fermi-LAT} collaboration, \emph{{Limits to Dark Matter Annihilation Cross-Section from a Combined Analysis of MAGIC and Fermi-LAT Observations of Dwarf Satellite Galaxies}}, \href{https://doi.org/10.1088/1475-7516/2016/02/039}{\emph{JCAP} {\bfseries 02} (2016) 039} [\href{https://arxiv.org/abs/1601.06590}{{\ttfamily 1601.06590}}].

\bibitem{ATLAS:2017bfj}
{\scshape ATLAS} collaboration, \emph{{Search for dark matter and other new phenomena in events with an energetic jet and large missing transverse momentum using the ATLAS detector}}, \href{https://doi.org/10.1007/JHEP01(2018)126}{\emph{JHEP} {\bfseries 01} (2018) 126} [\href{https://arxiv.org/abs/1711.03301}{{\ttfamily 1711.03301}}].

\bibitem{CMS:2017zts}
{\scshape CMS} collaboration, \emph{{Search for new physics in final states with an energetic jet or a hadronically decaying $W$ or $Z$ boson and transverse momentum imbalance at $\sqrt{s}=13\text{ }\text{ }\mathrm{TeV}$}}, \href{https://doi.org/10.1103/PhysRevD.97.092005}{\emph{Phys. Rev. D} {\bfseries 97} (2018) 092005} [\href{https://arxiv.org/abs/1712.02345}{{\ttfamily 1712.02345}}].

\bibitem{Gouttenoire:2022gwi}
Y.~Gouttenoire, \emph{{Beyond the Standard Model Cocktail}}, Springer Theses, Springer, Cham (2022), \href{https://doi.org/10.1007/978-3-031-11862-3}{10.1007/978-3-031-11862-3}, [\href{https://arxiv.org/abs/2207.01633}{{\ttfamily 2207.01633}}].

\bibitem{McDonald:1989jd}
J.~McDonald, \emph{{{WIMP} Densities in Decaying Particle Dominated Cosmology}}, \href{https://doi.org/10.1103/PhysRevD.43.1063}{\emph{Phys. Rev. D} {\bfseries 43} (1991) 1063}.

\bibitem{Moroi:1999zb}
T.~Moroi and L.~Randall, \emph{{Wino cold dark matter from anomaly mediated SUSY breaking}}, \href{https://doi.org/10.1016/S0550-3213(99)00748-8}{\emph{Nucl. Phys. B} {\bfseries 570} (2000) 455} [\href{https://arxiv.org/abs/hep-ph/9906527}{{\ttfamily hep-ph/9906527}}].

\bibitem{Visinelli:2009kt}
L.~Visinelli and P.~Gondolo, \emph{{Axion cold dark matter in non-standard cosmologies}}, \href{https://doi.org/10.1103/PhysRevD.81.063508}{\emph{Phys. Rev. D} {\bfseries 81} (2010) 063508} [\href{https://arxiv.org/abs/0912.0015}{{\ttfamily 0912.0015}}].

\bibitem{Erickcek:2015jza}
A.L.~Erickcek, \emph{{The Dark Matter Annihilation Boost from Low-Temperature Reheating}}, \href{https://doi.org/10.1103/PhysRevD.92.103505}{\emph{Phys. Rev. D} {\bfseries 92} (2015) 103505} [\href{https://arxiv.org/abs/1504.03335}{{\ttfamily 1504.03335}}].

\bibitem{Nelson:2018via}
A.E.~Nelson and H.~Xiao, \emph{{Axion Cosmology with Early Matter Domination}}, \href{https://doi.org/10.1103/PhysRevD.98.063516}{\emph{Phys. Rev. D} {\bfseries 98} (2018) 063516} [\href{https://arxiv.org/abs/1807.07176}{{\ttfamily 1807.07176}}].

\bibitem{Cheek:2023fht}
A.~Cheek, J.K.~Osi\'nski and L.~Roszkowski, \emph{{Extending preferred axion models via heavy-quark induced early matter domination}}, \href{https://doi.org/10.1088/1475-7516/2024/03/061}{\emph{JCAP} {\bfseries 03} (2024) 061} [\href{https://arxiv.org/abs/2310.16087}{{\ttfamily 2310.16087}}].

\bibitem{Cirelli:2018iax}
M.~Cirelli, Y.~Gouttenoire, K.~Petraki and F.~Sala, \emph{{Homeopathic Dark Matter, or how diluted heavy substances produce high energy cosmic rays}}, \href{https://doi.org/10.1088/1475-7516/2019/02/014}{\emph{JCAP} {\bfseries 02} (2019) 014} [\href{https://arxiv.org/abs/1811.03608}{{\ttfamily 1811.03608}}].

\bibitem{Gouttenoire:2019rtn}
Y.~Gouttenoire, G.~Servant and P.~Simakachorn, \emph{{BSM with Cosmic Strings: Heavy, up to EeV mass, Unstable Particles}}, \href{https://doi.org/10.1088/1475-7516/2020/07/016}{\emph{JCAP} {\bfseries 07} (2020) 016} [\href{https://arxiv.org/abs/1912.03245}{{\ttfamily 1912.03245}}].

\bibitem{Allahverdi:2020bys}
R.~Allahverdi et~al., \emph{{The First Three Seconds: a Review of Possible Expansion Histories of the Early Universe}},  \href{https://arxiv.org/abs/2006.16182}{{\ttfamily 2006.16182}}.

\bibitem{Allahverdi:2021grt}
R.~Allahverdi and J.K.~Osi\'nski, \emph{{Early matter domination from long-lived particles in the visible sector}}, \href{https://doi.org/10.1103/PhysRevD.105.023502}{\emph{Phys. Rev. D} {\bfseries 105} (2022) 023502} [\href{https://arxiv.org/abs/2108.13136}{{\ttfamily 2108.13136}}].

\bibitem{Allahverdi:2022zqr}
R.~Allahverdi, N.P.D.~Loc and J.K.~Osi\'nski, \emph{{Dark matter and baryogenesis from visible-sector long-lived particles}}, \href{https://doi.org/10.1103/PhysRevD.107.123510}{\emph{Phys. Rev. D} {\bfseries 107} (2023) 123510} [\href{https://arxiv.org/abs/2212.11303}{{\ttfamily 2212.11303}}].

\bibitem{Spokoiny:1993kt}
B.~Spokoiny, \emph{{Deflationary universe scenario}}, \href{https://doi.org/10.1016/0370-2693(93)90155-B}{\emph{Phys. Lett. B} {\bfseries 315} (1993) 40} [\href{https://arxiv.org/abs/gr-qc/9306008}{{\ttfamily gr-qc/9306008}}].

\bibitem{Joyce:1996cp}
M.~Joyce, \emph{{Electroweak Baryogenesis and the Expansion Rate of the Universe}}, \href{https://doi.org/10.1103/PhysRevD.55.1875}{\emph{Phys. Rev. D} {\bfseries 55} (1997) 1875} [\href{https://arxiv.org/abs/hep-ph/9606223}{{\ttfamily hep-ph/9606223}}].

\bibitem{Peebles:1998qn}
P.J.E.~Peebles and A.~Vilenkin, \emph{{Quintessential inflation}}, \href{https://doi.org/10.1103/PhysRevD.59.063505}{\emph{Phys. Rev. D} {\bfseries 59} (1999) 063505} [\href{https://arxiv.org/abs/astro-ph/9810509}{{\ttfamily astro-ph/9810509}}].

\bibitem{Poulin:2018dzj}
V.~Poulin, T.L.~Smith, D.~Grin, T.~Karwal and M.~Kamionkowski, \emph{{Cosmological implications of ultralight axionlike fields}}, \href{https://doi.org/10.1103/PhysRevD.98.083525}{\emph{Phys. Rev. D} {\bfseries 98} (2018) 083525} [\href{https://arxiv.org/abs/1806.10608}{{\ttfamily 1806.10608}}].

\bibitem{Gouttenoire:2021jhk}
Y.~Gouttenoire, G.~Servant and P.~Simakachorn, \emph{{Kination cosmology from scalar fields and gravitational-wave signatures}},  \href{https://arxiv.org/abs/2111.01150}{{\ttfamily 2111.01150}}.

\bibitem{Gouttenoire:2021wzu}
Y.~Gouttenoire, G.~Servant and P.~Simakachorn, \emph{{Revealing the Primordial Irreducible Inflationary Gravitational-Wave Background with a Spinning Peccei-Quinn Axion}},  \href{https://arxiv.org/abs/2108.10328}{{\ttfamily 2108.10328}}.

\bibitem{Co:2021lkc}
R.T.~Co, D.~Dunsky, N.~Fernandez, A.~Ghalsasi, L.J.~Hall, K.~Harigaya et~al., \emph{{Gravitational wave and CMB probes of axion kination}}, \href{https://doi.org/10.1007/JHEP09(2022)116}{\emph{JHEP} {\bfseries 09} (2022) 116} [\href{https://arxiv.org/abs/2108.09299}{{\ttfamily 2108.09299}}].

\bibitem{Heurtier:2022rhf}
L.~Heurtier, A.~Moursy and L.~Wacquez, \emph{{Cosmological imprints of SUSY breaking in models of sgoldstinoless non-oscillatory inflation}}, \href{https://doi.org/10.1088/1475-7516/2023/03/020}{\emph{JCAP} {\bfseries 03} (2023) 020} [\href{https://arxiv.org/abs/2207.11502}{{\ttfamily 2207.11502}}].

\bibitem{Chen:2025awt}
C.~Chen, S.~Jyoti~Das, K.~Dimopoulos and A.~Ghoshal, \emph{{Flipped Rotating Axion Non-minimally Coupled to Gravity: Baryogenesis and Dark Matter}},  \href{https://arxiv.org/abs/2502.08720}{{\ttfamily 2502.08720}}.

\bibitem{Guth:1980zk}
A.H.~Guth and E.J.~Weinberg, \emph{{A Cosmological Lower Bound on the Higgs Boson Mass}}, \href{https://doi.org/10.1103/PhysRevLett.45.1131}{\emph{Phys. Rev. Lett.} {\bfseries 45} (1980) 1131}.

\bibitem{Witten:1980ez}
E.~Witten, \emph{{Cosmological Consequences of a Light Higgs Boson}}, \href{https://doi.org/10.1016/0550-3213(81)90182-6}{\emph{Nucl. Phys. B} {\bfseries 177} (1981) 477}.

\bibitem{Creminelli:2001th}
P.~Creminelli, A.~Nicolis and R.~Rattazzi, \emph{{Holography and the electroweak phase transition}}, \href{https://doi.org/10.1088/1126-6708/2002/03/051}{\emph{JHEP} {\bfseries 03} (2002) 051} [\href{https://arxiv.org/abs/hep-th/0107141}{{\ttfamily hep-th/0107141}}].

\bibitem{Randall:2006py}
L.~Randall and G.~Servant, \emph{{Gravitational waves from warped spacetime}}, \href{https://doi.org/10.1088/1126-6708/2007/05/054}{\emph{JHEP} {\bfseries 05} (2007) 054} [\href{https://arxiv.org/abs/hep-ph/0607158}{{\ttfamily hep-ph/0607158}}].

\bibitem{Konstandin:2011dr}
T.~Konstandin and G.~Servant, \emph{{Cosmological Consequences of Nearly Conformal Dynamics at the TeV scale}}, \href{https://doi.org/10.1088/1475-7516/2011/12/009}{\emph{JCAP} {\bfseries 12} (2011) 009} [\href{https://arxiv.org/abs/1104.4791}{{\ttfamily 1104.4791}}].

\bibitem{Baratella:2018pxi}
P.~Baratella, A.~Pomarol and F.~Rompineve, \emph{{The Supercooled Universe}}, \href{https://doi.org/10.1007/JHEP03(2019)100}{\emph{JHEP} {\bfseries 03} (2019) 100} [\href{https://arxiv.org/abs/1812.06996}{{\ttfamily 1812.06996}}].

\bibitem{Baldes:2020kam}
I.~Baldes, Y.~Gouttenoire and F.~Sala, \emph{{String Fragmentation in Supercooled Confinement and Implications for Dark Matter}}, \href{https://doi.org/10.1007/JHEP04(2021)278}{\emph{JHEP} {\bfseries 04} (2021) 278} [\href{https://arxiv.org/abs/2007.08440}{{\ttfamily 2007.08440}}].

\bibitem{Baldes:2021aph}
I.~Baldes, Y.~Gouttenoire, F.~Sala and G.~Servant, \emph{{Supercool composite Dark Matter beyond 100 TeV}}, \href{https://doi.org/10.1007/JHEP07(2022)084}{\emph{JHEP} {\bfseries 07} (2022) 084} [\href{https://arxiv.org/abs/2110.13926}{{\ttfamily 2110.13926}}].

\bibitem{LIGOScientific:2016aoc}
{\scshape LIGO Scientific, Virgo} collaboration, \emph{{Observation of Gravitational Waves from a Binary Black Hole Merger}}, \href{https://doi.org/10.1103/PhysRevLett.116.061102}{\emph{Phys. Rev. Lett.} {\bfseries 116} (2016) 061102} [\href{https://arxiv.org/abs/1602.03837}{{\ttfamily 1602.03837}}].

\bibitem{LIGOScientific:2016sjg}
{\scshape LIGO Scientific, Virgo} collaboration, \emph{{GW151226: Observation of Gravitational Waves from a 22-Solar-Mass Binary Black Hole Coalescence}}, \href{https://doi.org/10.1103/PhysRevLett.116.241103}{\emph{Phys. Rev. Lett.} {\bfseries 116} (2016) 241103} [\href{https://arxiv.org/abs/1606.04855}{{\ttfamily 1606.04855}}].

\bibitem{Carilli:2004nx}
C.L.~Carilli and S.~Rawlings, \emph{{Science with the Square Kilometer Array: Motivation, key science projects, standards and assumptions}}, \href{https://doi.org/10.1016/j.newar.2004.09.001}{\emph{New Astron. Rev.} {\bfseries 48} (2004) 979} [\href{https://arxiv.org/abs/astro-ph/0409274}{{\ttfamily astro-ph/0409274}}].

\bibitem{Janssen:2014dka}
G.~Janssen et~al., \emph{{Gravitational wave astronomy with the SKA}}, \href{https://doi.org/10.22323/1.215.0037}{\emph{PoS} {\bfseries AASKA14} (2015) 037} [\href{https://arxiv.org/abs/1501.00127}{{\ttfamily 1501.00127}}].

\bibitem{Weltman:2018zrl}
A.~Weltman et~al., \emph{{Fundamental physics with the Square Kilometre Array}}, \href{https://doi.org/10.1017/pasa.2019.42}{\emph{Publ. Astron. Soc. Austral.} {\bfseries 37} (2020) e002} [\href{https://arxiv.org/abs/1810.02680}{{\ttfamily 1810.02680}}].

\bibitem{EPTA:2015qep}
{\scshape EPTA} collaboration, \emph{{European Pulsar Timing Array Limits On An Isotropic Stochastic Gravitational-Wave Background}}, \href{https://doi.org/10.1093/mnras/stv1538}{\emph{Mon. Not. Roy. Astron. Soc.} {\bfseries 453} (2015) 2576} [\href{https://arxiv.org/abs/1504.03692}{{\ttfamily 1504.03692}}].

\bibitem{EPTA:2015gke}
{\scshape EPTA} collaboration, \emph{{European Pulsar Timing Array Limits on Continuous Gravitational Waves from Individual Supermassive Black Hole Binaries}}, \href{https://doi.org/10.1093/mnras/stv2092}{\emph{Mon. Not. Roy. Astron. Soc.} {\bfseries 455} (2016) 1665} [\href{https://arxiv.org/abs/1509.02165}{{\ttfamily 1509.02165}}].

\bibitem{NANOGrav:2023gor}
{\scshape NANOGrav} collaboration, \emph{{The NANOGrav 15 yr Data Set: Evidence for a Gravitational-wave Background}}, \href{https://doi.org/10.3847/2041-8213/acdac6}{\emph{Astrophys. J. Lett.} {\bfseries 951} (2023) L8} [\href{https://arxiv.org/abs/2306.16213}{{\ttfamily 2306.16213}}].

\bibitem{NANOGrav:2023hvm}
{\scshape NANOGrav} collaboration, \emph{{The NANOGrav 15 yr Data Set: Search for Signals from New Physics}}, \href{https://doi.org/10.3847/2041-8213/acdc91}{\emph{Astrophys. J. Lett.} {\bfseries 951} (2023) L11} [\href{https://arxiv.org/abs/2306.16219}{{\ttfamily 2306.16219}}].

\bibitem{Grishchuk:1974ny}
L.P.~Grishchuk, \emph{{Amplification of gravitational waves in an istropic universe}}, {\emph{Zh. Eksp. Teor. Fiz.} {\bfseries 67} (1974) 825}.

\bibitem{Starobinsky:1979ty}
A.A.~Starobinsky, \emph{{Spectrum of relict gravitational radiation and the early state of the universe}}, {\emph{JETP Lett.} {\bfseries 30} (1979) 682}.

\bibitem{Rubakov:1982df}
V.A.~Rubakov, M.V.~Sazhin and A.V.~Veryaskin, \emph{{Graviton Creation in the Inflationary Universe and the Grand Unification Scale}}, \href{https://doi.org/10.1016/0370-2693(82)90641-4}{\emph{Phys. Lett. B} {\bfseries 115} (1982) 189}.

\bibitem{Guzzetti:2016mkm}
M.C.~Guzzetti, N.~Bartolo, M.~Liguori and S.~Matarrese, \emph{{Gravitational waves from inflation}}, \href{https://doi.org/10.1393/ncr/i2016-10127-1}{\emph{Riv. Nuovo Cim.} {\bfseries 39} (2016) 399} [\href{https://arxiv.org/abs/1605.01615}{{\ttfamily 1605.01615}}].

\bibitem{Seto:2003kc}
N.~Seto and J.~Yokoyama, \emph{{Probing the equation of state of the early universe with a space laser interferometer}}, \href{https://doi.org/10.1143/JPSJ.72.3082}{\emph{J. Phys. Soc. Jap.} {\bfseries 72} (2003) 3082} [\href{https://arxiv.org/abs/gr-qc/0305096}{{\ttfamily gr-qc/0305096}}].

\bibitem{Boyle:2005se}
L.A.~Boyle and P.J.~Steinhardt, \emph{{Probing the early universe with inflationary gravitational waves}}, \href{https://doi.org/10.1103/PhysRevD.77.063504}{\emph{Phys. Rev. D} {\bfseries 77} (2008) 063504} [\href{https://arxiv.org/abs/astro-ph/0512014}{{\ttfamily astro-ph/0512014}}].

\bibitem{Boyle:2007zx}
L.A.~Boyle and A.~Buonanno, \emph{{Relating gravitational wave constraints from primordial nucleosynthesis, pulsar timing, laser interferometers, and the CMB: Implications for the early Universe}}, \href{https://doi.org/10.1103/PhysRevD.78.043531}{\emph{Phys. Rev. D} {\bfseries 78} (2008) 043531} [\href{https://arxiv.org/abs/0708.2279}{{\ttfamily 0708.2279}}].

\bibitem{Kuroyanagi:2008ye}
S.~Kuroyanagi, T.~Chiba and N.~Sugiyama, \emph{{Precision calculations of the gravitational wave background spectrum from inflation}}, \href{https://doi.org/10.1103/PhysRevD.79.103501}{\emph{Phys. Rev. D} {\bfseries 79} (2009) 103501} [\href{https://arxiv.org/abs/0804.3249}{{\ttfamily 0804.3249}}].

\bibitem{Nakayama:2009ce}
K.~Nakayama and J.~Yokoyama, \emph{{Gravitational Wave Background and Non-Gaussianity as a Probe of the Curvaton Scenario}}, \href{https://doi.org/10.1088/1475-7516/2010/01/010}{\emph{JCAP} {\bfseries 01} (2010) 010} [\href{https://arxiv.org/abs/0910.0715}{{\ttfamily 0910.0715}}].

\bibitem{Kuroyanagi:2013ns}
S.~Kuroyanagi, C.~Ringeval and T.~Takahashi, \emph{{Early universe tomography with CMB and gravitational waves}}, \href{https://doi.org/10.1103/PhysRevD.87.083502}{\emph{Phys. Rev. D} {\bfseries 87} (2013) 083502} [\href{https://arxiv.org/abs/1301.1778}{{\ttfamily 1301.1778}}].

\bibitem{Jinno:2013xqa}
R.~Jinno, T.~Moroi and K.~Nakayama, \emph{{Inflationary Gravitational Waves and the Evolution of the Early Universe}}, \href{https://doi.org/10.1088/1475-7516/2014/01/040}{\emph{JCAP} {\bfseries 01} (2014) 040} [\href{https://arxiv.org/abs/1307.3010}{{\ttfamily 1307.3010}}].

\bibitem{Saikawa:2018rcs}
K.~Saikawa and S.~Shirai, \emph{{Primordial gravitational waves, precisely: The role of thermodynamics in the Standard Model}}, \href{https://doi.org/10.1088/1475-7516/2018/05/035}{\emph{JCAP} {\bfseries 05} (2018) 035} [\href{https://arxiv.org/abs/1803.01038}{{\ttfamily 1803.01038}}].

\bibitem{Nakayama:2008ip}
K.~Nakayama, S.~Saito, Y.~Suwa and J.~Yokoyama, \emph{{Space laser interferometers can determine the thermal history of the early Universe}}, \href{https://doi.org/10.1103/PhysRevD.77.124001}{\emph{Phys. Rev. D} {\bfseries 77} (2008) 124001} [\href{https://arxiv.org/abs/0802.2452}{{\ttfamily 0802.2452}}].

\bibitem{Nakayama:2008wy}
K.~Nakayama, S.~Saito, Y.~Suwa and J.~Yokoyama, \emph{{Probing reheating temperature of the universe with gravitational wave background}}, \href{https://doi.org/10.1088/1475-7516/2008/06/020}{\emph{JCAP} {\bfseries 06} (2008) 020} [\href{https://arxiv.org/abs/0804.1827}{{\ttfamily 0804.1827}}].

\bibitem{Kuroyanagi:2011fy}
S.~Kuroyanagi, K.~Nakayama and S.~Saito, \emph{{Prospects for determination of thermal history after inflation with future gravitational wave detectors}}, \href{https://doi.org/10.1103/PhysRevD.84.123513}{\emph{Phys. Rev. D} {\bfseries 84} (2011) 123513} [\href{https://arxiv.org/abs/1110.4169}{{\ttfamily 1110.4169}}].

\bibitem{Buchmuller:2013lra}
W.~Buchm\"uller, V.~Domcke, K.~Kamada and K.~Schmitz, \emph{{The Gravitational Wave Spectrum from Cosmological $B-L$ Breaking}}, \href{https://doi.org/10.1088/1475-7516/2013/10/003}{\emph{JCAP} {\bfseries 10} (2013) 003} [\href{https://arxiv.org/abs/1305.3392}{{\ttfamily 1305.3392}}].

\bibitem{Buchmuller:2013dja}
W.~Buchmuller, V.~Domcke, K.~Kamada and K.~Schmitz, \emph{{A Minimal Supersymmetric Model of Particle Physics and the Early Universe}},  \href{https://arxiv.org/abs/1309.7788}{{\ttfamily 1309.7788}}.

\bibitem{Jinno:2014qka}
R.~Jinno, T.~Moroi and T.~Takahashi, \emph{{Studying Inflation with Future Space-Based Gravitational Wave Detectors}}, \href{https://doi.org/10.1088/1475-7516/2014/12/006}{\emph{JCAP} {\bfseries 12} (2014) 006} [\href{https://arxiv.org/abs/1406.1666}{{\ttfamily 1406.1666}}].

\bibitem{Kuroyanagi:2014qza}
S.~Kuroyanagi, K.~Nakayama and J.~Yokoyama, \emph{{Prospects of determination of reheating temperature after inflation by DECIGO}}, \href{https://doi.org/10.1093/ptep/ptu176}{\emph{PTEP} {\bfseries 2015} (2015) 013E02} [\href{https://arxiv.org/abs/1410.6618}{{\ttfamily 1410.6618}}].

\bibitem{Maity:2024cpq}
S.~Maity and M.R.~Haque, \emph{{Probing the early universe with future GW observatories}},  \href{https://arxiv.org/abs/2407.18246}{{\ttfamily 2407.18246}}.

\bibitem{Haque_2021}
M.R.~Haque, D.~Maity, T.~Paul and L.~Sriramkumar, \emph{Decoding the phases of early and late time reheating through imprints on primordial gravitational waves}, \href{https://doi.org/10.1103/physrevd.104.063513}{\emph{Physical Review D} {\bfseries 104} (2021) }.

\bibitem{Chakraborty:2023ocr}
A.~Chakraborty, M.R.~Haque, D.~Maity and R.~Mondal, \emph{{Inflaton phenomenology via reheating in light of primordial gravitational waves and the latest BICEP/Keck data}}, \href{https://doi.org/10.1103/PhysRevD.108.023515}{\emph{Phys. Rev. D} {\bfseries 108} (2023) 023515} [\href{https://arxiv.org/abs/2304.13637}{{\ttfamily 2304.13637}}].

\bibitem{Berbig:2023yyy}
M.~Berbig and A.~Ghoshal, \emph{{Impact of high-scale Seesaw and Leptogenesis on inflationary tensor perturbations as detectable gravitational waves}}, \href{https://doi.org/10.1007/JHEP05(2023)172}{\emph{JHEP} {\bfseries 05} (2023) 172} [\href{https://arxiv.org/abs/2301.05672}{{\ttfamily 2301.05672}}].

\bibitem{Borboruah:2024eha}
Z.A.~Borboruah, A.~Ghoshal, L.~Malhotra and U.~Yajnik, \emph{{Inflationary Gravitational Wave Spectral Shapes as test for Low-Scale Leptogenesis}},  \href{https://arxiv.org/abs/2405.06603}{{\ttfamily 2405.06603}}.

\bibitem{Cheek:2025gvx}
A.~Cheek, A.~Ghoshal and D.~Paul, \emph{{Axion Dark Matter Archaeology with Primordial Gravitational Waves}},  \href{https://arxiv.org/abs/2505.04614}{{\ttfamily 2505.04614}}.

\bibitem{Borboruah:2024eal}
Z.A.~Borboruah, A.~Ghoshal and S.~Ipek, \emph{{Probing flavor violation and baryogenesis via primordial gravitational waves}}, \href{https://doi.org/10.1007/JHEP07(2024)228}{\emph{JHEP} {\bfseries 07} (2024) 228} [\href{https://arxiv.org/abs/2405.03241}{{\ttfamily 2405.03241}}].

\bibitem{Datta:2022tab}
S.~Datta and R.~Samanta, \emph{{Gravitational waves-tomography of Low-Scale-Leptogenesis}}, \href{https://doi.org/10.1007/JHEP11(2022)159}{\emph{JHEP} {\bfseries 11} (2022) 159} [\href{https://arxiv.org/abs/2208.09949}{{\ttfamily 2208.09949}}].

\bibitem{Datta:2023vbs}
S.~Datta and R.~Samanta, \emph{{Fingerprints of GeV scale right-handed neutrinos on inflationary gravitational waves and PTA data}}, \href{https://doi.org/10.1103/PhysRevD.108.L091706}{\emph{Phys. Rev. D} {\bfseries 108} (2023) L091706} [\href{https://arxiv.org/abs/2307.00646}{{\ttfamily 2307.00646}}].

\bibitem{Chianese:2024nyw}
M.~Chianese, S.~Datta, R.~Samanta and N.~Saviano, \emph{{Tomography of flavoured leptogenesis with primordial blue gravitational waves}}, \href{https://doi.org/10.1088/1475-7516/2024/11/051}{\emph{JCAP} {\bfseries 11} (2024) 051} [\href{https://arxiv.org/abs/2405.00641}{{\ttfamily 2405.00641}}].

\bibitem{Datta:2025yow}
S.~Datta, A.~Ghosal, A.~Ghoshal and G.~White, \emph{{Complementarity between Cosmic String Gravitational Waves and long lived particle searches in laboratory}},  \href{https://arxiv.org/abs/2501.03326}{{\ttfamily 2501.03326}}.

\bibitem{Borboruah:2025hai}
Z.A.~Borboruah, L.~Malhotra, F.F.~Deppisch and A.~Ghoshal, \emph{{Inflationary Gravitational Waves and Laboratory Searches as Complementary Probes of Right-handed Neutrinos}},  \href{https://arxiv.org/abs/2504.15374}{{\ttfamily 2504.15374}}.

\bibitem{Dutta:2009uf}
B.~Dutta, L.~Leblond and K.~Sinha, \emph{{Mirage in the Sky: Non-thermal Dark Matter, Gravitino Problem, and Cosmic Ray Anomalies}}, \href{https://doi.org/10.1103/PhysRevD.80.035014}{\emph{Phys. Rev. D} {\bfseries 80} (2009) 035014} [\href{https://arxiv.org/abs/0904.3773}{{\ttfamily 0904.3773}}].

\bibitem{Acharya:2009zt}
B.S.~Acharya, G.~Kane, S.~Watson and P.~Kumar, \emph{{A Non-thermal WIMP Miracle}}, \href{https://doi.org/10.1103/PhysRevD.80.083529}{\emph{Phys. Rev. D} {\bfseries 80} (2009) 083529} [\href{https://arxiv.org/abs/0908.2430}{{\ttfamily 0908.2430}}].

\bibitem{Baumann:2007zm}
D.~Baumann, P.J.~Steinhardt, K.~Takahashi and K.~Ichiki, \emph{{Gravitational Wave Spectrum Induced by Primordial Scalar Perturbations}}, \href{https://doi.org/10.1103/PhysRevD.76.084019}{\emph{Phys. Rev. D} {\bfseries 76} (2007) 084019} [\href{https://arxiv.org/abs/hep-th/0703290}{{\ttfamily hep-th/0703290}}].

\bibitem{Kohri:2018awv}
K.~Kohri and T.~Terada, \emph{{Semianalytic calculation of gravitational wave spectrum nonlinearly induced from primordial curvature perturbations}}, \href{https://doi.org/10.1103/PhysRevD.97.123532}{\emph{Phys. Rev. D} {\bfseries 97} (2018) 123532} [\href{https://arxiv.org/abs/1804.08577}{{\ttfamily 1804.08577}}].

\bibitem{Bhaumik:2025kuj}
A.~Bhaumik, T.~Papanikolaou and A.~Ghoshal, \emph{{Vector induced Gravitational Waves sourced by Primordial Magnetic Fields}},  \href{https://arxiv.org/abs/2504.10477}{{\ttfamily 2504.10477}}.

\bibitem{Planck:2018jri}
{\scshape Planck} collaboration, \emph{{Planck 2018 results. X. Constraints on inflation}}, \href{https://doi.org/10.1051/0004-6361/201833887}{\emph{Astron. Astrophys.} {\bfseries 641} (2020) A10} [\href{https://arxiv.org/abs/1807.06211}{{\ttfamily 1807.06211}}].

\bibitem{BICEP:2021xfz}
{\scshape BICEP, Keck} collaboration, \emph{{Improved Constraints on Primordial Gravitational Waves using Planck, WMAP, and BICEP/Keck Observations through the 2018 Observing Season}}, \href{https://doi.org/10.1103/PhysRevLett.127.151301}{\emph{Phys. Rev. Lett.} {\bfseries 127} (2021) 151301} [\href{https://arxiv.org/abs/2110.00483}{{\ttfamily 2110.00483}}].

\bibitem{BICEP2:2018kqh}
{\scshape BICEP2, Keck Array} collaboration, \emph{{BICEP2 / Keck Array x: Constraints on Primordial Gravitational Waves using Planck, WMAP, and New BICEP2/Keck Observations through the 2015 Season}}, \href{https://doi.org/10.1103/PhysRevLett.121.221301}{\emph{Phys. Rev. Lett.} {\bfseries 121} (2018) 221301} [\href{https://arxiv.org/abs/1810.05216}{{\ttfamily 1810.05216}}].

\bibitem{Liddle:1993fq}
A.R.~Liddle and D.H.~Lyth, \emph{{The Cold dark matter density perturbation}}, \href{https://doi.org/10.1016/0370-1573(93)90114-S}{\emph{Phys. Rept.} {\bfseries 231} (1993) 1} [\href{https://arxiv.org/abs/astro-ph/9303019}{{\ttfamily astro-ph/9303019}}].

\bibitem{Brandenberger:2006xi}
R.H.~Brandenberger, A.~Nayeri, S.P.~Patil and C.~Vafa, \emph{{Tensor Modes from a Primordial Hagedorn Phase of String Cosmology}}, \href{https://doi.org/10.1103/PhysRevLett.98.231302}{\emph{Phys. Rev. Lett.} {\bfseries 98} (2007) 231302} [\href{https://arxiv.org/abs/hep-th/0604126}{{\ttfamily hep-th/0604126}}].

\bibitem{Calcagni:2013lya}
G.~Calcagni, S.~Kuroyanagi, J.~Ohashi and S.~Tsujikawa, \emph{{Strong Planck constraints on braneworld and non-commutative inflation}}, \href{https://doi.org/10.1088/1475-7516/2014/03/052}{\emph{JCAP} {\bfseries 03} (2014) 052} [\href{https://arxiv.org/abs/1310.5186}{{\ttfamily 1310.5186}}].

\bibitem{Fujita:2018ehq}
T.~Fujita, S.~Kuroyanagi, S.~Mizuno and S.~Mukohyama, \emph{{Blue-tilted Primordial Gravitational Waves from Massive Gravity}}, \href{https://doi.org/10.1016/j.physletb.2018.12.025}{\emph{Phys. Lett. B} {\bfseries 789} (2019) 215} [\href{https://arxiv.org/abs/1808.02381}{{\ttfamily 1808.02381}}].

\bibitem{Cook:2011hg}
J.L.~Cook and L.~Sorbo, \emph{{Particle production during inflation and gravitational waves detectable by ground-based interferometers}}, \href{https://doi.org/10.1103/PhysRevD.85.023534}{\emph{Phys. Rev. D} {\bfseries 85} (2012) 023534} [\href{https://arxiv.org/abs/1109.0022}{{\ttfamily 1109.0022}}].

\bibitem{Mukohyama:2014gba}
S.~Mukohyama, R.~Namba, M.~Peloso and G.~Shiu, \emph{{Blue Tensor Spectrum from Particle Production during Inflation}}, \href{https://doi.org/10.1088/1475-7516/2014/08/036}{\emph{JCAP} {\bfseries 08} (2014) 036} [\href{https://arxiv.org/abs/1405.0346}{{\ttfamily 1405.0346}}].

\bibitem{Baldi:2005gk}
M.~Baldi, F.~Finelli and S.~Matarrese, \emph{{Inflation with violation of the null energy condition}}, \href{https://doi.org/10.1103/PhysRevD.72.083504}{\emph{Phys. Rev. D} {\bfseries 72} (2005) 083504} [\href{https://arxiv.org/abs/astro-ph/0505552}{{\ttfamily astro-ph/0505552}}].

\bibitem{Kobayashi:2010cm}
T.~Kobayashi, M.~Yamaguchi and J.~Yokoyama, \emph{{G-inflation: Inflation driven by the Galileon field}}, \href{https://doi.org/10.1103/PhysRevLett.105.231302}{\emph{Phys. Rev. Lett.} {\bfseries 105} (2010) 231302} [\href{https://arxiv.org/abs/1008.0603}{{\ttfamily 1008.0603}}].

\bibitem{Anber:2009ua}
M.M.~Anber and L.~Sorbo, \emph{{Naturally inflating on steep potentials through electromagnetic dissipation}}, \href{https://doi.org/10.1103/PhysRevD.81.043534}{\emph{Phys. Rev. D} {\bfseries 81} (2010) 043534} [\href{https://arxiv.org/abs/0908.4089}{{\ttfamily 0908.4089}}].

\bibitem{Adshead:2016omu}
P.~Adshead, E.~Martinec, E.I.~Sfakianakis and M.~Wyman, \emph{{Higgsed Chromo-Natural Inflation}}, \href{https://doi.org/10.1007/JHEP12(2016)137}{\emph{JHEP} {\bfseries 12} (2016) 137} [\href{https://arxiv.org/abs/1609.04025}{{\ttfamily 1609.04025}}].

\bibitem{Adshead:2017hnc}
P.~Adshead and E.I.~Sfakianakis, \emph{{Higgsed Gauge-flation}}, \href{https://doi.org/10.1007/JHEP08(2017)130}{\emph{JHEP} {\bfseries 08} (2017) 130} [\href{https://arxiv.org/abs/1705.03024}{{\ttfamily 1705.03024}}].

\bibitem{Caldwell:2017chz}
R.R.~Caldwell and C.~Devulder, \emph{{Axion Gauge Field Inflation and Gravitational Leptogenesis: A Lower Bound on B Modes from the Matter-Antimatter Asymmetry of the Universe}}, \href{https://doi.org/10.1103/PhysRevD.97.023532}{\emph{Phys. Rev. D} {\bfseries 97} (2018) 023532} [\href{https://arxiv.org/abs/1706.03765}{{\ttfamily 1706.03765}}].

\bibitem{Dimastrogiovanni:2018xnn}
E.~Dimastrogiovanni, M.~Fasiello, R.J.~Hardwick, H.~Assadullahi, K.~Koyama and D.~Wands, \emph{{Non-Gaussianity from Axion-Gauge Fields Interactions during Inflation}}, \href{https://doi.org/10.1088/1475-7516/2018/11/029}{\emph{JCAP} {\bfseries 11} (2018) 029} [\href{https://arxiv.org/abs/1806.05474}{{\ttfamily 1806.05474}}].

\bibitem{Oikonomou:2021kql}
V.K.~Oikonomou, \emph{{A refined Einstein\textendash{}Gauss\textendash{}Bonnet inflationary theoretical framework}}, \href{https://doi.org/10.1088/1361-6382/ac2168}{\emph{Class. Quant. Grav.} {\bfseries 38} (2021) 195025} [\href{https://arxiv.org/abs/2108.10460}{{\ttfamily 2108.10460}}].

\bibitem{Kuroyanagi:2020sfw}
S.~Kuroyanagi, T.~Takahashi and S.~Yokoyama, \emph{{Blue-tilted inflationary tensor spectrum and reheating in the light of NANOGrav results}}, \href{https://doi.org/10.1088/1475-7516/2021/01/071}{\emph{JCAP} {\bfseries 01} (2021) 071} [\href{https://arxiv.org/abs/2011.03323}{{\ttfamily 2011.03323}}].

\bibitem{Maggiore:1999vm}
M.~Maggiore, \emph{{Gravitational wave experiments and early universe cosmology}}, \href{https://doi.org/10.1016/S0370-1573(99)00102-7}{\emph{Phys. Rept.} {\bfseries 331} (2000) 283} [\href{https://arxiv.org/abs/gr-qc/9909001}{{\ttfamily gr-qc/9909001}}].

\bibitem{Cyburt:2015mya}
R.H.~Cyburt, B.D.~Fields, K.A.~Olive and T.-H.~Yeh, \emph{{Big Bang Nucleosynthesis: 2015}}, \href{https://doi.org/10.1103/RevModPhys.88.015004}{\emph{Rev. Mod. Phys.} {\bfseries 88} (2016) 015004} [\href{https://arxiv.org/abs/1505.01076}{{\ttfamily 1505.01076}}].

\bibitem{CORE:2017oje}
{\scshape CORE} collaboration, \emph{{Exploring cosmic origins with CORE: Survey requirements and mission design}}, \href{https://doi.org/10.1088/1475-7516/2018/04/014}{\emph{JCAP} {\bfseries 04} (2018) 014} [\href{https://arxiv.org/abs/1706.04516}{{\ttfamily 1706.04516}}].

\bibitem{SPT-3G:2014dbx}
{\scshape SPT-3G} collaboration, \emph{{SPT-3G: A Next-Generation Cosmic Microwave Background Polarization Experiment on the South Pole Telescope}}, \href{https://doi.org/10.1117/12.2057305}{\emph{Proc. SPIE Int. Soc. Opt. Eng.} {\bfseries 9153} (2014) 91531P} [\href{https://arxiv.org/abs/1407.2973}{{\ttfamily 1407.2973}}].

\bibitem{SimonsObservatory:2018koc}
{\scshape Simons Observatory} collaboration, \emph{{The Simons Observatory: Science goals and forecasts}}, \href{https://doi.org/10.1088/1475-7516/2019/02/056}{\emph{JCAP} {\bfseries 02} (2019) 056} [\href{https://arxiv.org/abs/1808.07445}{{\ttfamily 1808.07445}}].

\bibitem{Abazajian:2019eic}
K.~Abazajian et~al., \emph{{CMB-S4 Science Case, Reference Design, and Project Plan}},  \href{https://arxiv.org/abs/1907.04473}{{\ttfamily 1907.04473}}.

\bibitem{TopicalConvenersKNAbazajianJECarlstromATLee:2013bxd}
{\scshape Topical Conveners: K.N. Abazajian, J.E. Carlstrom, A.T. Lee} collaboration, \emph{{Neutrino Physics from the Cosmic Microwave Background and Large Scale Structure}}, \href{https://doi.org/10.1016/j.astropartphys.2014.05.014}{\emph{Astropart. Phys.} {\bfseries 63} (2015) 66} [\href{https://arxiv.org/abs/1309.5383}{{\ttfamily 1309.5383}}].

\bibitem{NASAPICO:2019thw}
{\scshape NASA PICO} collaboration, \emph{{PICO: Probe of Inflation and Cosmic Origins}},  \href{https://arxiv.org/abs/1902.10541}{{\ttfamily 1902.10541}}.

\bibitem{CMB-bharat}
{\scshape CMB Bharat Collaboration} collaboration, \emph{{CMB Bharat}}, .

\bibitem{CMB-HD:2022bsz}
{\scshape CMB-HD} collaboration, \emph{{Snowmass2021 CMB-HD White Paper}},  \href{https://arxiv.org/abs/2203.05728}{{\ttfamily 2203.05728}}.

\bibitem{Turner:1993vb}
M.S.~Turner, M.J.~White and J.E.~Lidsey, \emph{{Tensor perturbations in inflationary models as a probe of cosmology}}, \href{https://doi.org/10.1103/PhysRevD.48.4613}{\emph{Phys. Rev. D} {\bfseries 48} (1993) 4613} [\href{https://arxiv.org/abs/astro-ph/9306029}{{\ttfamily astro-ph/9306029}}].

\bibitem{Chongchitnan:2006pe}
S.~Chongchitnan and G.~Efstathiou, \emph{{Prospects for direct detection of primordial gravitational waves}}, \href{https://doi.org/10.1103/PhysRevD.73.083511}{\emph{Phys. Rev. D} {\bfseries 73} (2006) 083511} [\href{https://arxiv.org/abs/astro-ph/0602594}{{\ttfamily astro-ph/0602594}}].

\bibitem{Kuroyanagi:2014nba}
S.~Kuroyanagi, T.~Takahashi and S.~Yokoyama, \emph{{Blue-tilted Tensor Spectrum and Thermal History of the Universe}}, \href{https://doi.org/10.1088/1475-7516/2015/02/003}{\emph{JCAP} {\bfseries 02} (2015) 003} [\href{https://arxiv.org/abs/1407.4785}{{\ttfamily 1407.4785}}].

\bibitem{LIGOScientific:2017bnn}
{\scshape LIGO Scientific, VIRGO} collaboration, \emph{{GW170104: Observation of a 50-Solar-Mass Binary Black Hole Coalescence at Redshift 0.2}}, \href{https://doi.org/10.1103/PhysRevLett.118.221101}{\emph{Phys. Rev. Lett.} {\bfseries 118} (2017) 221101} [\href{https://arxiv.org/abs/1706.01812}{{\ttfamily 1706.01812}}].

\bibitem{LIGOScientific:2017vox}
{\scshape LIGO Scientific, Virgo} collaboration, \emph{{GW170608: Observation of a 19-solar-mass Binary Black Hole Coalescence}}, \href{https://doi.org/10.3847/2041-8213/aa9f0c}{\emph{Astrophys. J. Lett.} {\bfseries 851} (2017) L35} [\href{https://arxiv.org/abs/1711.05578}{{\ttfamily 1711.05578}}].

\bibitem{LIGOScientific:2017ycc}
{\scshape LIGO Scientific, Virgo} collaboration, \emph{{GW170814: A Three-Detector Observation of Gravitational Waves from a Binary Black Hole Coalescence}}, \href{https://doi.org/10.1103/PhysRevLett.119.141101}{\emph{Phys. Rev. Lett.} {\bfseries 119} (2017) 141101} [\href{https://arxiv.org/abs/1709.09660}{{\ttfamily 1709.09660}}].

\bibitem{LIGOScientific:2017vwq}
{\scshape LIGO Scientific, Virgo} collaboration, \emph{{GW170817: Observation of Gravitational Waves from a Binary Neutron Star Inspiral}}, \href{https://doi.org/10.1103/PhysRevLett.119.161101}{\emph{Phys. Rev. Lett.} {\bfseries 119} (2017) 161101} [\href{https://arxiv.org/abs/1710.05832}{{\ttfamily 1710.05832}}].

\bibitem{LIGOScientific:2014pky}
{\scshape LIGO Scientific} collaboration, \emph{{Advanced LIGO}}, \href{https://doi.org/10.1088/0264-9381/32/7/074001}{\emph{Class. Quant. Grav.} {\bfseries 32} (2015) 074001} [\href{https://arxiv.org/abs/1411.4547}{{\ttfamily 1411.4547}}].

\bibitem{LIGOScientific:2019lzm}
{\scshape LIGO Scientific, Virgo} collaboration, \emph{{Open data from the first and second observing runs of Advanced LIGO and Advanced Virgo}}, \href{https://doi.org/10.1016/j.softx.2021.100658}{\emph{SoftwareX} {\bfseries 13} (2021) 100658} [\href{https://arxiv.org/abs/1912.11716}{{\ttfamily 1912.11716}}].

\bibitem{Punturo_2010}
F.A.e.a.~M.~Punturo, M.~Abernathy, \emph{The einstein telescope: a third-generation gravitational wave observatory}, \href{https://doi.org/10.1088/0264-9381/27/19/194002}{\emph{Classical and Quantum Gravity} {\bfseries 27} (2010) 194002}.

\bibitem{Hild:2010id}
S.~Hild et~al., \emph{{Sensitivity Studies for Third-Generation Gravitational Wave Observatories}}, \href{https://doi.org/10.1088/0264-9381/28/9/094013}{\emph{Class. Quant. Grav.} {\bfseries 28} (2011) 094013} [\href{https://arxiv.org/abs/1012.0908}{{\ttfamily 1012.0908}}].

\bibitem{Reitze:2019iox}
D.~Reitze et~al., \emph{{Cosmic Explorer: The U.S. Contribution to Gravitational-Wave Astronomy beyond LIGO}}, {\emph{Bull. Am. Astron. Soc.} {\bfseries 51} (2019) 035} [\href{https://arxiv.org/abs/1907.04833}{{\ttfamily 1907.04833}}].

\bibitem{Sesana:2019vho}
A.~Sesana et~al., \emph{{Unveiling the gravitational universe at $\mu$-Hz frequencies}}, \href{https://doi.org/10.1007/s10686-021-09709-9}{\emph{Exper. Astron.} {\bfseries 51} (2021) 1333} [\href{https://arxiv.org/abs/1908.11391}{{\ttfamily 1908.11391}}].

\bibitem{amaroseoane2017laser}
P.~Amaro-Seoane, H.~Audley and S.B.~et~al., \emph{Laser interferometer space antenna},  2017.

\bibitem{Baker:2019nia}
J.~Baker et~al., \emph{{The Laser Interferometer Space Antenna: Unveiling the Millihertz Gravitational Wave Sky}},  \href{https://arxiv.org/abs/1907.06482}{{\ttfamily 1907.06482}}.

\bibitem{Corbin:2005ny}
V.~Corbin and N.J.~Cornish, \emph{{Detecting the cosmic gravitational wave background with the big bang observer}}, \href{https://doi.org/10.1088/0264-9381/23/7/014}{\emph{Class. Quant. Grav.} {\bfseries 23} (2006) 2435} [\href{https://arxiv.org/abs/gr-qc/0512039}{{\ttfamily gr-qc/0512039}}].

\bibitem{Harry_2006}
G.M.~Harry, P.~Fritschel, D.A.~Shaddock, W.~Folkner and E.S.~Phinney, \emph{Laser interferometry for the big bang observer}, \href{https://doi.org/10.1088/0264-9381/23/15/008}{\emph{Classical and Quantum Gravity} {\bfseries 23} (2006) 4887}.

\bibitem{Yagi:2011yu}
K.~Yagi, N.~Tanahashi and T.~Tanaka, \emph{{Probing the size of extra dimension with gravitational wave astronomy}}, \href{https://doi.org/10.1103/PhysRevD.83.084036}{\emph{Phys. Rev. D} {\bfseries 83} (2011) 084036} [\href{https://arxiv.org/abs/1101.4997}{{\ttfamily 1101.4997}}].

\bibitem{Seto:2001qf}
N.~Seto, S.~Kawamura and T.~Nakamura, \emph{{Possibility of direct measurement of the acceleration of the universe using 0.1-Hz band laser interferometer gravitational wave antenna in space}}, \href{https://doi.org/10.1103/PhysRevLett.87.221103}{\emph{Phys. Rev. Lett.} {\bfseries 87} (2001) 221103} [\href{https://arxiv.org/abs/astro-ph/0108011}{{\ttfamily astro-ph/0108011}}].

\bibitem{Kawamura_2006}
S.~Kawamura, T.~Nakamura and M.A.~et~al., \emph{The japanese space gravitational wave antenna—decigo}, \href{https://doi.org/10.1088/0264-9381/23/8/S17}{\emph{Classical and Quantum Gravity} {\bfseries 23} (2006) S125}.

\bibitem{Yagi:2011wg}
K.~Yagi and N.~Seto, \emph{{Detector configuration of DECIGO/BBO and identification of cosmological neutron-star binaries}}, \href{https://doi.org/10.1103/PhysRevD.83.044011}{\emph{Phys. Rev. D} {\bfseries 83} (2011) 044011} [\href{https://arxiv.org/abs/1101.3940}{{\ttfamily 1101.3940}}].

\bibitem{Kramer:2013kea}
{\scshape EPTA} collaboration, \emph{{The European Pulsar Timing Array and the Large European Array for Pulsars}}, \href{https://doi.org/10.1088/0264-9381/30/22/224009}{\emph{Class. Quant. Grav.} {\bfseries 30} (2013) 224009}.

\bibitem{Lentati:2015qwp}
L.~Lentati et~al., \emph{{European Pulsar Timing Array Limits On An Isotropic Stochastic Gravitational-Wave Background}}, \href{https://doi.org/10.1093/mnras/stv1538}{\emph{Mon. Not. Roy. Astron. Soc.} {\bfseries 453} (2015) 2576} [\href{https://arxiv.org/abs/1504.03692}{{\ttfamily 1504.03692}}].

\bibitem{Babak:2015lua}
S.~Babak et~al., \emph{{European Pulsar Timing Array Limits on Continuous Gravitational Waves from Individual Supermassive Black Hole Binaries}}, \href{https://doi.org/10.1093/mnras/stv2092}{\emph{Mon. Not. Roy. Astron. Soc.} {\bfseries 455} (2016) 1665} [\href{https://arxiv.org/abs/1509.02165}{{\ttfamily 1509.02165}}].

\bibitem{McLaughlin:2013ira}
M.A.~McLaughlin, \emph{{The North American Nanohertz Observatory for Gravitational Waves}}, \href{https://doi.org/10.1088/0264-9381/30/22/224008}{\emph{Class. Quant. Grav.} {\bfseries 30} (2013) 224008} [\href{https://arxiv.org/abs/1310.0758}{{\ttfamily 1310.0758}}].

\bibitem{NANOGRAV:2018hou}
{\scshape NANOGRAV} collaboration, \emph{{The NANOGrav 11-year Data Set: Pulsar-timing Constraints On The Stochastic Gravitational-wave Background}}, \href{https://doi.org/10.3847/1538-4357/aabd3b}{\emph{Astrophys. J.} {\bfseries 859} (2018) 47} [\href{https://arxiv.org/abs/1801.02617}{{\ttfamily 1801.02617}}].

\bibitem{Aggarwal:2018mgp}
K.~Aggarwal et~al., \emph{{The NANOGrav 11-Year Data Set: Limits on Gravitational Waves from Individual Supermassive Black Hole Binaries}}, \href{https://doi.org/10.3847/1538-4357/ab2236}{\emph{Astrophys. J.} {\bfseries 880} (2019) 2} [\href{https://arxiv.org/abs/1812.11585}{{\ttfamily 1812.11585}}].

\bibitem{Brazier:2019mmu}
A.~Brazier et~al., \emph{{The NANOGrav Program for Gravitational Waves and Fundamental Physics}},  \href{https://arxiv.org/abs/1908.05356}{{\ttfamily 1908.05356}}.

\bibitem{NANOGrav:2020bcs}
{\scshape NANOGrav} collaboration, \emph{{The NANOGrav 12.5 yr Data Set: Search for an Isotropic Stochastic Gravitational-wave Background}}, \href{https://doi.org/10.3847/2041-8213/abd401}{\emph{Astrophys. J. Lett.} {\bfseries 905} (2020) L34} [\href{https://arxiv.org/abs/2009.04496}{{\ttfamily 2009.04496}}].

\bibitem{Thrane:2013oya}
E.~Thrane and J.D.~Romano, \emph{{Sensitivity curves for searches for gravitational-wave backgrounds}}, \href{https://doi.org/10.1103/PhysRevD.88.124032}{\emph{Phys. Rev. D} {\bfseries 88} (2013) 124032} [\href{https://arxiv.org/abs/1310.5300}{{\ttfamily 1310.5300}}].

\bibitem{Caprini:2015zlo}
C.~Caprini et~al., \emph{{Science with the space-based interferometer eLISA. II: Gravitational waves from cosmological phase transitions}}, \href{https://doi.org/10.1088/1475-7516/2016/04/001}{\emph{JCAP} {\bfseries 04} (2016) 001} [\href{https://arxiv.org/abs/1512.06239}{{\ttfamily 1512.06239}}].

\bibitem{Gowling:2021gcy}
C.~Gowling and M.~Hindmarsh, \emph{{Observational prospects for phase transitions at LISA: Fisher matrix analysis}}, \href{https://doi.org/10.1088/1475-7516/2021/10/039}{\emph{JCAP} {\bfseries 10} (2021) 039} [\href{https://arxiv.org/abs/2106.05984}{{\ttfamily 2106.05984}}].

\bibitem{Dodelson:2003ft}
S.~Dodelson, \emph{{Modern Cosmology}}, Academic Press, Amsterdam (2003).

\bibitem{Bertone:2019irm}
G.~Bertone et~al., \emph{{Gravitational wave probes of dark matter: challenges and opportunities}}, \href{https://doi.org/10.21468/SciPostPhysCore.3.2.007}{\emph{SciPost Phys. Core} {\bfseries 3} (2020) 007} [\href{https://arxiv.org/abs/1907.10610}{{\ttfamily 1907.10610}}].

\bibitem{Miller:2025yyx}
A.L.~Miller, \emph{{Gravitational wave probes of particle dark matter: a review}},  \href{https://arxiv.org/abs/2503.02607}{{\ttfamily 2503.02607}}.

\bibitem{HESS:2021zzm}
{\scshape H.E.S.S.} collaboration, \emph{{Search for dark matter annihilation in the Wolf-Lundmark-Melotte dwarf irregular galaxy with H.E.S.S.}}, \href{https://doi.org/10.1103/PhysRevD.103.102002}{\emph{Phys. Rev. D} {\bfseries 103} (2021) 102002} [\href{https://arxiv.org/abs/2105.04325}{{\ttfamily 2105.04325}}].

\bibitem{HESS:2020zwn}
{\scshape H.E.S.S.} collaboration, \emph{{Search for dark matter signals towards a selection of recently detected DES dwarf galaxy satellites of the Milky Way with H.E.S.S.}}, \href{https://doi.org/10.1103/PhysRevD.102.062001}{\emph{Phys. Rev. D} {\bfseries 102} (2020) 062001} [\href{https://arxiv.org/abs/2008.00688}{{\ttfamily 2008.00688}}].

\bibitem{HESS:2015cda}
{\scshape H.E.S.S.} collaboration, \emph{{Constraints on an Annihilation Signal from a Core of Constant Dark Matter Density around the Milky Way Center with H.E.S.S.}}, \href{https://doi.org/10.1103/PhysRevLett.114.081301}{\emph{Phys. Rev. Lett.} {\bfseries 114} (2015) 081301} [\href{https://arxiv.org/abs/1502.03244}{{\ttfamily 1502.03244}}].

\bibitem{HESS:2014zqa}
{\scshape H.E.S.S.} collaboration, \emph{{Search for dark matter annihilation signatures in H.E.S.S. observations of Dwarf Spheroidal Galaxies}}, \href{https://doi.org/10.1103/PhysRevD.90.112012}{\emph{Phys. Rev. D} {\bfseries 90} (2014) 112012} [\href{https://arxiv.org/abs/1410.2589}{{\ttfamily 1410.2589}}].

\bibitem{MAGIC:2021mog}
{\scshape MAGIC} collaboration, \emph{{Combined searches for dark matter in dwarf spheroidal galaxies observed with the MAGIC telescopes, including new data from Coma Berenices and Draco}}, \href{https://doi.org/10.1016/j.dark.2021.100912}{\emph{Phys. Dark Univ.} {\bfseries 35} (2022) 100912} [\href{https://arxiv.org/abs/2111.15009}{{\ttfamily 2111.15009}}].

\bibitem{MAGIC:2020ceg}
{\scshape MAGIC} collaboration, \emph{{A search for dark matter in Triangulum II with the MAGIC telescopes}}, \href{https://doi.org/10.1016/j.dark.2020.100529}{\emph{Phys. Dark Univ.} {\bfseries 28} (2020) 100529} [\href{https://arxiv.org/abs/2003.05260}{{\ttfamily 2003.05260}}].

\bibitem{Aleksic:2013xea}
J.~Aleksi\'c et~al., \emph{{Optimized dark matter searches in deep observations of Segue 1 with MAGIC}}, \href{https://doi.org/10.1088/1475-7516/2014/02/008}{\emph{JCAP} {\bfseries 02} (2014) 008} [\href{https://arxiv.org/abs/1312.1535}{{\ttfamily 1312.1535}}].

\bibitem{HAWC:2018eaa}
{\scshape HAWC} collaboration, \emph{{Search for Dark Matter Gamma-ray Emission from the Andromeda Galaxy with the High-Altitude Water Cherenkov Observatory}}, \href{https://doi.org/10.1088/1475-7516/2018/06/043}{\emph{JCAP} {\bfseries 06} (2018) 043} [\href{https://arxiv.org/abs/1804.00628}{{\ttfamily 1804.00628}}].

\bibitem{HAWC:2017mfa}
{\scshape HAWC} collaboration, \emph{{Dark Matter Limits From Dwarf Spheroidal Galaxies with The HAWC Gamma-Ray Observatory}}, \href{https://doi.org/10.3847/1538-4357/aaa6d8}{\emph{Astrophys. J.} {\bfseries 853} (2018) 154} [\href{https://arxiv.org/abs/1706.01277}{{\ttfamily 1706.01277}}].

\bibitem{DAMPE:2021hsz}
{\scshape DAMPE} collaboration, \emph{{Search for gamma-ray spectral lines with the DArk Matter Particle Explorer}}, \href{https://doi.org/10.1016/j.scib.2021.12.015}{\emph{Sci. Bull.} {\bfseries 67} (2022) 679} [\href{https://arxiv.org/abs/2112.08860}{{\ttfamily 2112.08860}}].

\bibitem{Thorpe-Morgan:2020czg}
C.~Thorpe-Morgan, D.~Malyshev, C.-A.~Stegen, A.~Santangelo and J.~Jochum, \emph{{Annihilating dark matter search with 12 yr of Fermi LAT data in nearby galaxy clusters}}, \href{https://doi.org/10.1093/mnras/stab208}{\emph{Mon. Not. Roy. Astron. Soc.} {\bfseries 502} (2021) 4039} [\href{https://arxiv.org/abs/2010.11006}{{\ttfamily 2010.11006}}].

\bibitem{Hoof:2018hyn}
S.~Hoof, A.~Geringer-Sameth and R.~Trotta, \emph{{A Global Analysis of Dark Matter Signals from 27 Dwarf Spheroidal Galaxies using 11 Years of Fermi-LAT Observations}}, \href{https://doi.org/10.1088/1475-7516/2020/02/012}{\emph{JCAP} {\bfseries 02} (2020) 012} [\href{https://arxiv.org/abs/1812.06986}{{\ttfamily 1812.06986}}].

\bibitem{Fermi-LAT:2016uux}
{\scshape Fermi-LAT, DES} collaboration, \emph{{Searching for Dark Matter Annihilation in Recently Discovered Milky Way Satellites with Fermi-LAT}}, \href{https://doi.org/10.3847/1538-4357/834/2/110}{\emph{Astrophys. J.} {\bfseries 834} (2017) 110} [\href{https://arxiv.org/abs/1611.03184}{{\ttfamily 1611.03184}}].

\bibitem{Fermi-LAT:2015xij}
{\scshape Fermi-LAT} collaboration, \emph{{Search for extended gamma-ray emission from the Virgo galaxy cluster with Fermi-LAT}}, \href{https://doi.org/10.1088/0004-637X/812/2/159}{\emph{Astrophys. J.} {\bfseries 812} (2015) 159} [\href{https://arxiv.org/abs/1510.00004}{{\ttfamily 1510.00004}}].

\bibitem{Fermi-LAT:2015att}
{\scshape Fermi-LAT} collaboration, \emph{{Searching for Dark Matter Annihilation from Milky Way Dwarf Spheroidal Galaxies with Six Years of Fermi Large Area Telescope Data}}, \href{https://doi.org/10.1103/PhysRevLett.115.231301}{\emph{Phys. Rev. Lett.} {\bfseries 115} (2015) 231301} [\href{https://arxiv.org/abs/1503.02641}{{\ttfamily 1503.02641}}].

\bibitem{CherenkovTelescopeArray:2023aqu}
{\scshape Cherenkov Telescope Array} collaboration, \emph{{Sensitivity of the Cherenkov Telescope Array to TeV photon emission from the Large Magellanic Cloud}}, \href{https://doi.org/10.1093/mnras/stad1576}{\emph{Mon. Not. Roy. Astron. Soc.} {\bfseries 523} (2023) 5353} [\href{https://arxiv.org/abs/2305.16707}{{\ttfamily 2305.16707}}].

\bibitem{ANTARES:2022aoa}
{\scshape ANTARES} collaboration, \emph{{Search for secluded dark matter towards the Galactic Centre with the ANTARES neutrino telescope}}, \href{https://doi.org/10.1088/1475-7516/2022/06/028}{\emph{JCAP} {\bfseries 06} (2022) 028} [\href{https://arxiv.org/abs/2203.06029}{{\ttfamily 2203.06029}}].

\bibitem{Albert:2016emp}
A.~Albert et~al., \emph{{Results from the search for dark matter in the Milky Way with 9 years of data of the ANTARES neutrino telescope}}, \href{https://doi.org/10.1016/j.physletb.2017.03.063}{\emph{Phys. Lett. B} {\bfseries 769} (2017) 249} [\href{https://arxiv.org/abs/1612.04595}{{\ttfamily 1612.04595}}].

\bibitem{KM3NeT:2024xca}
{\scshape KM3NeT} collaboration, \emph{{First searches for dark matter with the KM3NeT neutrino telescopes}}, \href{https://doi.org/10.1088/1475-7516/2025/03/058}{\emph{JCAP} {\bfseries 03} (2025) 058} [\href{https://arxiv.org/abs/2411.10092}{{\ttfamily 2411.10092}}].

\bibitem{2023PhRvD.108j2004A}
{IceCube Collaboration}, \emph{{Search for neutrino lines from dark matter annihilation and decay with IceCube}}, \href{https://doi.org/10.1103/PhysRevD.108.102004}{\emph{Phys. Rev. D} {\bfseries 108} (2023) 102004} [\href{https://arxiv.org/abs/2303.13663}{{\ttfamily 2303.13663}}].

\bibitem{2017EPJC...77..627A}
{IceCube Collaboration}, \emph{{Search for neutrinos from dark matter self-annihilations in the center of the Milky Way with 3 years of IceCube/DeepCore: IceCube Collaboration}}, \href{https://doi.org/10.1140/epjc/s10052-017-5213-y}{\emph{Eur. Phys. J. C} {\bfseries 77} (2017) 627} [\href{https://arxiv.org/abs/1705.08103}{{\ttfamily 1705.08103}}].

\bibitem{Super-Kamiokande:2020sgt}
{\scshape Super-Kamiokande} collaboration, \emph{{Indirect search for dark matter from the Galactic Center and halo with the Super-Kamiokande detector}}, \href{https://doi.org/10.1103/PhysRevD.102.072002}{\emph{Phys. Rev. D} {\bfseries 102} (2020) 072002} [\href{https://arxiv.org/abs/2005.05109}{{\ttfamily 2005.05109}}].

\bibitem{Frankiewicz:2015zma}
{\scshape Super-Kamiokande} collaboration, \emph{{Searching for Dark Matter Annihilation into Neutrinos with Super-Kamiokande}},  in \emph{{Meeting of the APS Division of Particles and Fields}}, 10, 2015 [\href{https://arxiv.org/abs/1510.07999}{{\ttfamily 1510.07999}}].

\bibitem{Dasgupta:2022isg}
A.~Dasgupta, P.S.B.~Dev, A.~Ghoshal and A.~Mazumdar, \emph{{Gravitational wave pathway to testable leptogenesis}}, \href{https://doi.org/10.1103/PhysRevD.106.075027}{\emph{Phys. Rev. D} {\bfseries 106} (2022) 075027} [\href{https://arxiv.org/abs/2206.07032}{{\ttfamily 2206.07032}}].

\bibitem{Bhaumik:2022pil}
N.~Bhaumik, A.~Ghoshal and M.~Lewicki, \emph{{Doubly peaked induced stochastic gravitational wave background: testing baryogenesis from primordial black holes}}, \href{https://doi.org/10.1007/JHEP07(2022)130}{\emph{JHEP} {\bfseries 07} (2022) 130} [\href{https://arxiv.org/abs/2205.06260}{{\ttfamily 2205.06260}}].

\bibitem{Barman:2022yos}
B.~Barman, D.~Borah, A.~Dasgupta and A.~Ghoshal, \emph{{Probing high scale Dirac leptogenesis via gravitational waves from domain walls}}, \href{https://doi.org/10.1103/PhysRevD.106.015007}{\emph{Phys. Rev. D} {\bfseries 106} (2022) 015007} [\href{https://arxiv.org/abs/2205.03422}{{\ttfamily 2205.03422}}].

\bibitem{Ghoshal:2022jdt}
A.~Ghoshal and P.~Saha, \emph{{Detectable gravitational waves from preheating probes nonthermal dark matter}}, \href{https://doi.org/10.1103/PhysRevD.109.023526}{\emph{Phys. Rev. D} {\bfseries 109} (2024) 023526} [\href{https://arxiv.org/abs/2203.14424}{{\ttfamily 2203.14424}}].

\bibitem{Dunsky:2021tih}
D.I.~Dunsky, A.~Ghoshal, H.~Murayama, Y.~Sakakihara and G.~White, \emph{{GUTs, hybrid topological defects, and gravitational waves}}, \href{https://doi.org/10.1103/PhysRevD.106.075030}{\emph{Phys. Rev. D} {\bfseries 106} (2022) 075030} [\href{https://arxiv.org/abs/2111.08750}{{\ttfamily 2111.08750}}].

\bibitem{Ghoshal:2020vud}
A.~Ghoshal and A.~Salvio, \emph{{Gravitational waves from fundamental axion dynamics}}, \href{https://doi.org/10.1007/JHEP12(2020)049}{\emph{JHEP} {\bfseries 12} (2020) 049} [\href{https://arxiv.org/abs/2007.00005}{{\ttfamily 2007.00005}}].

\bibitem{Chowdhury:2022pnv}
D.~Chowdhury, G.~Tasinato and I.~Zavala, \emph{{Response of the Einstein Telescope to Doppler anisotropies}}, \href{https://doi.org/10.1103/PhysRevD.107.083516}{\emph{Phys. Rev. D} {\bfseries 107} (2023) 083516} [\href{https://arxiv.org/abs/2209.05770}{{\ttfamily 2209.05770}}].

\end{thebibliography}\endgroup
\end{document}